\documentclass[a4paper,11pt]{article}
\pdfoutput=1 

\usepackage{jheppub} 

\usepackage[T1]{fontenc} 
\usepackage{graphicx}
\usepackage{lipsum}
\usepackage{amsmath}
\usepackage{bm}
\usepackage[normalem]{ulem}
\usepackage{slashed}
\usepackage{epsfig}
\usepackage{epstopdf}
\usepackage{amsfonts}
\usepackage{subfigure}
\usepackage{xcolor}
\usepackage[outermarks]{titlesec}

\usepackage{bbding}
\usepackage{pifont}
\usepackage{wasysym}
\usepackage{amssymb}

\DeclareMathAlphabet\bfcal{OMS}{cmsy}{b}{n}

\newcommand{\nslash}{n\!\!\!/}
\newcommand{\nbslash}{{\bar n}\!\!\!/}
\newcommand{\lpslash}{{l \!\!/}_\perp}
\newcommand{\kpslash}{{k \!\!\!/}_\perp}
\newcommand{\pjslash}{{p \!\!\!/}_{j\perp}}
\newcommand{\pqslash}{{p \!\!\!/}_{q\perp}}
\newcommand{\ppslash}{{p \!\!\!/}_{\perp}}

\newcommand{\nbar}{{\bar n}}

\newcommand{\beq}{\begin{equation}}
\newcommand{\eeq}{\end{equation}}
\newcommand{\bea}{\begin{eqnarray}}
\newcommand{\eea}{\end{eqnarray}}

\newcommand{\nn}{\nonumber}

\newcommand{\beal}{\begin{align}}
\newcommand{\eeal}{\end{align}}
\newcommand{\bspl}{\begin{split}}
\newcommand{\espl}{\end{split}}
\def\cm{{\cal M}}

\newcommand\M[2]{\left|{\cal{M}}^{#1}_{#2}\right|^2}
\def\ket#1{|{#1}\rangle}
\def\bra#1{\langle{#1}|}

\newcommand{\Graph}[2]{\vcenter{\hbox{\includegraphics[scale=#1]{#2}}}}

\title{Single Inclusive Jet Production in $pA$ Collisions at NLO in the small-$x$ regime}

\author[a,b]{Hao-yu Liu,}
\author[c]{Kexin Xie,}
\author[d,e,f]{Zhong-Bo Kang,} 
\author[a,b,g]{and Xiaohui Liu}

\affiliation[a]{Center of Advanced Quantum Studies, Beijing Normal University,Beijing, 100875, China}
\affiliation[b]{Department of Physics, Beijing Normal University, Beijing, 100875, China}
\affiliation[c]{College of Arts \& Sciences, 
Beijing Normal University, Zhuhai, Guangdong, 519087, China}
\affiliation[d]{Department of Physics and Astronomy, University of California, Los Angeles, CA 90095, USA}
\affiliation[e]{Mani L. Bhaumik Institute for Theoretical Physics, University of California, Los Angeles, CA 90095, USA}
\affiliation[f]{Center for Frontiers in Nuclear Science, Stony Brook University, Stony Brook, NY 11794, USA}
\affiliation[g]{Center for High Energy Physics, Peking University, Beijing 100871, China}

 \emailAdd{zkang@ucla.edu}
 \emailAdd{xiliu@bnu.edu.cn}

\abstract{We present the first complete next-to-leading-order (NLO) prediction with full jet algorithm implementation for the single inclusive jet production in $pA$ collisions at forward rapidities within the color glass condensate (CGC) effective theory. Our prediction is fully differential over the final state physical kinematics, which allows the implementation of any infra-red safe observable including the jet clustering procedure. The NLO calculation is 
organized with the aid of the observable originated power counting proposed in~\cite{Kang:2019ysm} which gives rise to the novel soft contributions in the CGC factorization. We achieve the fully-differential calculation by constructing suitable subtraction terms to handle the singularities in the real corrections. The subtraction contributions can be exactly integrated analytically. 
We present the NLO cross section with the jets constructed using the anti-$k_T$ algorithm. 
The NLO calculation demonstrates explicitly the validity of the CGC factorization in jet production. Furthermore, as a byproduct of the subtraction method, we also derive the fully analytic cross section for the forward jet production in the small-$R$ limit. We show that in the small-$R$ approximation, the forward jet cross section can be factorized into a semi-hard 
cross section that produces a parton and the semi-inclusive jet functions (siJFs).
We argue that this feature holds for generic jet production and jet substructure observables in the CGC framework. Last, we show numerical analyses of the derived formula to validate our calculations. We justify when the small-$R$ approximation is appropriate. Like forward hadron production, the obtained NLO result also exhibits the negativity of the cross section in the large jet transverse regime, which 
signals the need for the threshold resummation. A sketch of the threshold resummation in the CGC framework is presented based on the multiple emission picture and it is found to agree with the approach using the rapidity renormalization group equation developed in~\cite{Liu:2020mpy}.
}

\begin{document}
\maketitle
\flushbottom


\section{Introduction}
Hard probes are among the most fundamental tools to understand the internal structure of hadrons. 
The partonic fields bound within the hadron can fluctuate and radiate virtual quanta that could live for a short time period, therefore probes such as (virtual) photons, will snapshot the hadron interior at different time resolution scales~\cite{Accardi:2012qut,AbdulKhalek:2021gbh}. In the high-energy limit, the probe gets dramatically boosted, revealing smaller and smaller values of the Bjorken-$x$ and observing a substantially growing gluon density. In the sufficiently small-$x$ region, the gluons are so close to each other, initiating the recombination in which two gluons annihilate into one. In the small-$x$ domain, the gluon recombination is equally important to the splitting process and the evolution of the partonic fields enters the regime where the nonlinear B-JIMWLK equation (or its infinite color approximation, the BK equation)~\cite{Mueller:1989st,Mueller:1993rr,Balitsky:1995ub, Kovchegov:1999yj, Jalilian-Marian:1997qno, Jalilian-Marian:1997jhx, Iancu:2000hn, Ferreiro:2001qy} takes over the linear parton evolution equations~\cite{Balitsky:1978ic}. The non-linear evolution predicts gluon saturation~\cite{Gribov:1983ivg,Mueller:1985wy} as a consequence of the balance between recombination and gluon bremsstrahlung. The proper framework to describe the small-$x$ dynamics is the color glass condensate (CGC) effective theory~\cite{McLerran:1993ni,McLerran:1993ka,McLerran:1994vd,Gelis:2010nm}, in which the gluon saturation is featured by the saturation scale $Q_s$ which grows as $x$ decreases. When $Q_s \gg \Lambda_{\rm QCD}$, perturbative QCD can be applied. Tremendous efforts have been devoted in searching for the emergent phenomena of the gluon saturation in the ultra-dense regime~\cite{Albacete:2014fwa,Blaizot:2016qgz,Morreale:2021pnn}. 
There exist experimental hints~\cite{Golec-Biernat:1998zce,BRAHMS:2004xry,PHENIX:2004nzn,STAR:2006dgg,PHENIX:2005veb,Braidot:2010ig,ALICE:2018vhm} that are compatible with saturation-model predictions, but we still lack deterministic evidence to claim a discovery. In the future, dedicated measurements at the Electron-Ion
Collider (EIC) will provide further insight on the regime of gluon saturation~\cite{Boer:2011fh,Accardi:2012qut,Proceedings:2020eah,AbdulKhalek:2021gbh,Aschenauer:2017jsk}.

One piece of the hints of the gluon saturation comes from the suppression of the nuclear modification factor $R_{dAu}$~\cite{BRAHMS:2004xry,STAR:2006dgg} in single forward hadron production in deuteron-nucleus collisions, $d \,+A\to h( y \, , p_\perp) +X$~\cite{Kovchegov:1998bi, Dumitru:2002qt, Dumitru:2005kb, Albacete:2010bs, Levin:2010dw, Fujii:2011fh, Albacete:2012xq, Albacete:2013ei, Altinoluk:2011qy, Chirilli:2011km, Chirilli:2012jd, Stasto:2013cha, Lappi:2013zma, vanHameren:2014lna, Stasto:2014sea, Kang:2014lha,Altinoluk:2014eka, Watanabe:2015tja, Stasto:2016wrf, Iancu:2016vyg, Ducloue:2016shw, Ducloue:2017dit} where tagging a hadron $h$ with intermediate transverse momentum $p_\perp$ in the forward large rapidity $y$ forces the projectile deuteron to probe the small-$x$ component of the nuclear target~\cite{Dumitru:2002qt, Chirilli:2011km, Chirilli:2012jd, Dominguez:2010xd, Dominguez:2011wm}. The theoretical prediction is made out of the hybrid factorization theorem, in which the incoming nucleon (proton or deuteron) 
is regarded as a dilute system described by the collinear parton distribution functions (PDFs) while the nuclear target is treated as a strong classical field described by the CGC effective theory. The leading-order (LO) predictions have been presented in~\cite{Blaizot:2004wu, Blaizot:2004wv, Albacete:2010bs, Levin:2010dw, Fujii:2011fh, Albacete:2012xq, Lappi:2013zma, vanHameren:2014lna, Bury:2017xwd}, and the approximations to the next-to-leading order (NLO) corrections have been studied in~\cite{Dumitru:2005gt, Altinoluk:2011qy}. The complete analytical NLO predictions were first carried out in~\cite{Chirilli:2011km, Chirilli:2012jd} and later confirmed by~\cite{Liu:2019iml,Kang:2019ysm}. The full calculation demonstrates the validity of the hybrid factorization at ${\cal O}(\alpha_s)$. Meanwhile the threshold resummation was also carried out~\cite{Liu:2020mpy,Shi:2021hwx} to resolve the negativity of the cross section in which the NLO $p_\perp$ spectrum turns negative promptly when the hadron $p_\perp$ slightly exceeds the saturation scale $Q_s$~\cite{Stasto:2014sea}. 

Despite the fact that the hadron production is conceptually simple, the extraction of the CGC nucleus distribution out of this process could be more involved due to the additional dependence on the non-perturbative fragmentation functions. On the other hand, in order to perform a global extraction of the gluon saturation phenomenon, we need other processes that could probe into the small-$x$ regime. Among the probes of the small-$x$ gluon saturation, jets have drawn tremendous attention in recent years. In comparison with hadron production, jets have the advantage that their behaviour can be predicted perturbatively and therefore one could have better theoretical control than with hadrons. Initially, two forward jets with a large rapidity separation, known as the 
Mueller-Navelet jet~\cite{Mueller:1986ey,Colferai:2010wu,Ducloue:2013hia,Ducloue:2013bva}, was introduced to probe the linear BFKL eovlution in $pp$ collisions. The diffractive dijet
production in deep inelastic scattering (DIS) was later proposed and its phenomenology has been intensively discussed in the small-$x$ physics~\cite{Nikolaev:1994cd,Bartels:1996ne,Bartels:1996tc,Boussarie:2014lxa,Boussarie:2016ogo,Boussarie:2016bkq,Boussarie:2019ero}, which has shown the feasibility in the study of the small-$x$ gluon tomography~\cite{Hatta:2016dxp}. The semi-inclusive dijet processes in DIS were also investigated in the scenario of 
extraction of the Weizs\"{a}cker-Williams gluon distribution~\cite{Dumitru:2015gaa,Dumitru:2018kuw} and the quadrupole correlator of Wilson lines at small $x$~\cite{Mantysaari:2019hkq,Boussarie:2021ybe}. 

Modern jets are usually constructed out of recursive jet clustering algorithms with radius parameter $R$, such as the anti-$k_T$ jet algorithm~\cite{Cacciari:2008gp} which is the most frequently used at the LHC and RHIC. 
However, the recursive clustering procedure imposes non-trivial constraints on the phase space and dramatically complicates the theoretical calculations. So far, no calculation with actual jet algorithm dependence is known within the CGC framework. Recently, there have been pioneered attempts to obtain the CGC jet or photon-jet processes at the NLO~\cite{Caucal:2021ent,Roy:2019hwr,Kolbe:2020tlq} using the small-cone approximation~\cite{Mukherjee:2012uz, Liu:2012sz, Ivanov:2012ms}. The main focus of these attempts is to demonstrate the infrared finiteness of the jet cross section. However, to have an apple-to-apple comparison of the CGC theory with the experimental results, it is necessary to include in the calculation the jet clustering procedure that strictly follows the experimental analyses. On the other hand, a complete NLO calculation with full jet algorithm dependence can be used to justify the reliability of the small-cone approximation to the full cross section.    

{\color{orange}{In this work, for the first time, we present the NLO calculation with the complete jet algorithm implemented. We walk through the detailed calculation of the forward inclusive jet production in $pA$ collisions to introduce our computational framework.}} The LO jet production in $pA$ collisions is already presented in~\cite{Dumitru:2002qt} and the recent phenomenological analysis based on the LO formalism is investigated in~\cite{Mantysaari:2019nnt}. Meanwhile, efforts to go beyond the LO small-$x$ eikonal approximation are also carried through in~\cite{Jalilian-Marian:2018iui,Altinoluk:2020oyd}. 
Remarkable efforts to march toward NLO predictions are
set out to compute the real corrections for dijet production~\cite{Iancu:2020mos} and LO trijets~\cite{Iancu:2018hwa} in $pA$ collisions. A missing component of these calculations is to yet take into account real jet algorithms. 
{{\color{orange}{The jet clustering procedure that decides when the partons are clustered together and when they form jets separately dramatically complicates the phase space integration compared to the hadron production. Further complications come from the experimental threshold used to select the jet events, for instance, the cuts made to the jet rapidities and transverse momenta.  
However, it is these procedures that are used experimentally to determine the jet multiplicities and to generate the distributions. Therefore, it is of utmost importance to have a NLO computational scheme to allow for a restricted jet procedure in order to perform the reliable revelation of the CGC phenomena out of the jet data.}}

We applied the recently developed computational scheme~\cite{Kang:2019ysm,Liu:2020mpy} within the CGC effective theory to compute the full NLO predictions for the single inclusive jet production in $pA$ collisions. In order to implement the full jet clustering procedure, we developed a ``subtraction scheme''~\footnote{To avoid confusion, we clarify that the subtraction scheme developed here is conceptually different from the subtraction scheme used as a solution to the negative cross section problem in~\cite{Kang:2014lha,Xiao:2014uba,Ducloue:2016shw}. Here the subtraction scheme refers to the NLO technique to isolate the poles in the phase space integrals, in which ``we subtract and we add back'' and therefore, every contribution is kept in the calculation and nothing is subtracted away.} for the NLO real corrections. We present sufficient details on how the calculation is set up. We also examine the small-jet cone approximation in CGC and build up an interesting connection to the small-jet cone approximation in collinear factorization~\cite{Kang:2016mcy,Dai:2016hzf} through the small-$R$ factorization theorem derived in this work.  

\paragraph{Outline of this manuscript} 

The manuscript is organized as the follows: 

In Section~\ref{sec:CGC-review}, we briefly review the CGC effective theory. We go through in detail the theoretical set-ups in Section~\ref{sec:theo-setup}, which cover the power counting, the soft and collinear modes as well as the Feynman rules in Section~\ref{subsec:powercounting}, the rapidity regulator in Section~\ref{subsec:rapidityreg} and the color charge operator in Section~\ref{subsec:colorcharge}. Examples 
on applying the set-ups are given in Section~\ref{subsec:examples}. 

The LO jet production results are summarized in Section~\ref{sec:LOsec}, followed by the detailed 
descriptions of the NLO calculations in Section~\ref{sec:NLO}, where we review the anti-$k_T$ jet algorithm~\cite{Cacciari:2008gp} in Section~\ref{subsec:antikt}. We show how the phase space is parameterized for the real emission in Section~\ref{subsec:kin} and discuss various kinematics limits. The subtraction strategy will be elaborated in Section~\ref{subsubsect:subtraction} and the explicit subtraction term will be presented in Section~\ref{subsubsec:coll-sub} and Section~\ref{subsec:nlosoft} for both the collinear and the soft contribution, respectively. The full NLO results are given in Section~\ref{subsec:fullNLO}. The results are fully differential and can be adapted to other experimental observables. 

In Section~\ref{sec:smallR}, we present the full analytic NLO cross section for the jet production in the small-$R$ limit. A factorized form was found in the result, and we argue that the factorization will hold for any other jet/jet-like observables in their small value limits. A sketch of the threshold resummation for the jet production is given in Section~\ref{sec:threshold}. Except for additional Sudakov logarithms to be resummed in the jet case, the resummation structure is similar to the hadron production ~\cite{Liu:2020mpy}, which we reproduced in the work using a different approach. 

We use $q\to q$ channel to highlight the calculations, all the other channels can be found in the appendix~\ref{sec:allchannels}. 
Numerical analyses are presented in Section~\ref{sec:numerics}. We conclude in Section~\ref{sec:conclusion}.

We note that all the results presented in this work are in the large $N_C$ limit while our methodology is not limited to the approximation but can be applied to more general cases with finite $N_C$.

\section{A Brief Review of the CGC effective theory}\label{sec:CGC-review}
We briefly go through the basic ideas of the color glass condensate (CGC) effective theory in the section. We refer the readers to Ref.~\cite{Iancu:2003xm,Gelis:2010nm,Kovchegov:2012mbw} for comprehensive reviews.

\subsection{Basic Picture}
In high energy collisions, when nuclei get boosted, the phase space for bremsstrahlung radiations grows. 
The radiations generate more gluons with small longitudinal momentum fraction $x$. As we keep on probing smaller values of the $x$, the gluon density will continue growing, until it reaches a sufficiently small $x$ region, where the gluon density is so high that the wave functions of $2$ gluons start to overlap and the $2$ gluons can recombine into one. The gluon bremsstrahlung and the recombination will eventually balance with each other to reach equilibrium, which is known as the gluon saturation. The gluon saturation happens typically at the saturation scale $Q_s$, which features the order of the gluon transverse momentum within the nucleus. The saturation scale $Q_s$ grows as $x\to 0$.
 \begin{figure}[htbp]
	\begin{center}
		\includegraphics*[scale=0.6]{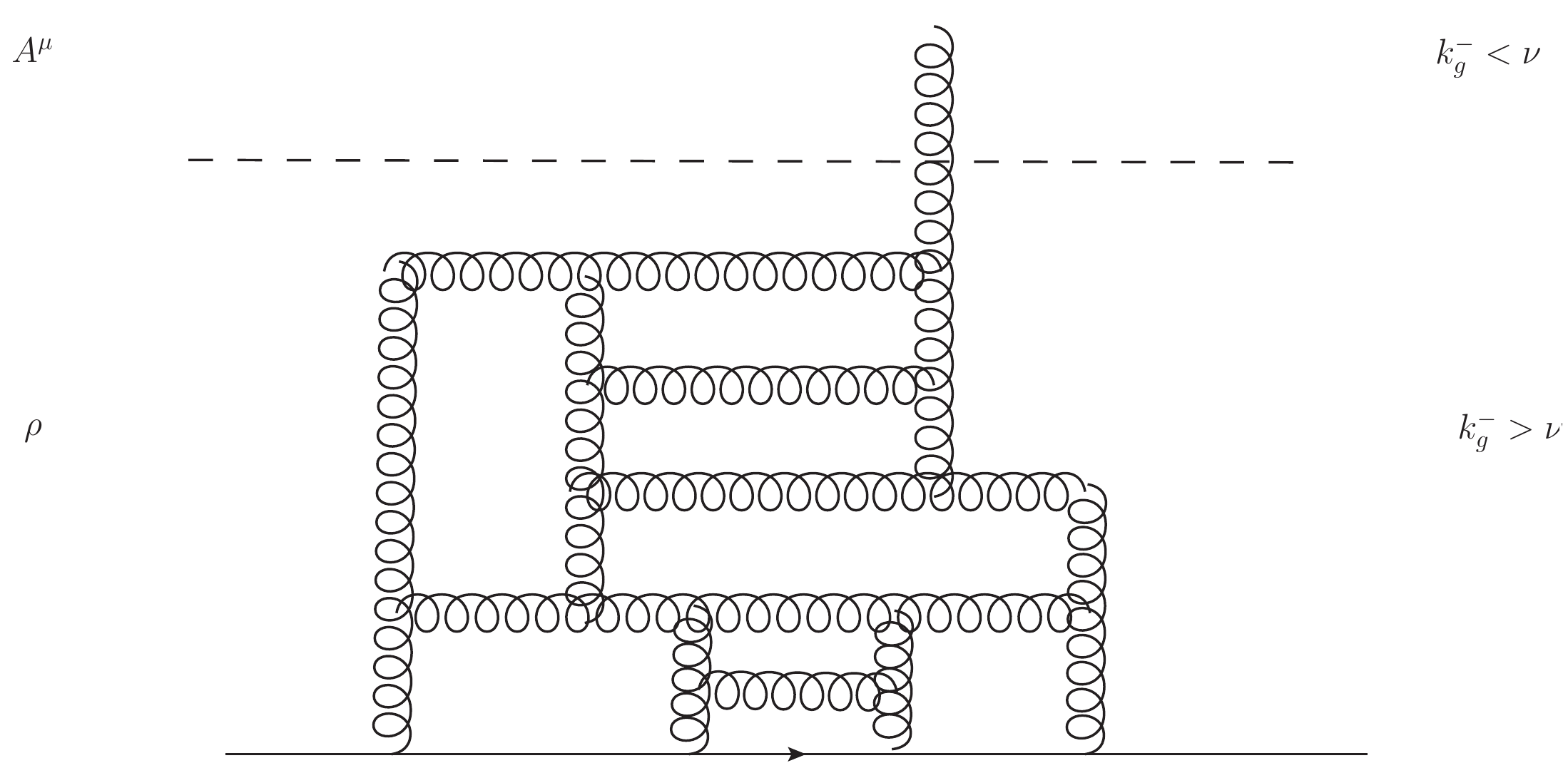}
		\caption{A small-$x$ gluon with $k_g^- < \nu$ enters the semi-hard interaction (above the dashed line), originated from the radiation and recombination of a large-$x$ parton (as the solid line) from the nucleus. }
		\label{fig:cgcpic}
	\end{center}
\end{figure}

When $Q_s \gg \Lambda_{\rm QCD}$, the small-$x$ constituents of the nucleus could be studied perturbatively, while the partons with large momentum fraction remain non-perturbative. The CGC effective theory is able to disentangle the perturbative physics and the non-perturbative effects, by introducing a cut-off scale $\nu$ (or $\Lambda$ as in many CGC literature~\cite{Gelis:2010nm}) in the longitudinal momentum $k_g^-$ along the nuclei propagating direction~\footnote{We assume the nucleus are moving along the direction $\nbar=(1,0,0,-1)$ with negative $z$-component momentum, and we define $k_g^- \equiv n \cdot k_g=  k_g^0 - k_g^3$, with the light-like vector $n = (1,0,0,1)$. Note that for the gluon from the nucleus $k_g^3<0$.}, to separate the small ($k_g^- < \nu $) and the large $x$ ($k_g^{-} > \nu$) degrees of freedom. During a collision, the large $x$ partons at the scale $\Lambda_{\rm QCD}$ will not participate directly the semi-hard interaction at ${\cal O}(Q_s)$, but instead they will gradually reduce the $x$ and elevate the scale through the radiation and recombination. As $k_g^- < \nu$ and the scale reaches $Q_s$, the gluon enters the semi-hard interaction, as illustrated in fig.~\ref{fig:cgcpic}. The CGC integrates out those gluon fields with $k_g^- > \nu$, left with a random color source remainder for the active small-$x$ gluons that take part in the semi-hard interaction. One can also defines the cut-off momentum fraction $X_f$ for the gluon in the nucleus through $\nu = X_f p_A$ where $p_A$ is the momentum of the highly boosted nucleus.

\subsection{The CGC Wilson Line (the Interaction with the Shock Wave)}

In a real collision, instead of scattering just once with the small-$x$ gluon as in fig.~\ref{fig:cgcpic}, a probe parton (from the proton for instance) will experience multi-interactions coherently with the CGC small-$x$ fields. One thus needs to include the multiple scattering effects, which in the CGC, is written as a sum of the small-$x$ gluons
\bea 
W=W_0+W_1+W_2+W_3+\cdots\,,
\eea 
as shown in fig.~\ref{fig:cgcmult}. 
\begin{figure}[htbp]
	\begin{center}
		\includegraphics*[scale=0.4]{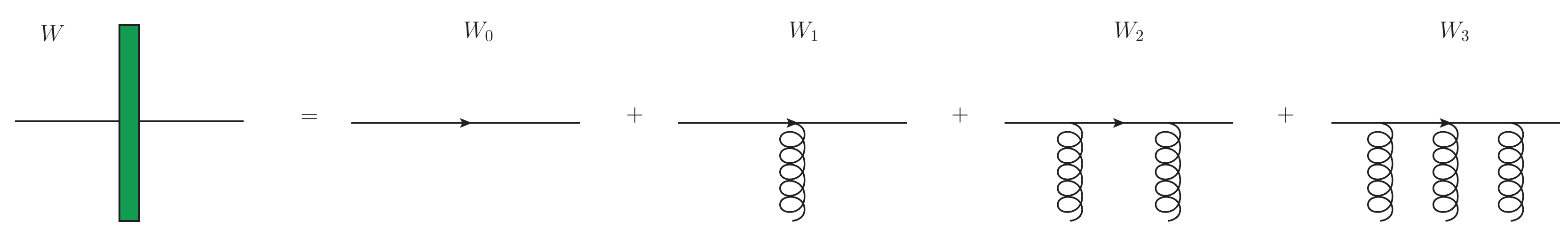}
		\caption{This figure shows the multi-interaction between the parton and the CGC field}
		\label{fig:cgcmult}
	\end{center}
\end{figure}
The summation of the small-$x$ gluon fields gives rise to the CGC Wilson line, which represents the interaction between the projectile parton and the nucleus field {\it shock wave}. 
\bea \label{eq:wilsonq}
	 W_{\alpha\beta}(x_\perp)=\mathcal{P}\exp\left\{ig\int_{-\infty}^{+\infty}dx^+T_{\alpha\beta}^cA_c^-(x^+,\bm{x}_\perp)\right\},
\eea 
where ${\cal P}$ denotes the path ordering. $A_c^-(x^+,\bm{x}_\perp)$ is the small-$x$ gluonic field with color $c$. The color generator $T_{\alpha\beta}^c$ encodes the color message of a single interaction between the probe and the gluon. For a ferminoic probe, $T_{\alpha\beta}^c = t_{\alpha\beta}^c$ the fundamental representation of the $SU(N_C)$ group, while $T_{\alpha\beta}^c = if^{\alpha c \beta}$ the adjoint representation, if a gluon acts as the probe. The relation between the fundamental and adjoint color generators suggests the following connection between the Wilson lines in different representations,
\bea
W_{ab}(\bm{x}_\perp)=2{\rm Tr}[t^a W(\bm{x}_\perp)t^b W^\dagger(\bm{x}_\perp)]\,.
\eea

\section{Theoretical Set-ups}\label{sec:theo-setup}
In this section, we introduce the theoretical framework we developed for the NLO calculation and the resummation. 
\subsection{Light-cone Coordinate}
We work in the frame where the proton is moving forward along the $+z$-axis while the nucleus in the $-z$ direction. It is then natural to introduce the light-cone coordinates $n^\mu = (1,0,0,1)$ and $\nbar^\mu = (1,0,0,-1)$, which satisfy $n\cdot \nbar = 2$. Any four vector $p^\mu$ can be decomposed as 
\bea 
p^\mu = p^+ \frac{n^\mu}{2} + p^- \frac{\nbar}{2} + p_{\perp }^\mu \,, 
\eea 
known as the Sudakov decomposition, where
\bea 
p^+ = \nbar \cdot p = p^0 + p^3 \,, \quad \quad 
p^- = n \cdot p = p^0 - p^3 \,, 
\eea 
and $p_\perp^\mu = (0,p^1,p^2,0)$ is the transverse momentum with respect to the beam. In light-cone coordinates, we may also denote a vector $p^\mu$ by its light-cone components as $p^\mu = (p^+,p^-,p^\mu_\perp)$. 
\subsection{Power counting, Modes, and Feynman Rules}\label{subsec:powercounting}

In this work, we are interested in forward inclusive jet production in proton-nucleus collisions, $p(p_P)A \to J(p_{J})X$, as illustrated in fig.~\ref{fig:pA-illustrate}. We assume the jet $J$ is constructed by the anti-$k_T$ jet algorithm~\cite{Cacciari:2008gp}, but {\color{orange}{we emphasize that our calculation is fully differential and is applicable to any jet algorithms.}} 
\begin{figure}[htbp]
	\begin{center}
		\includegraphics*[width=0.7\textwidth]{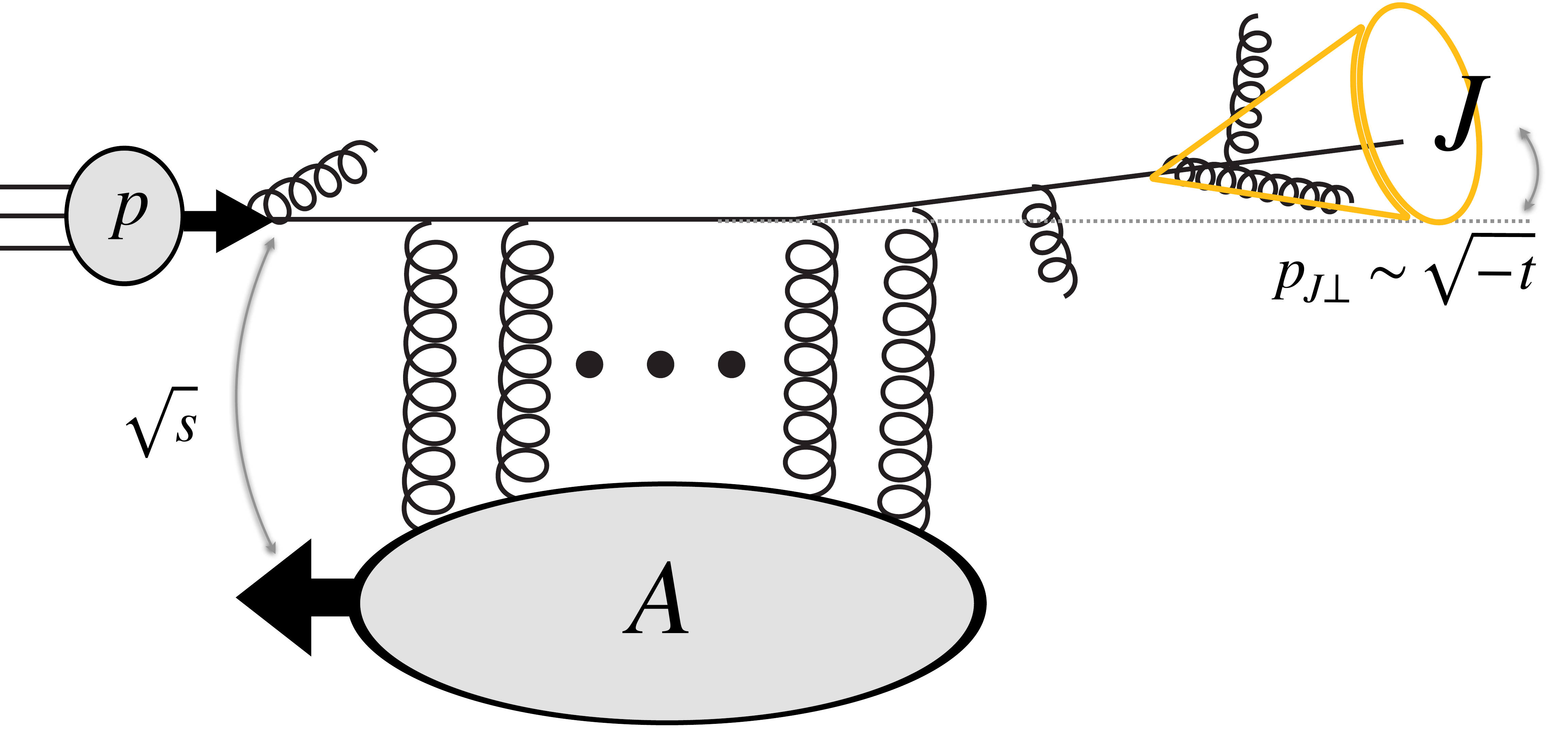}
		\caption{An illustration of the $pA \to JX$ process. }
		\label{fig:pA-illustrate}
	\end{center}
\end{figure}
%


Given that the proton is moving along the direction $n^\mu = (1,0,0,1)$ while the nucleus along $\nbar^\mu = (1,0,0,-1)$, the forward scattering is then defined by the momentum hierarchy 
\bea 
\lambda \equiv \frac{p_{J\perp}}{\nbar\cdot p_i} 
\sim e^{-|y_J|} \sim \sqrt{ - t/s} \ll 1  \,, 
\,
\eea 
where we have introduce the power counting parameter $\lambda$ and $p_i$ is the momentum of the collinear parton coming out of the proton and $p_{J\perp}$ the transverse momentum of the jet with respect to the beam. 
We measure the jet rapidity $y_J$ and transverse momentum $p_{J\perp}$ to select the forward jet events where $y_J$ is very large. We note that to probe the gluon saturation, $p_{J\perp} \sim Q_s$.

The existing hierarchy $\lambda \ll 1$ in the observables and the kinematics yields several different modes with different momentum scaling that simultaneously contribute to the leading region of the cross section~\footnote{Rigorously speaking, there is also the anti-collinear mode scaling as $({\nbar}\cdot p,n\cdot p,p_\perp) \sim n\cdot p_A(\lambda^2,1,\lambda)$ for the highly boosted particles moving along the nucleus direction. The power 
expansion in the limit $\lambda \to 0$ is equivalent to the infinite boost approximation. In our calculation, this mode is described directly using the CGC effective theory.}${}^,$~\footnote{In the jet production, the jet radius $R$ can introduce another scale hierarchy if $R\ll 1$. This leads to the small-$R$ limit of the jet production and will be discussed in section~\ref{sec:smallR}.} 
 \begin{itemize}  
\item  the collinear mode whose momentum scales as $k_c = (k_c^+ , k_c^-, k_{c\perp}) \sim  \nbar\cdot p_i (1, \lambda^2, \lambda)$, and  
\item the soft mode with 
$k_s \sim   \nbar\cdot p_i (\lambda,\lambda,\lambda)$. 
\end{itemize} 
Both modes will couple to the CGC Wilson line made up of the 
\begin{itemize} 
\item glauber gluons with 
$k_G \sim  \nbar\cdot p_i (0,0,\lambda)$ 
coming from the nucleus, which provide the transverse kick through the potential 
$1/k_G^2 \sim - 1/k_{G,\perp}^2$. 
 \end{itemize}
  We note that the virtuality of all the modes 
 are ${\cal O}(p_{J\perp})\sim {\cal O}(Q_s)$ and they all contribute at leading power in $\lambda$. The collinear and the CGC Wilson line exist in all known CGC calculations, while in CGC, the soft mode was first identified in~\cite{Kang:2019ysm}, which is found crucial to correctly produce the rapidity poles to cancel with the nucleus distribution and to automatically and systematically give rise to the kinematic constraint~\cite{Kang:2019ysm} which was introduced by hand in~\cite{Watanabe:2015tja}.  The soft mode compensates the relieved phase space restrictions due to the power expansion in $\lambda$. Its appearance should not be too surprising given that it already arises and plays an important role under similar circumstances such as the transverse momentum dependent (TMD) studies~\cite{Collins:2011zzd}. The necessity of the soft mode in the perturbative calculation can be seen as a direct consequence of the method of regions (or strategy of regions) technique~\cite{Beneke:1997zp}.

The collinear and the soft modes are governed by their own Lagrangian which is obtained by appropriate power expansion of the QCD Lagrangian in $\lambda$. The procedure is exactly how the soft collinear effective theory (SCET) Lagrangian~\cite{Bauer:2000ew, Bauer:2000yr, Bauer:2001ct,Bauer:2001yt,Beneke:2002ph} is constructed. The difference is that now we need to include in the Lagrangian the interaction between the CGC shock wave and the collinear and the soft fields through the CGC Wilson line. The detailed derivation of the Lagrangian and its comparison with the conventional CGC approach will be given in another work~\cite{kang:future1}. Here we go through the Feynman rules that will be directly used in our calculation to get the NLO jet cross section. Throughout the work, we stick to the light-cone gauge in which $\nbar\cdot A = A^+ = 0$. 

The collinear Feynman rules are given by 
\bea\label{eq:quarkprop}
\Graph{0.3}{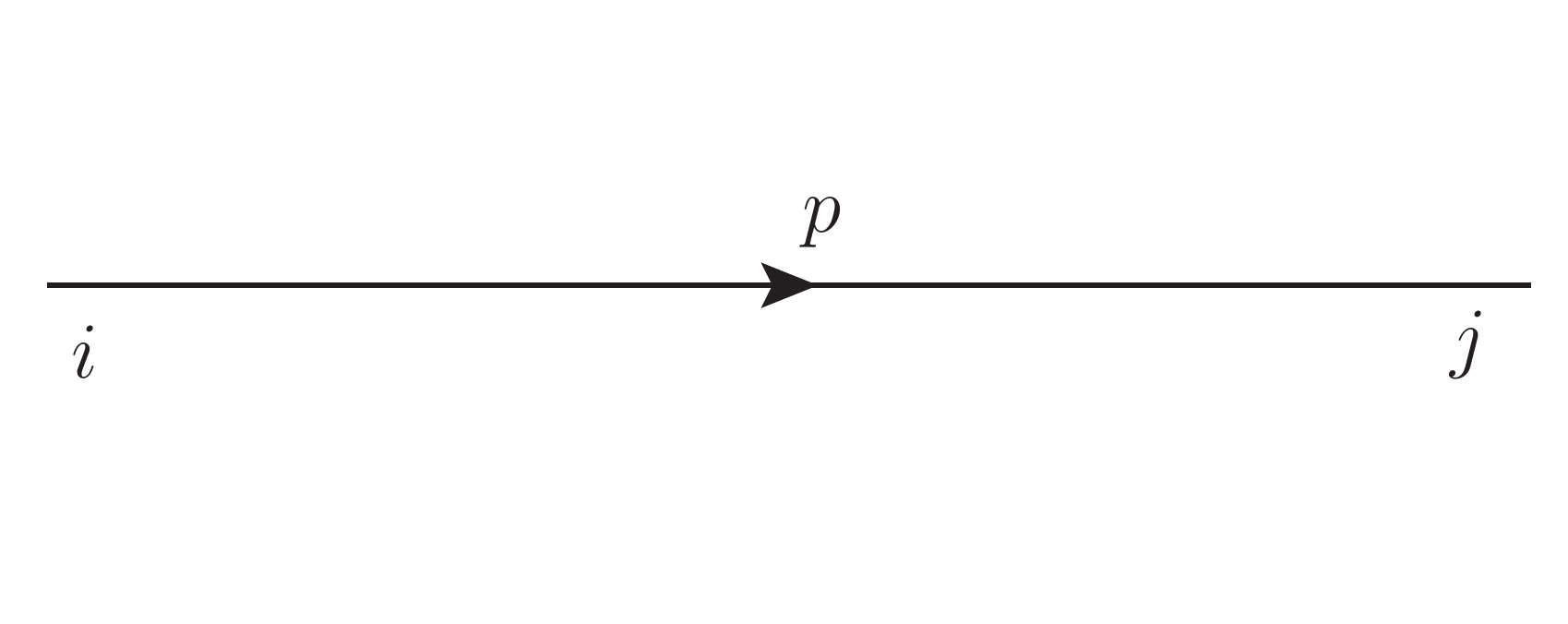}  \, 
=  \, \delta_{ij}  \, \frac{i \, p^+  }{p^2+i\epsilon} \frac{\nslash}{2} \,, 
\eea 
for the collinear quark propagator moving along the $n$ direction, where the power expansion has been performed and only the leading contribution in $\lambda$ is kept~\cite{Bauer:2000yr}. Here $i$ ($j$) is the color index and $p$ is the momentum carried by the quark.

The collinear gluon is represented by a spring line with a straight line going through
\bea\label{eq:gluonprop}
\Graph{0.3}{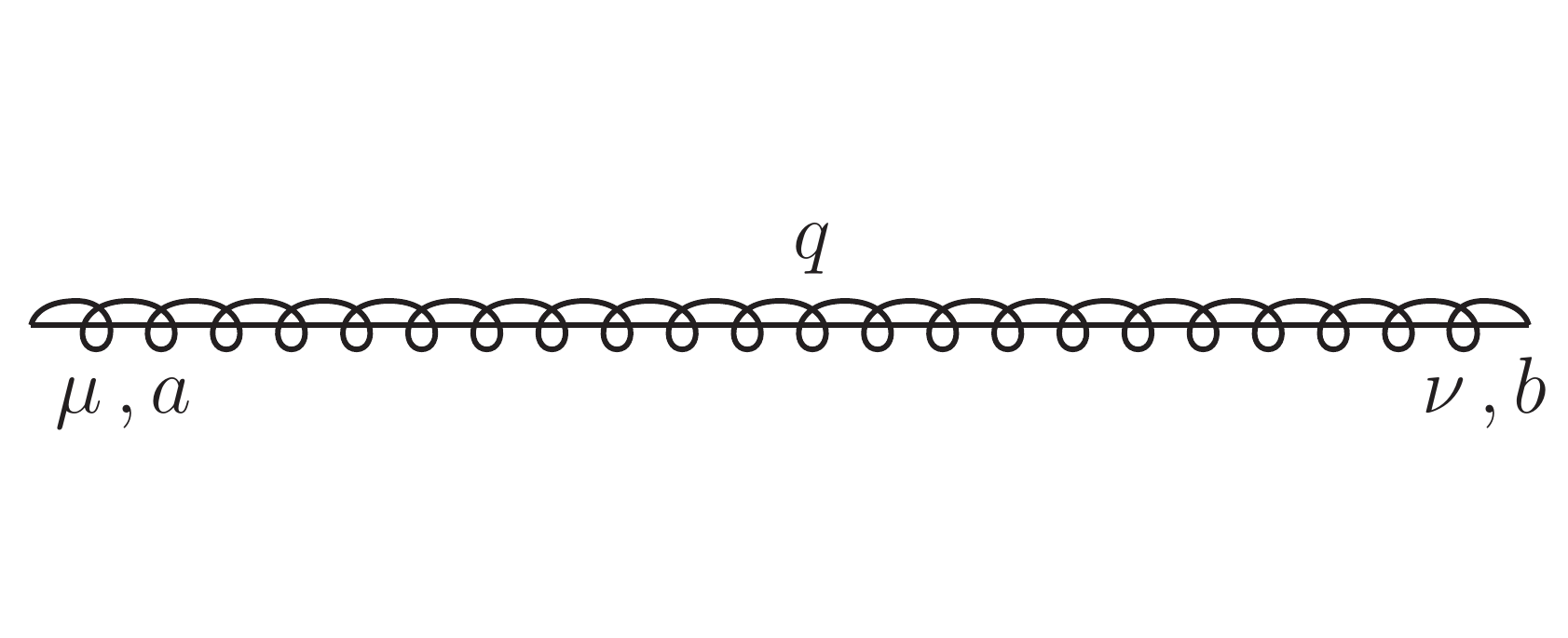}  \, 
= 
\frac{ i \, \delta_{ab}  }{q^2+i\epsilon}  
\left( -
g_{\mu\nu} + \frac{ \nbar_\mu q_\nu + q_\mu \nbar_\nu}{ q^+ } 
\right)
\,, 
\eea
where $a$ ($b$) is the color index and $q$ is the gluon momentum, and the form of the gluon polarization sum arises from the fact that we work in the light-cone gauge $\nbar\cdot A = 0$ as we have mentioned above.

The interaction between the colliner quarks and the collinear gluon was derived within the SCET framework before~\cite{Bauer:2000yr} and is given by 
\bea\label{eq:qqg}
\Graph{0.3}{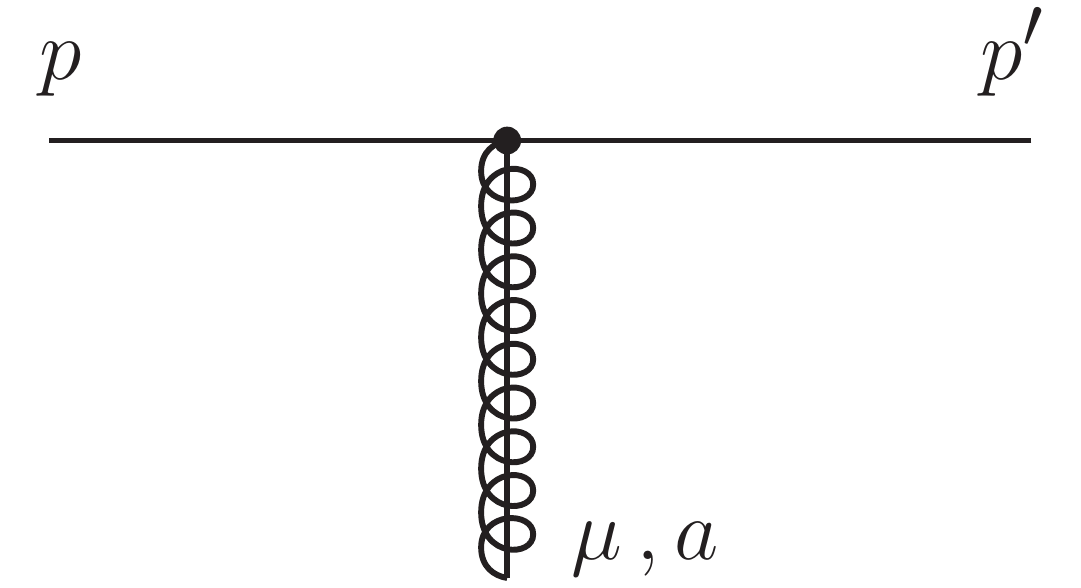}
= ig_s t^a \left( n^\mu + 
\frac{\gamma_\perp^\mu  \ppslash }{p^+} + 
\frac{ \ppslash' \gamma_\perp^\mu}{{p'}^+} 
\right) \,,
\eea 
where we have used the gauge condition $\nbar \cdot A = 0$. One can recognize that the collinear interaction also exists in the light-front perturbation theory, when one expresses the so-called bad component of a spinor field in terms of the good component, see for instance~\cite{Beuf:2016wdz}.    

Now we include the interactions between the CGC shock wave and the collinear modes via the CGC Wilson line. \textcolor{orange}{We note that these vertices are identical to the ones in the conventional CGC calculations~\cite{McLerran:1994vd,Balitsky:1995ub,Ayala:1995hx,Balitsky:2001mr,Ayala:1995kg,Caucal:2021ent}, but now derived via a different approach~\cite{kang:future1}.} 

For the collinear quark, we can derive
\bea\label{eq:qshock}
\Graph{0.3}{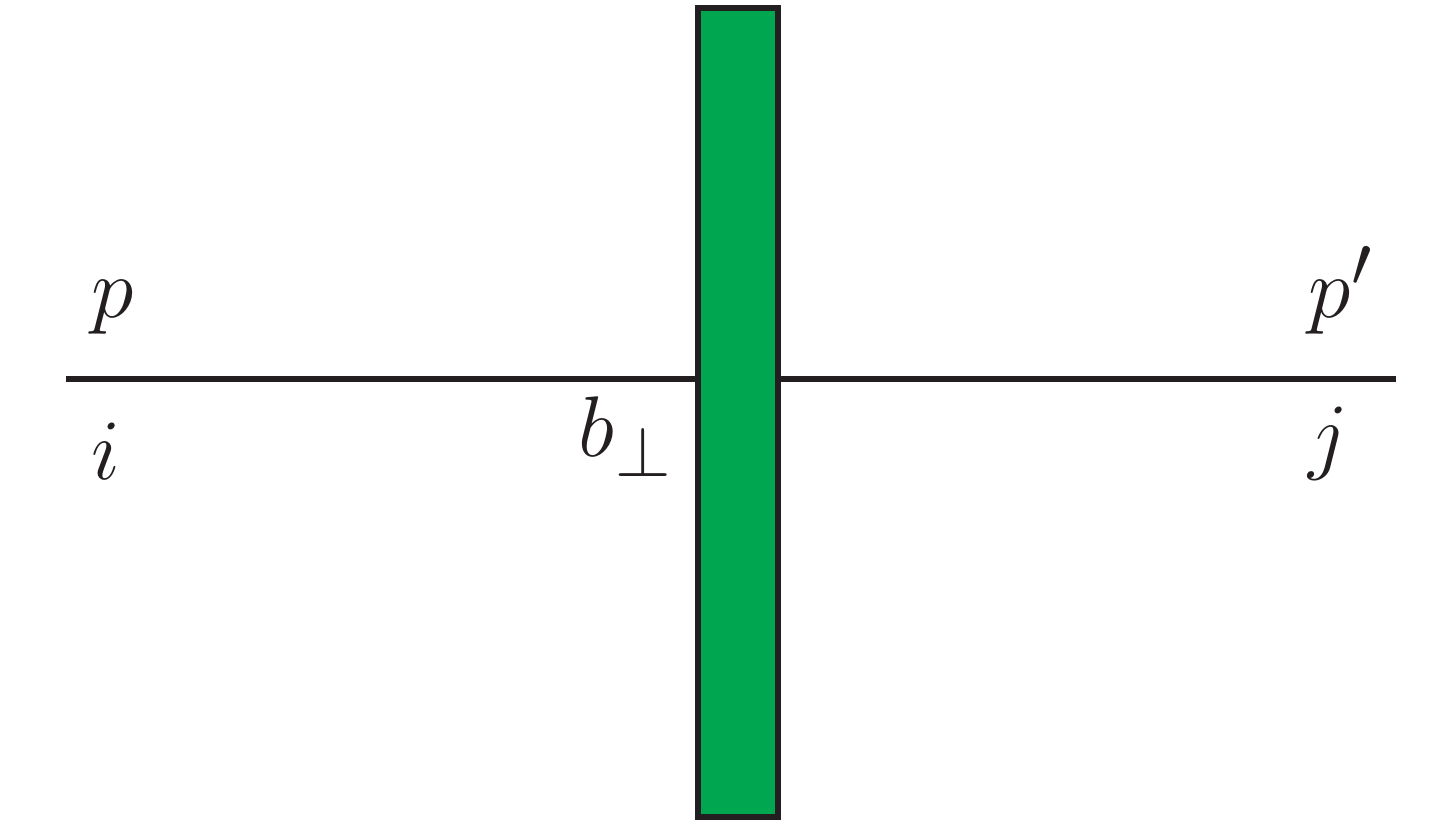}
= 
(4\pi)\delta({p'}^+ - p^+) 
\frac{\nbslash}{2} 
\int d^2b_\perp e^{-i(p'_\perp - p_\perp ) \cdot b_\perp } 
W_{ji}(b_\perp) \,, 
\eea 
where the interaction is represented by the green box and $W_{ji}$ is the corresponding Wilson line in the fundamental representation. In comparison with~\cite{Caucal:2021ent}, we have a factor of $4\pi$ instead of $2\pi$ in Eq.~\eqref{eq:qshock} above. This is because we use $p^{\pm} = p^0\pm p^3$ as our convention for the $\pm$-component of a $4$-vector, while $p^{\pm} = (p^0\pm p^3)/\sqrt{2}$ in~\cite{Caucal:2021ent}. {\color{orange}{\it We note that the form of the interaction indicates the $3$-momentum conservation in the `$+$' and the `$\perp$' components, while the `$-$' component is not conserved which is a consequence of the homogeneous power expansion in $\lambda$~\cite{kang:future1} as required by CGC.}} The gluon-shock-wave interaction is 
\bea\label{eq:gshock}
\Graph{0.3}{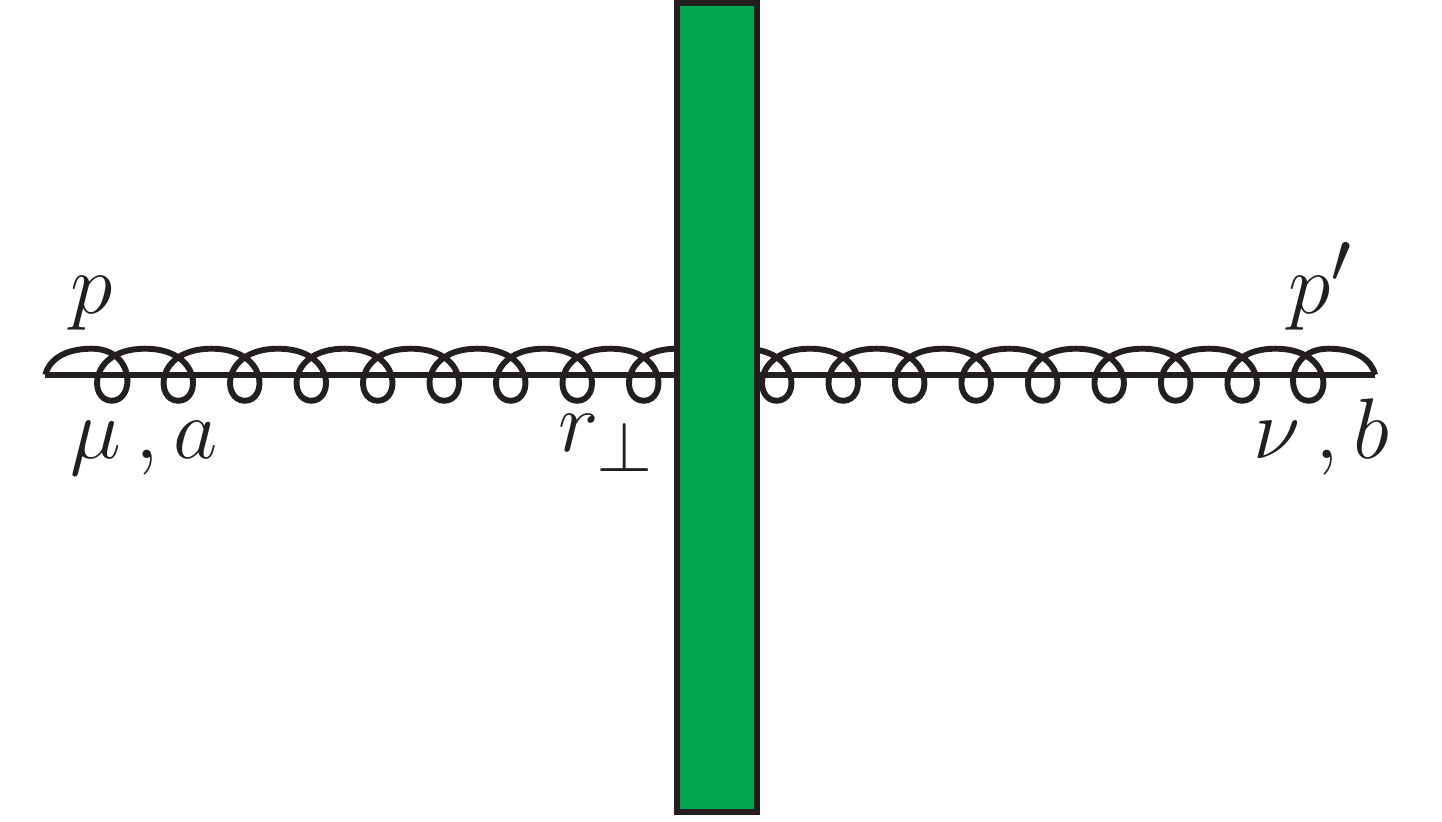}
= -g_{\mu\nu} 
(4\pi)\delta({p'}^+ - p^+)
 p^+ 
\int d^2b_\perp e^{-i(p'_\perp - p_\perp ) \cdot r_\perp } 
W_{ba}(r_\perp)
\,, 
\eea 
where $W_{ba}$ is the Wilson line in  the adjoint representation. 

The soft Feynman rule for emitting a soft gluon is given by 
\bea\label{eq:eikonal-feyn-raw}
\Graph{0.46}{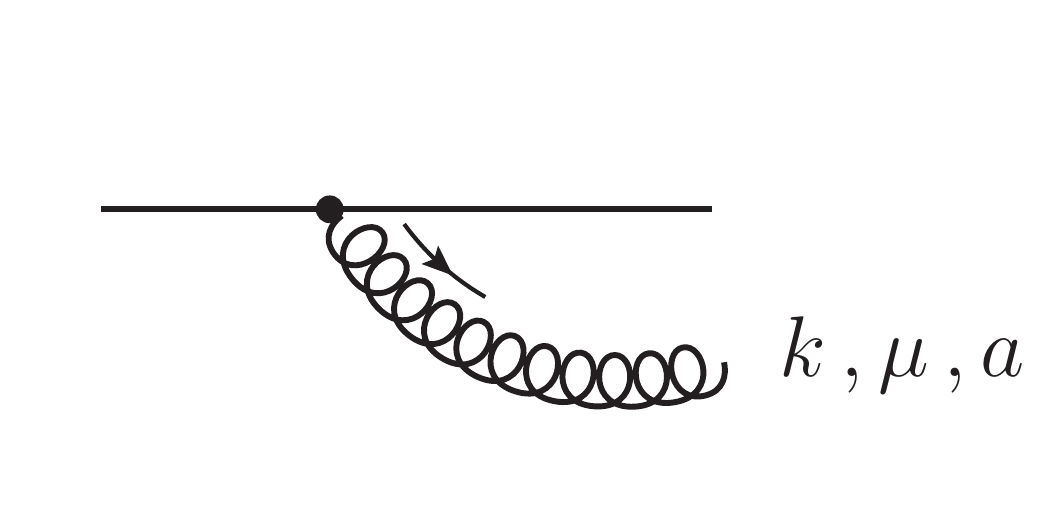}
= -g_s T^a_{\alpha\beta} \frac{n^\mu}{k^-} 
\,, 
\eea  
which is nothing but the eikonal approximation. 
$T^a_{\alpha\beta} = t^a_{\alpha\beta}$ if the soft gluon is radiated from a quark with $\alpha$ and $\beta$ the color indices for the quark after and before the radiation, otherwise $T^a_{\alpha\beta} = if^{\alpha a \beta}$ if it is from a gluon. 

There is also the interaction between the soft gluon and the shock wave through the CGC Wilson line, which is found identical to the collinear gluon interaction in Eq.~(\ref{eq:gshock})
\bea\label{eq:eikonal-shockwave}
\hspace{-2.ex}\Graph{0.3}{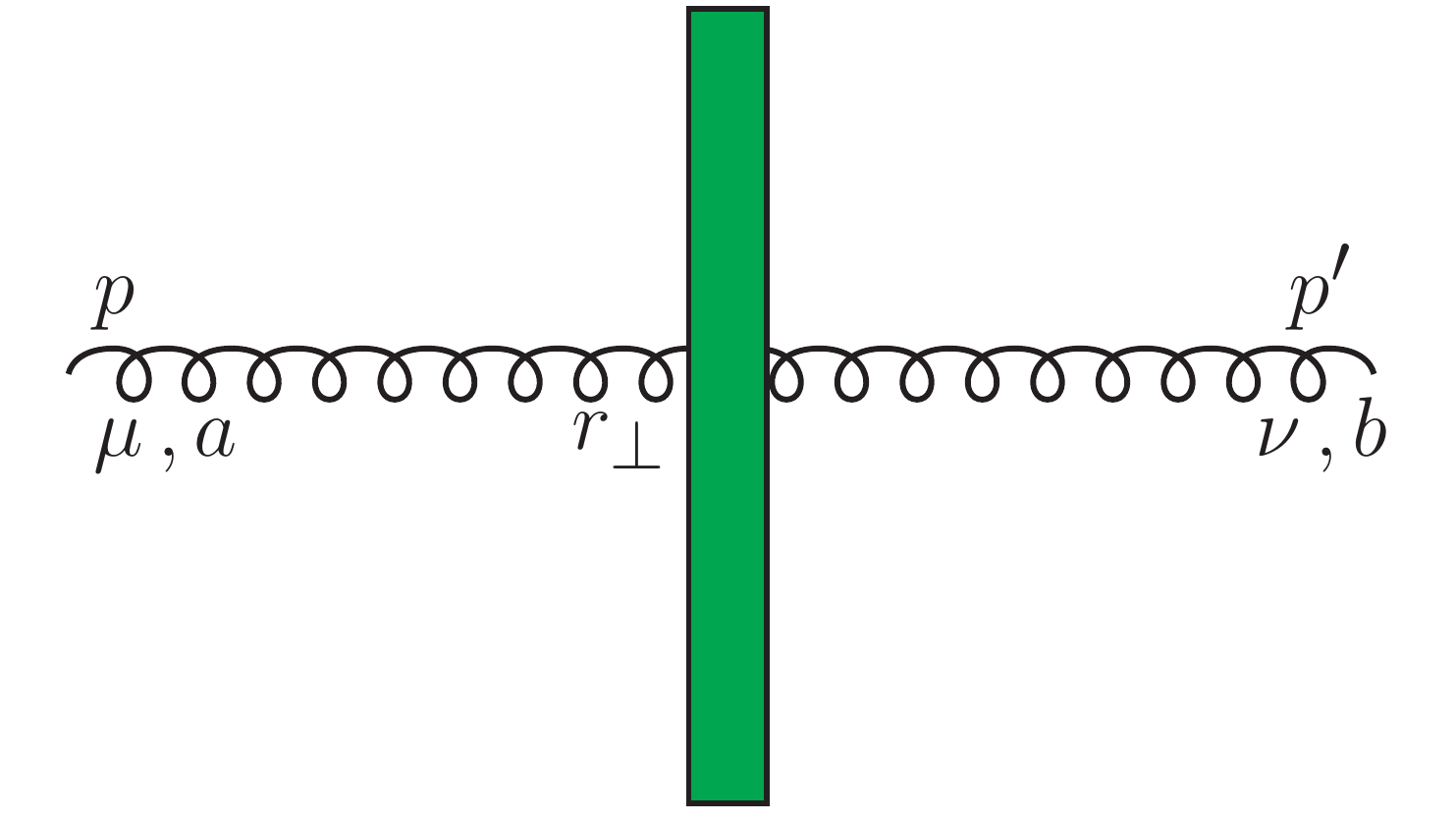}
= -g_{\mu\nu} 
(4\pi)\delta({p'}^+ - p^+)
 p^+ 
\int d^2b_\perp e^{-i(p'_\perp - p_\perp ) \cdot r_\perp } 
W_{ba}(r_\perp)
\,. \quad 
\eea

\subsection{Rapidity Regulator}\label{subsec:rapidityreg}
In this work, dimensional regularization $4\to D = 4-2\epsilon$ will be used to deal with the collinear and infrared divergences. However, in the effective framework, we could encounter divergences which are not regularized by dimensional regularization. For instance we could have integrals of the form 
\bea 
I_{rap.} = \int_0\, \frac{dk^+}{k^+} \,, 
\eea 
which is divergent when $k^+ \to 0$ and is known as the rapidity divergence. The rapidity divergence has been extensively discussed in the scenario of the TMD physics~\cite{Ji:2004wu,Hautmann:2007uw,Bozzi:2007pn,Collins:2008ht,Becher:2011dz,Collins:2011ca,Echevarria:2011epo,Chiu:2012ir,Echevarria:2012js,Li:2016axz}, where the divergence occurs due to the homogeneous power expansion in the phase space kinematics, see for instance~\cite{Collins:2011ca,Chiu:2012ir}. Various rapidity regulators have been proposed to regulate the divergence~\cite{Ji:2004wu,Hautmann:2007uw,Bozzi:2007pn,Collins:2008ht,Becher:2011dz,Collins:2011ca,Collins:2011zzd,Echevarria:2011epo,Chiu:2012ir,Echevarria:2012js,Li:2016axz}.

In CGC, one will face the same rapidity divergence~\cite{Watanabe:2015tja,Iancu:2016vyg,Beuf:2016wdz,Beuf:2017bpd,Liu:2019iml,Kang:2019ysm,Liu:2020mpy,Caucal:2021ent,Beuf:2021srj}, which needs to be properly regulated. {\color{orange}{\it In general, an introduced regulator better not 
 destroy the properties necessary to establish all order factorization theorems, for example, the eikonal identities needed for exponentiation of the soft emissions into Wilson lines~\cite{Collins:2011zzd,Collins:2011ca}. Moreover a regulator should not break the power counting, otherwise
one could lose control over the non-perturbative power corrections to make the perturbatively calculable part ambiguous and thus sacrifices the predictive power of a perturbative calculation. We refer readers to Ref.~\cite{Becher:2011dz,Collins:2011ca,Chiu:2012ir} for more discussions. On the computational side, we better utilize a unified and conceptually well-defined regulator for both real and virtual contributions and could be easily applied to higher order corrections.}} For these purposes, throughout the work, we implement the regulator proposed in~\cite{Chiu:2012ir}. Consequently, 
\begin{itemize} 
\item
for the ${\cal O}(\alpha_s)$ soft eikonal current in Eq.~(\ref{eq:eikonal-feyn-raw}), we add the rapidity regulator factor as 
\bea\label{eq:eikonal-feyn}
-g_s T^a_{\alpha\beta} \frac{n^\mu}{k^-}  
\to -g_s T^a_{\alpha\beta} \frac{n^\mu}{k^-} 
\left| \frac{\nu}{k^+ - k^-} \right|^{\eta/2}
\,, 
\eea  
where the factor $\left| \frac{\nu}{k^+-k^-} \right|^{\eta/2}$ is the regulator introduced for the rapidity divergence, where $\nu$ is a rapidity scale and $\eta$ is a small parameter to be expanded by the end of a calculation. We will show that $\nu$ is associated to the cut-off scale in the CGC. We note that since the rapidity divergence only arises when $k^+ \to \infty \gg k^- $ or $k^- \to \infty \gg k^+$, in our case, we could simplify the calculation if we make the following change~\cite{Liu:2013hba} 
\bea\label{eq:softrapchange}
\left| \frac{\nu}{k^+-k^-} \right|^{\eta/2}
\to 
\left(
\frac{\nu}{k_\perp} 
\right)^{\frac{\eta}{2}}
e^{-\frac{\eta}{2}|\eta_k|} \,,
\eea 
where $\eta_k = \frac{1}{2}\ln\left(k^+/k^-\right)$ is the rapidity of $k^\mu$ and we have used $k^\pm = k_\perp e^{\pm \eta_k}$. To see this, we note that 
\bea 
\left| \frac{\nu}{k^+-k^-} \right|^{\eta/2}
\to 
\left| \frac{\nu}{k^+} \right|^{\eta/2} 
= \left( \frac{\nu}{k_\perp} \right)^{\eta/2} 
e^{-\frac{\eta}{2}\eta_k}\,, \quad \text{for $k^+\gg k^-$ and $\eta_k > 0$}\,,
\eea 
and 
\bea 
\left| \frac{\nu}{k^+-k^-} \right|^{\eta/2}
\to 
\left| \frac{\nu}{k^-} \right|^{\eta/2} 
= \left( \frac{\nu}{k_\perp} \right)^{\eta/2} 
e^{\frac{\eta}{2} \eta_k}\,, \quad \text{for $k^+\ll k^-$ and $\eta_k < 0$}\,.
\eea 

\item For the collinear contributions, we apply the following replacement
\bea\label{eq:collrap} 
\int \frac{dk^+}{k^+ }
\to \int \frac{dk^+}{k^+} \left( \frac{\nu}{k^+}\right)^\eta  \,, 
\eea 
where $k^+$ is the momentum of the gluon. 
\end{itemize} 
At the end of a calculation, one should carry out the $\eta$ expansion before the $\epsilon$ expansion~\cite{Chiu:2012ir}. 
The rapidity regulator will then turn the rapidity divergence to $\eta$-poles. 

\subsection{Color Charge Operator}\label{subsec:colorcharge}
Throughout the work, we will use the color charge operator ${\bf T}^a_i$ introduced by Catani, et al. in~\cite{Catani:1996vz}. Although the notation is not necessary for fixed order calculations, we find it quite helpful in unifying the notation in different channels~\cite{Liu:2020mpy}. Meanwhile the notation helps to sort out the origins of different types of logarithms in the cross section and is thus beneficial for deriving the threshold resummation in Section~\ref{sec:threshold}. 

In order to carry out the threshold resummation in CGC, we need to deal with multiple emissions that interact with the shock wave, which dramatically complicates the color flow and mixes the LO and higher order color structures. For instance, we can see that for the jet production (as in the hadron case), the LO matrix element squared possesses the color structure of the form
\bea\label{eq:LOsqcolor}
  \Graph{0.23}{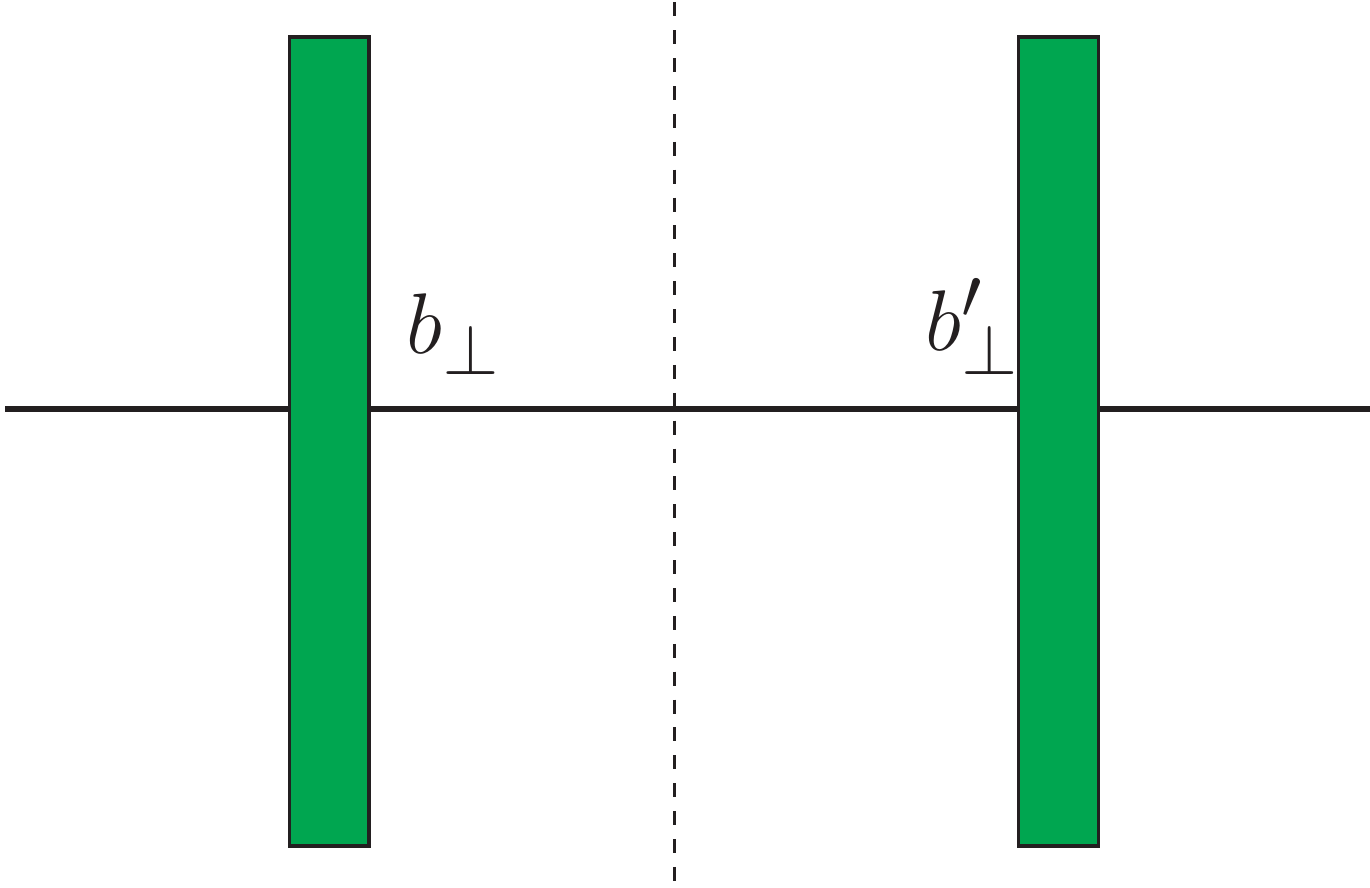} 
= |{\cal M}_0|^2 \propto {\rm Tr}[W^\dagger(b_\perp')W(b_\perp)] \,, 
 \eea 
 where $W$ is the CGC Wilson line in Eq.~(\ref{eq:wilsonq}). The vertical dashed line cuts through the final state particles. Here we pay our attention only to the color structures but ignore the details on the kinematics. 
 
 At higher orders with more gluons emitted, 
one could encounter
\bea\label{eq:color-example-nlo} 
&&  \Graph{0.23}{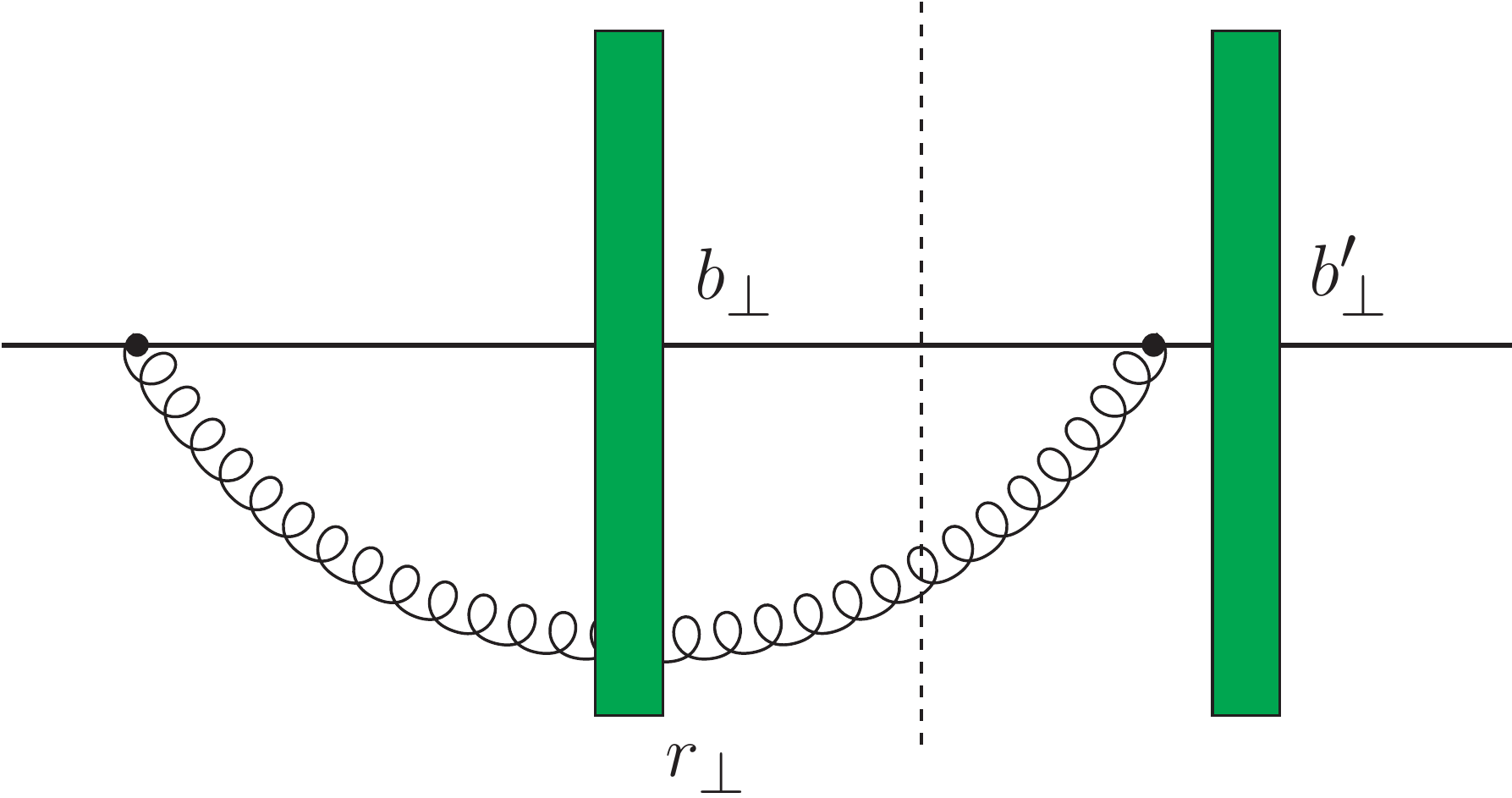} 
\propto 
\int d^2r_\perp 
{\rm Tr}\left[
W^\dagger (b_\perp') 
  t^{a}
W(b_\perp) t^c   
 \right]
 W_{ac}(r_\perp )  \,, 
\eea 
for the color structures at NLO, and the NNLO correction will involve contributions such as
\bea\label{eq:color-example-nnlo} 
&&  \Graph{0.23}{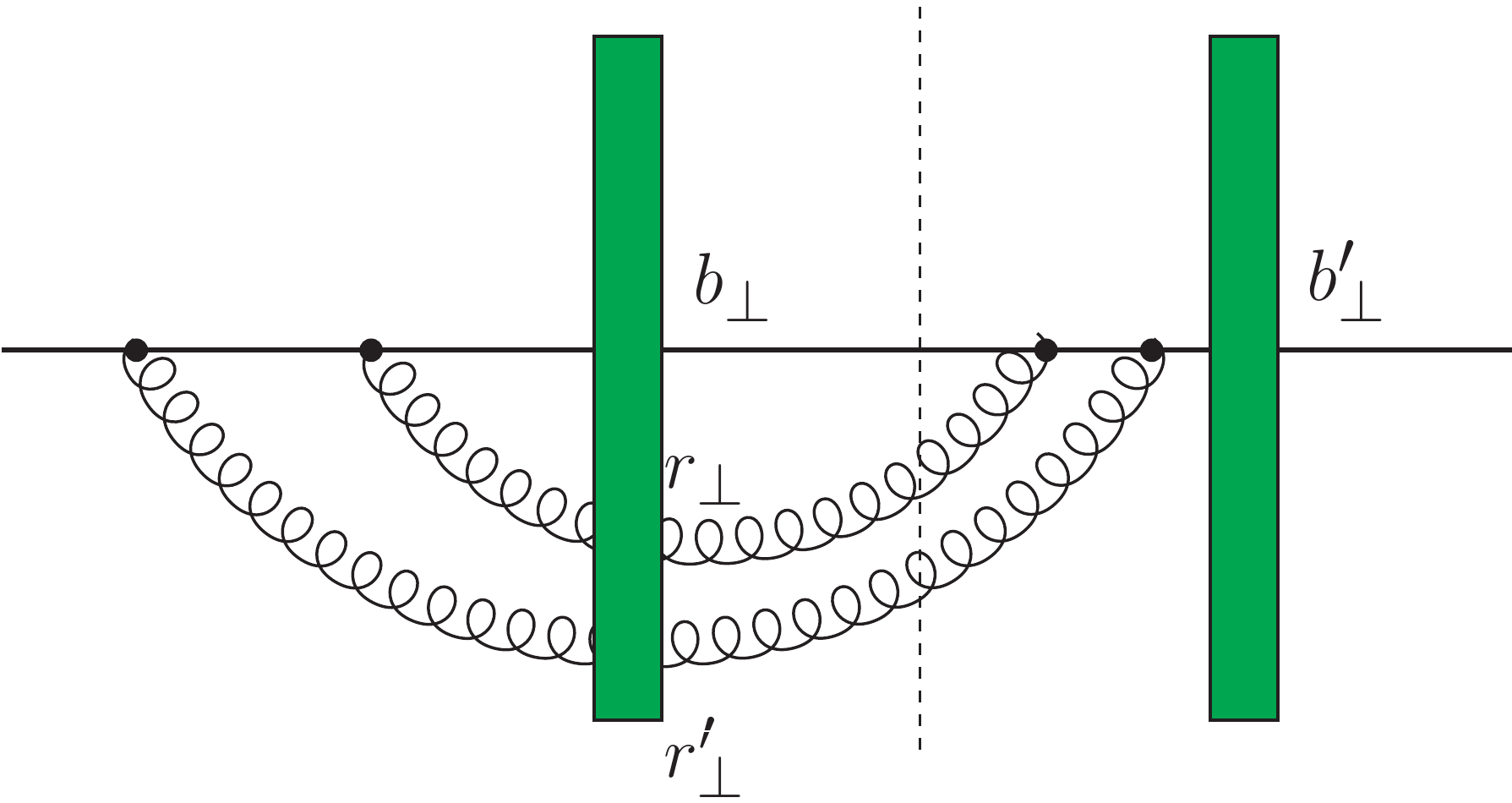} 
\propto 
\int d^2r_\perp d^2r_\perp'
{\rm Tr}\left[
W^\dagger (b_\perp') 
 t^{b}  t^{a}
W(b_\perp) t^c  t^{d}
 \right]
 W_{ac}(r_\perp )
 W_{bd}(r'_\perp ) \,,  \nn \\
\eea 
where $W_{ac}$ and $W_{bd}$ are the CGC Wilson lines in the adjoint representation. 
We see that 
\begin{itemize} 
\item the NLO and NNLO color structures are completely entangled with the structures,  $W_{ij}({b_\perp'})$ and $W_{ij}(b_\perp)$;
\item the NLO and NNLO digrams showed above exhibit some kind of common features, however the color structures seem quite different. 
\end{itemize} 

However, resummation usually requires the factorization of the higher order corrections from the LO structures to some extent. Also, realizing a resummation actually means that one can identify in the higher order corrections a certain pattern as the repetition of the lower order results. Therefore, one can resum the lower order calculations to obtain the all order results, see for instance Ref.~\cite{Banfi:2012jm}. 
For these purposes, instead of tracing over color indices, we need to keep track of the color correlation when squaring the matrix element~\footnote{Similar situation happens to the non-global logarithms (NGLs), where one also maintains the color correlation information in order to achieve the resummation~\cite{Caron-Huot:2015bja}. Given the similarities between NGLs and small-x evolution, it is perhaps not too surprise that in CGC certain resummations will also rely on the color correlation.}. 
The color charge operator offers a neat way to handle the color correlation and allows us to achieve the factorization in which the higher order corrections can be seen as the color currents acting on LO matrix element, and in turn has been shown to help establish a compact formalism for the resummation~\cite{Liu:2020mpy}.

To understand the notation, let us consider processes that involve $m$ final-state QCD partons at the tree-level, $i=1,2,\dots,m$, each carrying a momentum $p_i$. The LO matrix element takes the general structure
\bea
\cm_{p_1^+,\dots,\, p_m^+}^{c_1,\dots,\, c_m;\, s_1,\dots,\, s_m}(p_{1\perp}, \dots,\,p_{m\perp})
\eea
where the subscript $\{p^+_1, \cdots,\, p^+_m\}$ denotes the large `$+$'-component of the momenta. The superscripts $\{c_1, ...,\, c_m\}$ are the color indices
($\alpha =1, \dots ,N_C$ different colors for each quark or antiquark, $a= 1, \dots, N_C^2-1$ for gluons, where $N_C$ is the number of colors for the quark) and the $\{s_1, ...,\, s_m\}$ are the 
spin indices ($s=1,2$ for fermions and $s=1,...,D-2$ for gluons, with $D$ the space-time dimension), respectively. 
\begin{figure}[htbp]
	\begin{center}
		\includegraphics*[width=0.4\textwidth]{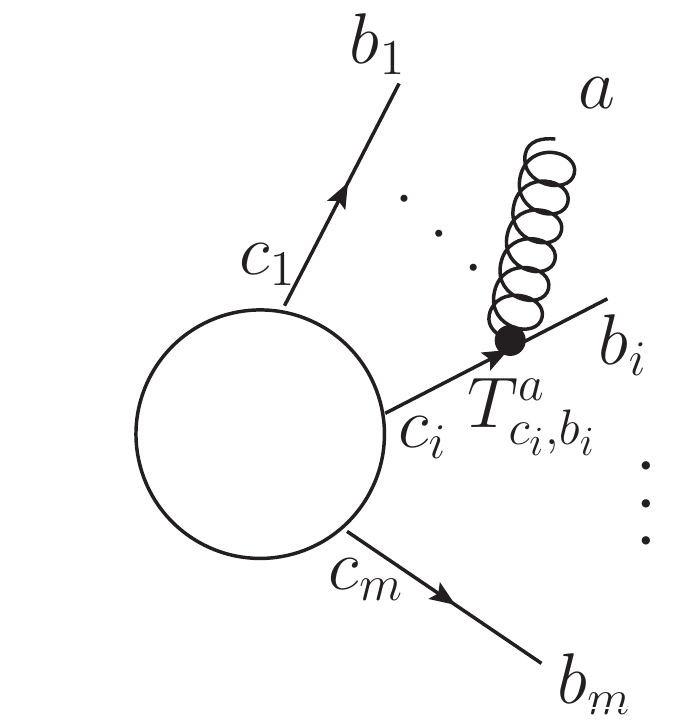}
		\caption{A gluon is emitted from parton-$i$,  which rotates the color basis by $T_{c_i,b_i}^a$. }
		\label{fig:color-T}
	\end{center}
\end{figure}

A basis in the spin-color space 
$\{ \ket{c_1,...,c_m} \otimes \ket{s_1,...,s_m} \}$ was introduced in~\cite{Catani:1996vz}, such that
the matrix element can be written as the inner product of the basis and a vector in the spin-color space, which is 
\bea\label{cmmdef}
\cm_{p_1^+,\dots,p_m^+}^{c_1,\dots,c_m; s_1,\dots,s_m}(p_{1\perp}, \dots,p_{m\perp})
 \equiv
\Bigl( \bra{c_1,\dots,c_m} \otimes \bra{s_1,\dots,s_m} \Bigr) \ket{\cm(p_{1\perp}, \dots,p_{m\perp})}_{p_i^+} \,,
\eea
where $\ket{\cm(p_{1\perp}\,, \dots \,, p_{m,\perp})}_{p_i^+}$ is a vector in the spin-color space. We will suppress the subscript to denote it as $\ket{\cm(p_{1\perp}, \dots,p_{m,\perp})}$. And the matrix element squared after summing over all the color and the spin indices is then
\bea
\M{}{p_1^+,\dots p_m^+} =  \langle {\cal M}(p_{1\perp}, \dots,p_{m\perp})|{\cal M}(p_{1\perp}, \dots,p_{m\perp}) \rangle \;.
\eea

At higher orders, once a gluon with color index $a$ is emitted from parton $i$, as illustrated in  fig.~\ref{fig:color-T}, it will generate a color charge ${\bf T}_i^a$ which acts on the $i$-th parton color index and rotates the color basis in a way that
\bea
\bra{c_1, \dots , c_i, \dots , c_m; a} {\bf T}^a_i
\ket{b_1, \dots , b_i, \dots , b_m} = \delta_{c_1 b_1} \dots 
T_{c_i b_i}^a  \dots   \delta_{c_m b_m} \;\;,
\eea
where $T_{c b}^a \equiv i f_{cab}$ if the emitting particle $i$
is a gluon and $T_{\alpha \beta}^a \equiv t^a_{\alpha \beta}$ 
if the emitting parton $i$ is a quark (in the case of an antiquark $T_{\alpha \beta}^a \equiv {\bar t}^a_{\alpha \beta}
= - t^a_{\beta \alpha }$). 

For the processes with initial QCD partons, the color operator rules can be obtained by crossing, such that if $i$ is an initial gluon, $T_{c b}^a = i f_{cab}$; while for intial quark we have $T_{\alpha \beta}^a = {\bar t}^a_{\alpha \beta}
= - t^a_{\beta \alpha }$ and anti-quark with $T_{\alpha \beta}^a = t^a_{\alpha \beta}$. 

The dot product of the color charge operator is given by 
\bea 
{\bf T}_i \cdot {\bf T}_j
= \sum_a {\bf T}_i^a {\bf T}_j^a  \,, 
\eea 
where we note that the subscripts $i$ and $j$ denote the $i$- and $j$-th partons that the color charge operators are acting on. 
The color operator rule immediately leads to the color algebra that 
\bea
{\bf T}_i \cdot {\bf T}_j ={\bf T}_j \cdot {\bf T}_i \;\;\;\;{\rm if}
\;\;i \neq j; \;\;\;\;\;\;{\bf T}_i^2= C_i,
\eea
where $C_i= N_C$ if $i$ is a gluon and $C_i=C_F=(N_C^2-1)/2N_C$ if $i$ is a quark or antiquark, which reduces to $C_i = \frac{N_C}{2}$ in the large $N_C$ limit. Also for any QCD process, we have the color charge conservation 
\bea 
\sum_{i=1}^m {\bf T}^a_i = 0 \,, 
\eea 
 when acting on the vector $|{\cal M} \rangle$.
 
Using this notation, the square of color-correlated amplitudes is given by  
\bea
\label{colam}
|{\cal{M}}^{i,k}|^2 &\equiv&
\!\!\langle {\cal M}| \,{\bf T}_i \cdot {\bf T}_k \,|{\cal M} \rangle 
\nonumber \\
&=&
\left[ {\cal M}^{a_1.. b_i ... b_k ... a_m}(p_{1\perp},...,p_{m\perp}) \right]^*
\; T_{b_ia_i}^c \, T_{b_ka_k}^c
\; {\cal M}^{a_1.. a_i ... a_k ... a_m}(p_{1\perp},...,p_{m\perp}) \,. \quad 
\eea
Again, the $i$ and $k$ here denote the $i$-th and the $k$-th parton, respectively. For instance, in the jet production, we will have 
\bea\label{eq:colorforresum} 
  \Graph{0.23}{lo-square.pdf} 
&=& \langle {\cal M}_0 |  {\cal M}_0 \rangle  \,,
\nn \\ 
  \Graph{0.23}{real-soft-1-isr.pdf} 
&\propto& 
\int d^2r_\perp  
\langle {\cal M}_0 | 
  {\bf T}_j^{a}
 W_{ac}(r_\perp ){\bf T}_i^c \, 
 |{\cal M}_0 \rangle  
 \,,
\nn \\ 
  \Graph{0.23}{real-soft-multi-isr.pdf} 
&\propto& 
\int d^2r_\perp d^2r_\perp'
\langle {\cal M}_0 | 
  {\bf T}_j^{a}
 W_{ac}(r_\perp ){\bf T}_i^c \, 
 {\bf T}_j^{b}
 W_{bd}(r'_\perp ) {\bf T}_i^{d}
 |{\cal M}_0 \rangle  \nn \\ 
 &=& 
 \langle {\cal M}_0 | 
 \left( \int d^2r_\perp
 {\bf T}^a_iW_{ab}(r_\perp) {\bf T}^b_j
 \right)^2 |{\cal M}_0\rangle
 \,,
\eea 
where we have labeled the incoming quark as the $i$-th particle and the outgoing quark the $j$-th parton and again we have only paid attention to the colors but ignored the kinematic details. We have used  
$ 
|{\cal M}_0\rangle 
\propto W(b_\perp) 
$, 
as in Eqs.~\eqref{eq:qshock} and~\eqref{eq:LOsqcolor} and will also derive in detail in Eq.~(\ref{eq:LOM}). Now apparently the higher orders are obtained by color charge operator acting on the LO vector $|{\cal M}_0\rangle$ and {\color{orange}{\it it is also clear that this specific NNLO term is the square of the NLO in terms of the colors.}}~\footnote{We note the difference between the ``trace-after-square''
\bea 
\langle {\cal M}_0 | 
 \left( \int d^2r_\perp
 {\bf T}^a_iW_{ab}(r_\perp) {\bf T}^b_j
 \right)^2 |{\cal M}_0\rangle \,, 
\eea 
in Eq.~(\ref{eq:colorforresum}), 
and the ``square-after-trace''
\bea 
\left( 
\int d^2r_\perp  
\langle {\cal M}_0 | 
  {\bf T}_j^{a}
 W_{ac}(r_\perp ){\bf T}_i^c \, 
 |{\cal M}_0 \rangle  
 \right)^2  \,, 
 \quad \text{or} \quad
\int d^2r_\perp  \left(
\langle {\cal M}_0 | 
  {\bf T}_j^{a}
 W_{ac}(r_\perp ){\bf T}_i^c \, 
 |{\cal M}_0 \rangle  
 \right)^2  \,. 
\eea 
The former would reduce to the color structure in Eq.~(\ref{eq:color-example-nnlo}) while the latters are the square of the traced NLO in Eq.~(\ref{eq:color-example-nlo}) before or after the $r_\perp$-integration, respectively. Neither of the latters corresponds to the color structure arises at NNLO.
} This feature could explain the non-linearity of the BK evolution in Eq.~(\ref{eq:LOBK}) and the threshold resummation in the CGC framework~\cite{Liu:2020mpy}. It is also crucial for the threshold resummation for the jet production in section~\ref{sec:threshold}.

\subsection{Examples}\label{subsec:examples}
As an example of the application of the Feynman rules and the color operator ${\bf T}^a$, 
we derive the initial state radiation (ISR) collinear current, which will be used in the following sections to derive the collinear cross section.
By ISR collinear current, we mean a collinear gluon is radiated before the incoming parton interacts with the shock wave, which is described by
\bea
&& \Graph{0.33}{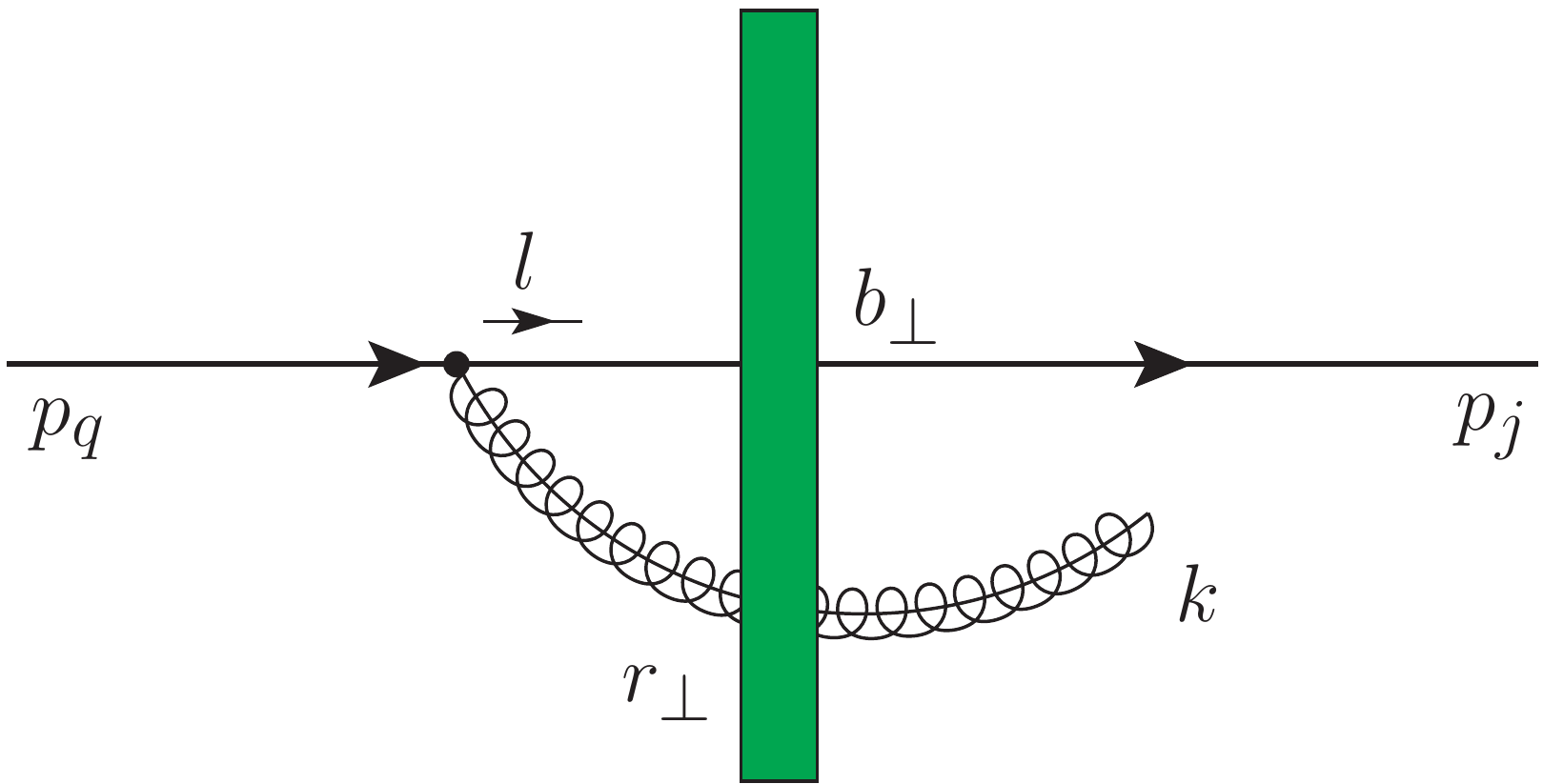}  \nn \\
&=& ig_s \mu^{2\epsilon} \int \frac{d^Dl}{(2\pi)^D}
 \frac{il^+}{l^2+i\epsilon} 
\frac{\nslash}{2}\, 
\left( n^\beta + \frac{\lpslash \gamma_\beta^\perp}{l^+} \,
+ \frac{\gamma_\beta^\perp \pqslash }{p_q^+}
\right)  \frac{\nbslash}{2} \, 
\frac{i
d^{\beta\beta'}(p_q-l) 
}{(p_q-l)^2+i\epsilon} 
\, \nn \\ 
&\times& \left[ - 
g_{\alpha\beta'}
k^+ (4\pi)\delta(l^+-p_j^+)
\right] \int d r_\perp 
e^{-i(k_\perp+l_\perp) \cdot r_\perp} 
W_{ab}(r_\perp) {\bf T}^b |{\cal M}_0(l_\perp,p_{j\perp}) \rangle  \, \nn \\ 
&=& -ig_s 
\mu^{2\epsilon}
\int \frac{dl_- d^{D-2}l_\perp }{(2\pi)^{D-1}}
 \frac{p_j^+}{l^2+i\epsilon} 
\frac{
1 
}{(p_q-l)^2+i\epsilon} 
\left( 
2l_\perp^\alpha 
+ \frac{k^+}{p_j^+} \lpslash \gamma_\alpha^\perp
\right)
\, \nn \\ 
&\times& \int d r_\perp 
e^{-i(k_\perp+l_\perp) \cdot r_\perp} 
W_{ab}(r_\perp) {\bf T}_i^b |{\cal M}_0(l_\perp,p_{j\perp}) \rangle  \,  \nn \\
&=& - g_s 
\mu^{2\epsilon}
\int d r_\perp\, \frac{ d^{D-2}l_\perp }{(2\pi)^{D-2}}
  \frac{
  e^{-i (l_\perp+k_\perp) \cdot r_\perp} 
  }{l_\perp^2} 
\left( 
2l_\perp^\alpha \frac{p_j^+}{p_q^+}
+ \frac{k^+}{p_q^+} \lpslash \gamma_\alpha^\perp
\right)
W_{ab}(r_\perp) {\bf T}_i^b |{\cal M}_0(l_\perp,p_{j\perp}) \rangle  \,,  \nn \\   
\eea 
where $
d_{\alpha\beta}(q)
= -g_{\alpha\beta} + \frac{{\bar n}_{\alpha} q_\beta + q_\alpha {\bar n}_{\beta}}{q^+}$. 
Here we have carried out the $l^-$-integration by contour integral, see Eq.~(\ref{eq:contour}). 

Similar steps lead to the ISR soft current which can also be derived directly by power counting the above result assuming $k^+ \sim p_q^+\lambda $, which gives 
\bea 
&& \Graph{0.33}{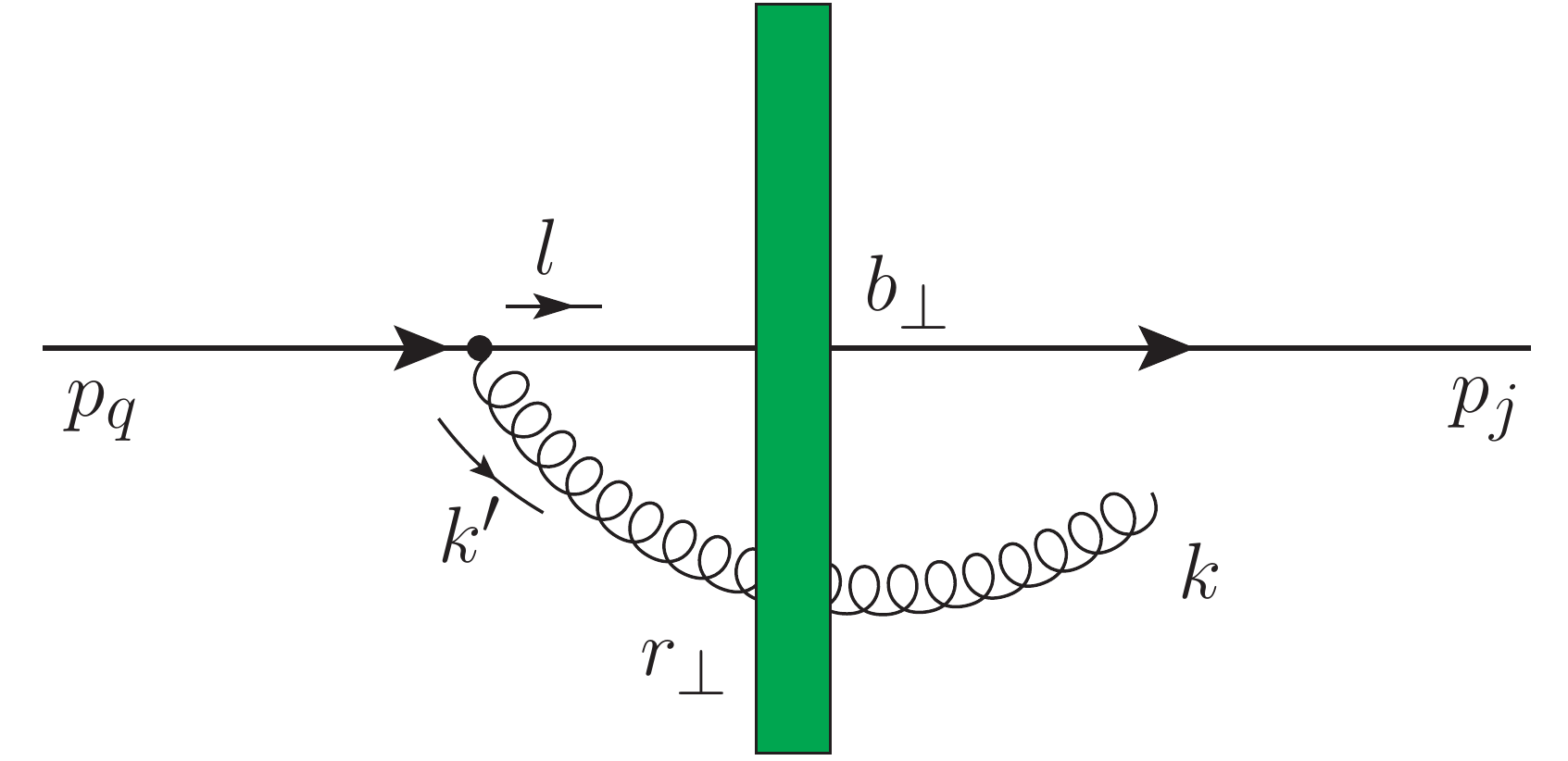}  \nn \\
&=& - g_s 
\mu^{2\epsilon}
e^{-\frac{\eta}{2} |\eta_k|}
\int d r_\perp \, \frac{ d^{D-2}l_\perp }{(2\pi)^{D-2}}
  \frac{2l_\perp^\alpha \,
  e^{-i l_\perp \cdot r_\perp} 
  }{l_\perp^2}  
 \left(  \frac{\nu}{l_\perp} 
 \right)^{\frac{\eta}{2}} 
\, 
e^{-i k_\perp \cdot r_\perp} 
W_{ab}(r_\perp) {\bf T}_i^b |{\cal M}_0(l_\perp,p_{j\perp}) \rangle  \,, \nn \\
\eea 
where the rapidity regulator follows Eq.~(\ref{eq:eikonal-feyn}). We have made the change in Eq.~(\ref{eq:softrapchange}) and use the fact that $-l^\mu_\perp = {k_\perp'}^\mu$ and for very large rapidities, $\eta_{k'} \to \eta_k$. To see this, we note that since $k^+ = {k^+}'$, we will have $k_\perp e^{\eta_k} = k_\perp' e^{\eta_k }e^{\delta \eta_k }$, where $\delta \eta_k = \eta_{k'} - \eta_k$ characterizes their difference. When $\eta_k$ and thus $\eta_{k'} \to \infty$, $\delta \eta_k = \frac{k_\perp}{k_\perp'} \sim {\cal O}(1) \ll |\eta_k|$, and $\eta_{k'} \to \eta_k$ for very large rapidities. It is therefore justified to replace $e^{-\eta |\eta_{k'}|}$ by $e^{-\eta |\eta_k|}$ to regulate the singular behavior at very large rapidities.

\subsection{Multiple point correlator}
In CGC, the incoming nucleus is described by the correlators of the Wilson lines. The nucleus dipole distribution is defined as
\bea\label{eq:dipole}
S_{X_f}^{(2)}(b_\perp,b_\perp') = \frac{1}{N_C}{\rm Tr}\left[W(b_\perp) W^\dagger(b_\perp') \right]
\,. 
\eea 
The subscript $X_f$ represents the scale to evaluate the dipole, and is given by the typical momentum fraction of the nucleus carried by the gluon. 

We will also encounter the triple-pole distribution which is 
\bea\label{eq:S3} 
S^{(3)}_{X_f}(b_\perp,r_\perp,b_\perp') 
= \frac{2}{N_C^2} 
{\rm Tr} \left[ 
W^\dagger(b_\perp') t^a
W(b_\perp) t^b
\right] W_{ab}(r_\perp) \,, 
\eea 
where $W_{ab}$ is the Wilson line in the adjoint representation.
The triple-pole distribution can be further related to the dipole by the Fiertz identity in Eq.~(\ref{eq:WWWab}) as 
\bea 
S^{(3)}_{X_f}(b_\perp,r_\perp,b_\perp')
=  
S_{X_f}^{(2)}(b_\perp,r_\perp)
S_{X_f}^{(2)}(r_\perp,b'_\perp)
- \frac{1}{N^2_C} S_{X_f}^{(2)}(b_\perp,b'_\perp)\,,
\eea 
which in the large $N_C$ limit reduces to the product of $2$ dipoles
\bea 
S^{(3)}_{X_f}(b_\perp,r_\perp,b_\perp')
\approx 
S_{X_f}^{(2)}(b_\perp,r_\perp)
S_{X_f}^{(2)}(r_\perp,b'_\perp)\,.
\eea 

In the jet production, we will also need the $6$-point correlator, defined as 
\bea \label{eq:6-point} 
S^{(6)}_{X_f}(b_\perp,r_\perp,b_\perp',r_\perp')
= \frac{2}{N_C^2}{\rm Tr}\left[ t^{b'} 
W^\dagger(b_\perp')
W(b_\perp) t^b
\right]
W^\dagger_{b'a}(r_\perp')
W_{ab}(r_\perp)\,. 
\eea 
The $6$-point correlator occurs at NLO when we have $2$ distinguished jets. 

In the large $N_C$ limit, the $S_{X_f}^{(6)}$ can be written as a product of the $4$-point function and the dipole such that 
\bea 
S_{X_f}^{(6)} \approx 
S_{X_f}^{(4)}(b_\perp,b_\perp',r_\perp',r_\perp) S_{X_f}^{(2)}(r_\perp,r_\perp') \,,
\eea 
which is originated from the color identity in Eq.~(\ref{eq:color-identity}), while
%
the quadru-pole distribution is given by 
\bea 
S_{X_f}^{(4)}(b_\perp,b_\perp',r_\perp',r_\perp)
= \frac{1}{N_C}
{\rm Tr}[W(b_\perp)W^\dagger(b_\perp')W(r_\perp')W^\dagger(r_\perp)] \,.
\eea

For later use, we also introduce the nucleus distributions in the momentum space which are 
\bea\label{eq:dipole-fourier} 
(2\pi)^{D-2}S_\perp 
{\cal F}_F(k_\perp;X_f)
 = \int d^{D-2}b_\perp \int d^{D-2}b_\perp' e^{-ik_\perp\cdot (b_\perp' - b_\perp)} S_{X_f}^{(2)}(b_\perp,b_\perp')\,, 
\eea 
for the dipoles, where $S_\perp$ is the transverse area of the nucleus and ${\cal F}_F$ is the distribution in the momentum space. Here we have applied the translational invariance of the dipole distribution. In the large $N_C$ limit, the momentum space bilinear distribution for the $3$-point function is given by
\begin{align}
(2\pi)^{D-2}
S_\perp {\cal F}_F(k_\perp,X_f){\cal F}_F(l_\perp,X_f) 
=& \int d^{D-2} b_\perp d^{D-2}b_\perp' d^{D-2}r_\perp 
e^{-i k_\perp \cdot x_\perp}
\nonumber\\
&\times e^{-i l_\perp \cdot y_\perp} \, S^{(3)}_{X_f}(b_\perp,r_\perp,b_\perp') \,, 
\end{align} 
where $x_\perp = b_\perp - r_\perp$ and $y_\perp = r_\perp - b_\perp'$. We also define $z_\perp \equiv b_\perp - b_\perp' = x_\perp + y_\perp$. The dipole distribution satisfies the BK-evolution, which at LO gives  
\bea\label{eq:LOBK} 
\frac{d}{d \ln X_f} S^{(2)}_{X_f}(b_\perp,b_\perp')
=- \frac{\alpha_s}{\pi} \frac{N_C}{2} \int \frac{d^2r_\perp}{\pi}
\left( \frac{z_\perp^2}{x_\perp^2 \, y_\perp^2 } \right)_+
S_{X_f}^{(3)}(b_\perp,r_\perp,b_\perp') \,, 
\eea
where we have introduced the `$+$'-prescription for a  distribution ${\cal D}(x)$ defined as 
\bea\label{eq:plus-pres}
\int d^2 x \, {\cal D}_+(x)\,  f(x) 
= \lim_{\epsilon \to 0} \int d^{D-2} x \, {\cal D}(x,\epsilon)\, \Big( f(x) - f(0) \Big) \,,
\eea 
with the requirement that $\lim_{|x|\to \infty} {\cal D}(x,\epsilon) \to 0$. 

The BK-equation is non-linear in the color structures, which can be seen by rewriting the equation using the color charge operator notation such that 
\bea\label{eq:BK-T} 
&&\frac{d}{d \ln X_f}   | 
{\cal M}_0(b_\perp) \rangle_{X_f} {}_{X_f}\langle{\cal M}_0(b_\perp')| \nn \\
&& = \frac{\alpha_s}{\pi} \int \frac{d^2r_\perp}{\pi}
\left( \frac{z_\perp^2}{x_\perp^2 \, y_\perp^2 } \right)_+
 W_{ac}(r_\perp ){\bf T}_i^c
|
{\cal M}_0(b_\perp) \rangle_{X_f} {}_{X_f}  
\langle{\cal M}_0(b_\perp') |{\bf T}_j^a  
\,,   
\eea 
where we add the subscript $X_f$ to remind the rapidity scale dependence of the matrix. 
The equation can be formally solved to find  
\bea\label{eq:BK-T-solve} 
&& |{\cal M}_0(b_\perp) \rangle_{X_f} {}_{X_f} \langle{\cal M}_0(b_\perp')| \nn \\ 
&=& \exp\left[\frac{\alpha_s}{\pi}\ln\frac{X_f}{X_0}\int 
\frac{d^2 r_\perp}{\pi} 
\left( \frac{z_\perp^2}{x_\perp^2 \, y_\perp^2 } \right)_+
{\bf T}_j^a W_{ac}(r_\perp ){\bf T}_i^c 
\right] 
|{\cal M}_0(b_\perp) \rangle_{X_0} {}_{X_0} \langle{\cal M}_0(b_\perp')| \,.
\eea 
The exponentiation of the color charge operator ${\bf T}^a$ manifests the non-linearity of the BK-evolution in the color structure. The non-linearity is originated from the multiple emission picture in Eq.~(\ref{eq:colorforresum}) where for each emission, an additional Wilson line $W$ with respect to the lowest order is generated. 
Eqs.~(\ref{eq:BK-T}) and~(\ref{eq:BK-T-solve}) also hold for the gluon case. 

\section{Leading Order Cross Section}\label{sec:LOsec}
We start with the LO cross section. Throughout the work, we take the quark channel as an example to walk through our calculation in detail. Other channels can be obtained in a similar way and the results are given in the Appendix~\ref{sec:allchannels}.  
\begin{figure}[htbp]
	\begin{center}
		\includegraphics*[width=0.65\textwidth]{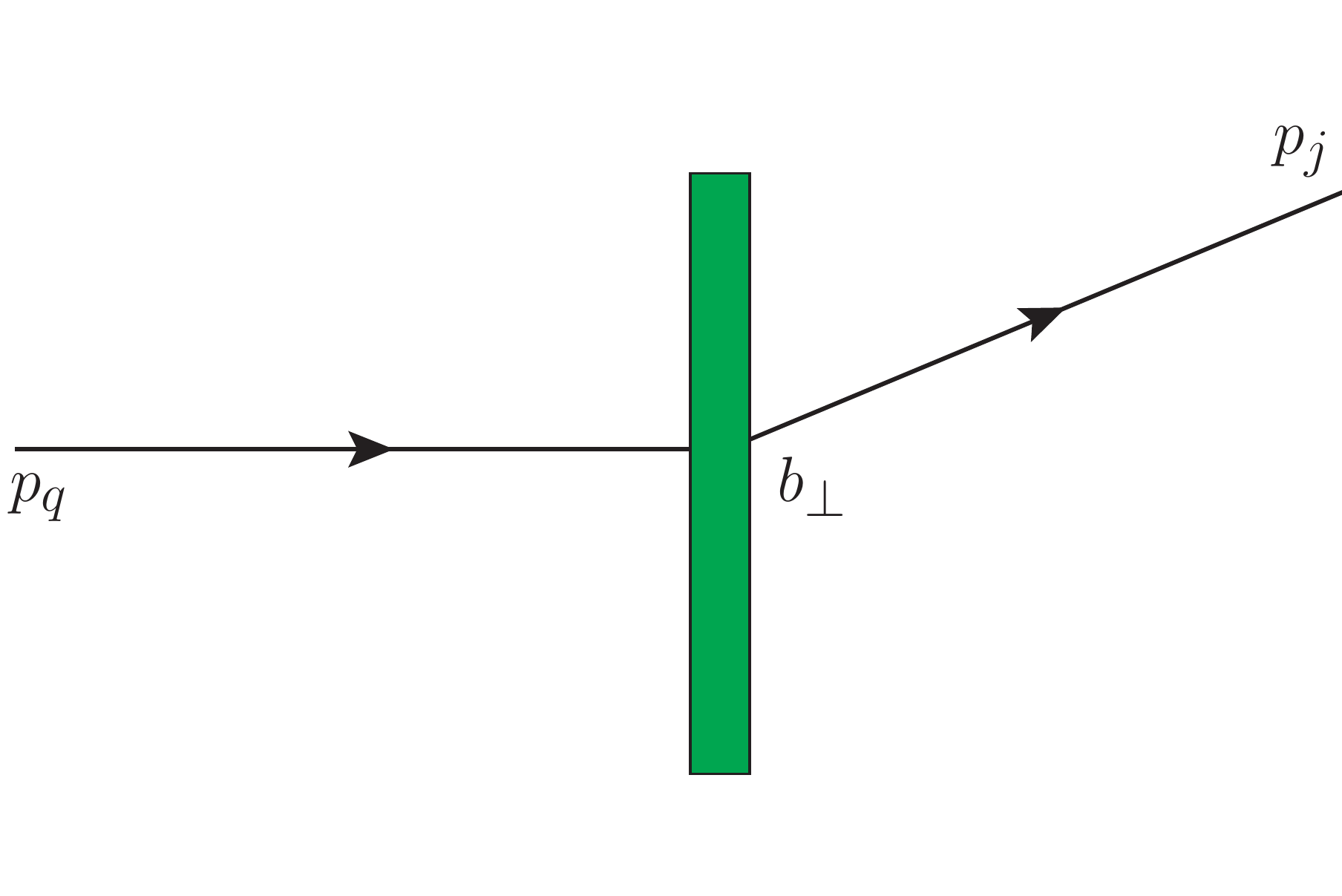}
		\caption{LO Feynman diagram for the quark channel, where an incoming quark carries momentum $p_q$, kicked by the Wilson line (green box) to produce an out-going quark with momentum $p_j$. }
		\label{fig:feyn-lo}
	\end{center}
\end{figure}

At LO, there is only one diagram that contributes to the jet production in the quark channel, where $q(p_q) \to q(q_j)$, as shown in fig.~\ref{fig:feyn-lo}, which gives the amplitude
\bea\label{eq:LOM} 
|{\cal M}_0(p_{q\perp},p_{j\perp}) \rangle = {\bar \chi}^i_{p^+_j} \,  \frac{\nbslash}{2} \, 
\chi^j_{p^+_q} \int d^{D-2} b_\perp  e^{-i (p_{j\perp}-p_{q\perp}) \cdot b_\perp }
W_{ij}(b_\perp)  \,.
\eea 
Square the matrix element and sum over all the spins and average over the initial quark color, one finds
\bea
\langle {\cal M}_0|
{\cal M}_0 \rangle 
= 2 (p_q^+)^2 
\int d^{D-2}b_\perp d^{D-2}b_\perp' 
e^{-ip_{j\perp} \cdot z_\perp }
\frac{1}{N_C}
{\rm Tr}\left[W(b_\perp) W^\dagger(b_\perp')\right] \,, 
\eea 
The cross section is then given by 
\bea\label{eq:LO-phase-integral}
\sigma^{(0)} = \frac{1}{2} \, \int d^4p_J \, 
 \frac{d x}{p_q^+} \,f(x)\, 
 \frac{d p_j^+}{4\pi p_j^+}
\frac{d^{D-2} p_{j\perp}}{(2\pi)^{D-2}} \, 
2 (2\pi) \delta(p_j^+ - p_q^+) \, \langle {\cal M}_0 |{\cal M}_0\rangle  \, \delta^{(4)}(p_J-p_j)
\,,  \quad 
\eea 
where we have averaged over the proton spin which gives the overall factor $1/2$ and including the partonic flux factor $1/(p_q^+)$. Here $f(x)$ is the quark parton distribution function (PDF) and the momentum fraction satisfies $p_q^+ = x p_p^+$ with $p_p^+$ the proton momentum. We have worked out the $p_j$ on-shell condition and thus $p_j^- = p_{j\perp}^2/p_{j}^+$. The unity $\int d^4p_J \delta^{(4)}(p_J-p_j)$ is inserted to define the jet momentum. At LO, since we only have one parton in the final state, the jet algorithm acts trivially on the $p_j$ phase space. The situation is different at NLO, when there is one additional radiated parton and the jet clustering procedure imposes non-trivial restriction on the two-parton phase space, which dramatically complicates the calculation. We have suppressed possible additional experimental cuts and the distribution bin separation placed on the jet momentum $p_J$ in Eq.~(\ref{eq:LO-phase-integral}).

Plugging the $\langle {\cal M}_0|{\cal M}_0 \rangle$ into Eq.~(\ref{eq:LO-phase-integral}), we find 
\bea 
\sigma^{(0)} =  \tau 
  \,f(\tau)\, 
\int \frac{d p_J^+}{ p_J^+}
\frac{d^{D-2} p_{J\perp}}{(2\pi)^{D-2}} \, 
   \, 
\int d^{D-2}b_\perp d^{D-2}b_\perp' 
e^{-ip_{J\perp} \cdot z_\perp }
S^{(2)}_{X_f}(b_\perp,b_\perp') \, 
\eea 
where $\tau = p^+_J/p_p^+$. As we have mentioned above, the subscript $X_f$ in the dipole $S^{(2)}_{X_f}(b_\perp,b_\perp')$ denotes the scale to evaluate the dipole distribution. At LO, the scale choice is arbitrary, however the CGC framework demands that $X_f \sim {\cal O}(X_A)$ to separate the fast and the slow moving partons, where $X_A$ is the momentum fraction carried by the gluon from the nucleus. We will see that the NLO calculation will determine the $X_f$ by minimizing the logarithms in the NLO corrections, whose value produces the CGC requirement. 

In the dipole momentum space, by using Eq.~(\ref{eq:dipole-fourier}), the differential cross section at LO can also be written as 
\bea
d\sigma^{(0)} \equiv 
\frac{d \sigma^{(0)}}{d\eta_J d^2p_{J\perp}} = S_\perp \tau f(\tau)
\mathcal{F}_F(p_{J\perp};X_f), 
\label{eq:LOsec}
\eea 
where we have used $d \eta_J = dp_J^+/p_J^+$. We note that in practice we still use $p_J^+$ to generate the phase space instead of the rapidity $\eta_J$. It is merely a matter of choice at LO, but crucial in order to completely determine the phase space of the real emission at NLO, as described in Section~\ref{subsec:kin}. For later convenience, for the rest of this work, we introduce $d \sigma$ through  
\bea 
\sigma =   
\int  \frac{d p_J^+}{p_J^+} 
d^2 p_{J\perp}  \,  
 d \sigma \,.
\eea 

\section{Next-to-Leading Order Correction}\label{sec:NLO}

\subsection{Virtual Corrections}
Now we move on to evaluate the NLO corrections to the inclusive jet cross section. We start with the virtual corrections, which receive contributions from both the collinear virtual correction and the soft virtual correction, respectively elaborated in~\ref{sec:coll-virtual} and in~\ref{sec:soft-virtual}. Since the virtual corrections share the LO kinematics, the jet algorithm plays no effect to the calculation, and the partnoic results are identical to the forward hadron production~\cite{Chirilli:2011km,Chirilli:2012jd,Liu:2019iml,Kang:2019ysm}. 
\begin{figure}[htbp]
    \centering
    \subfigure{\includegraphics[width=0.325\textwidth]{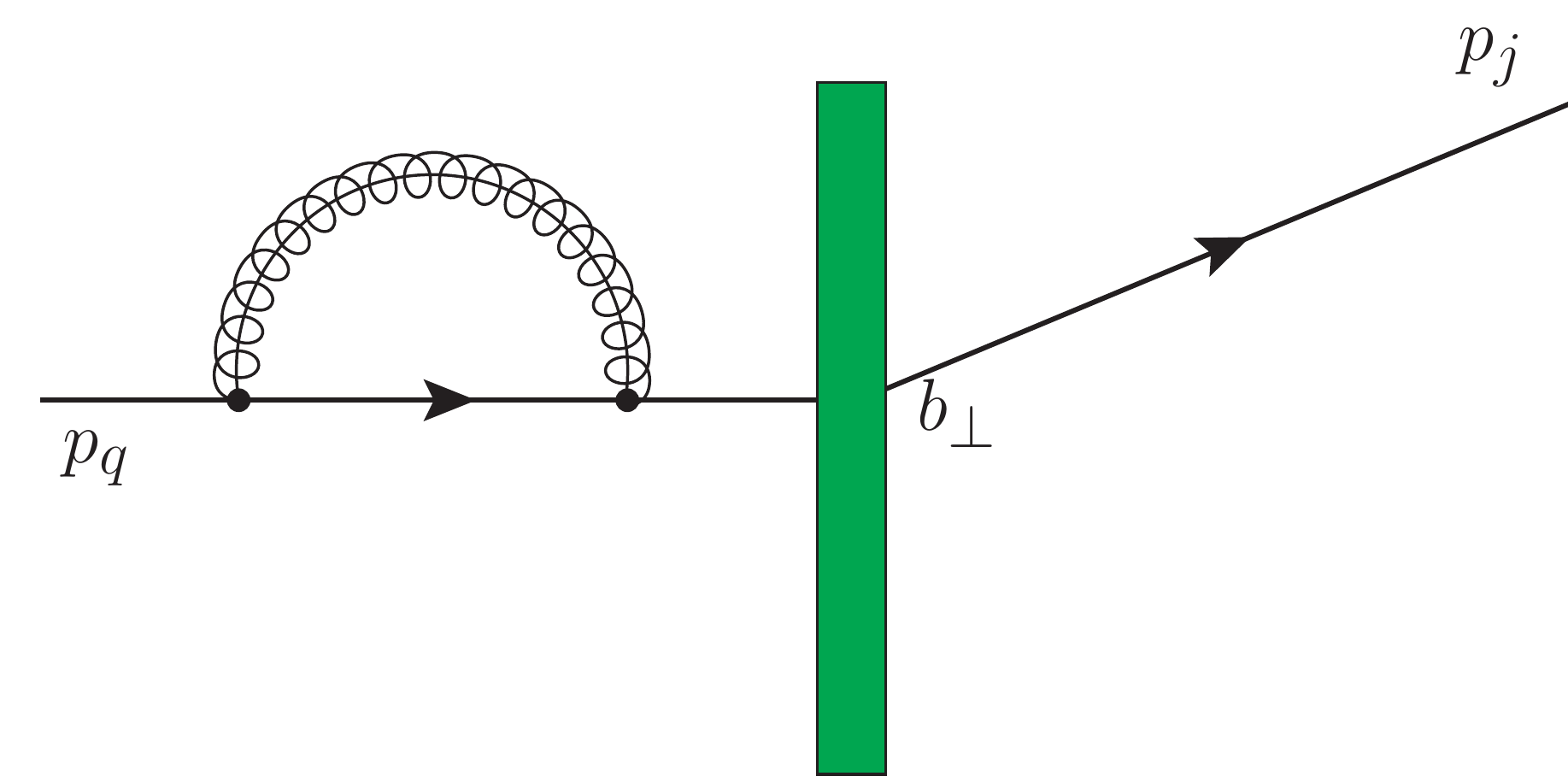}} 
    \subfigure{\includegraphics[width=0.325\textwidth]{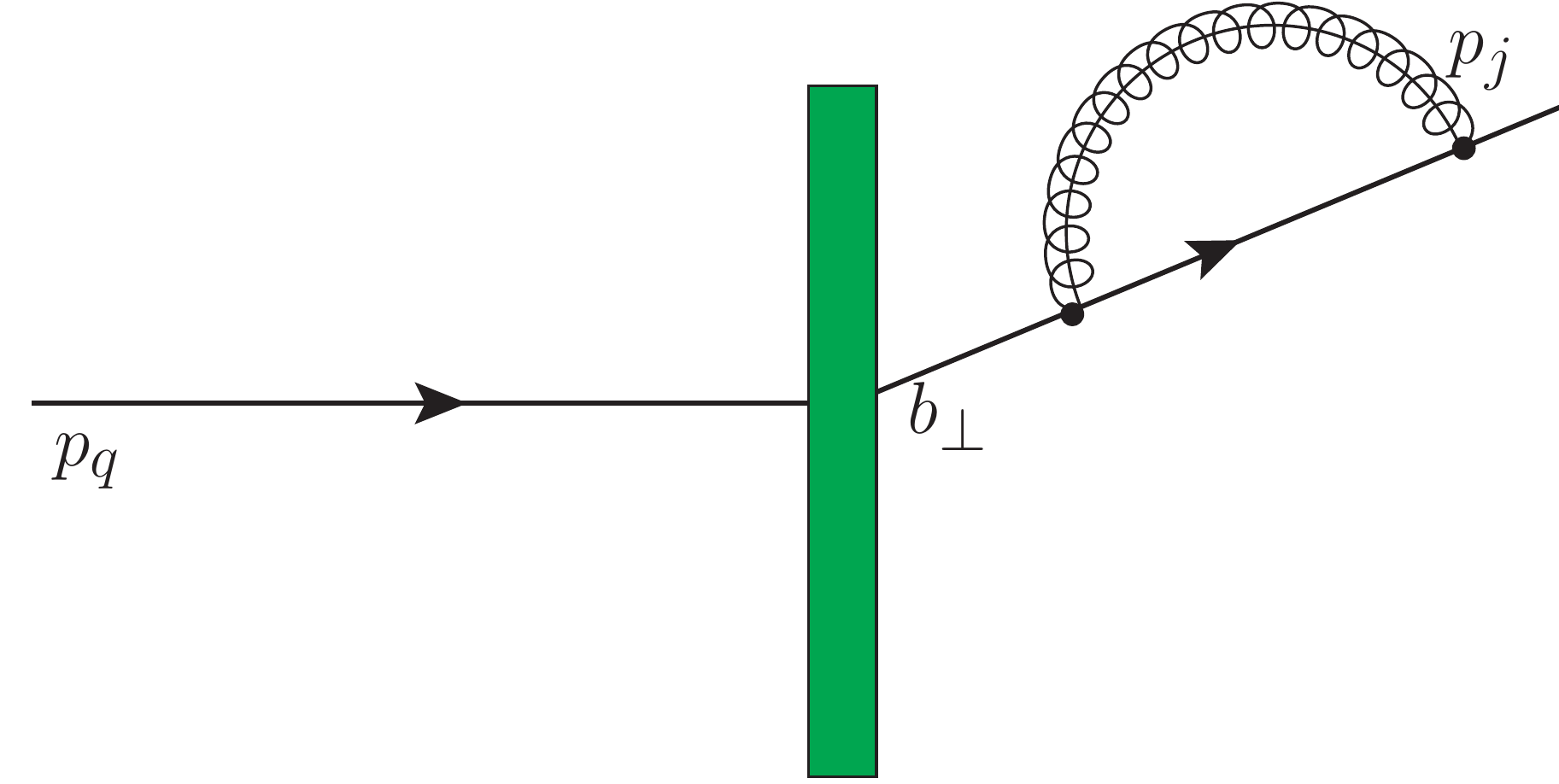}} 
    \subfigure{\includegraphics[width=0.325\textwidth]{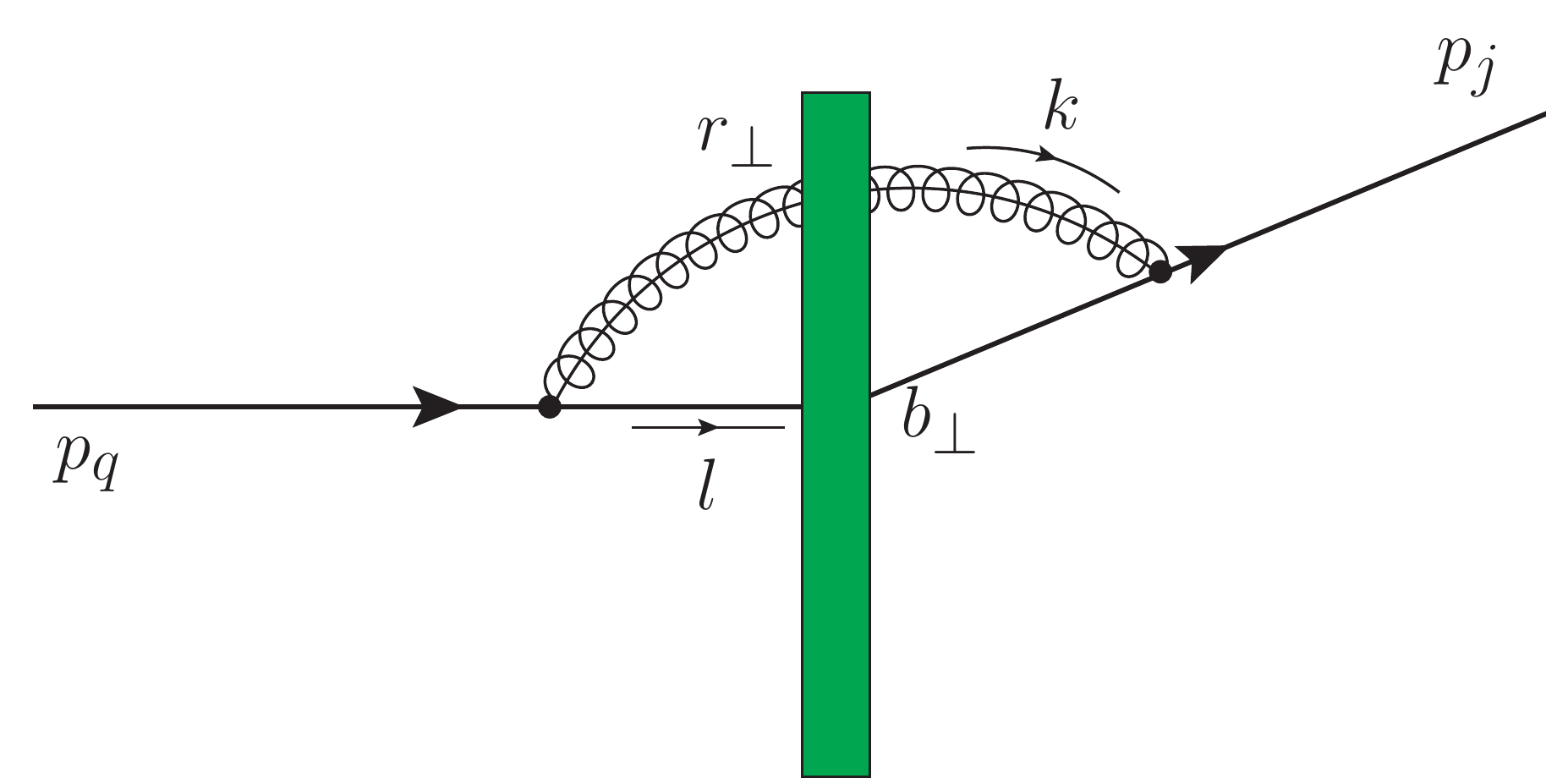}} 
    \caption{Feynman diagrams for collinear virtual corrections: (a) initial quark self energy correction (b) final quark self energy correction (c) the only non-vanishing contribution. }
    \label{fig:feyn-virt}
\end{figure}

\subsection{The collinear virtual correction}~\label{sec:coll-virtual}
Three loop configurations can occur at NLO, as shown in fig.~\ref{fig:feyn-virt}, while the first two vanish due to scaleless integrals and only the last one survives, which gives 
\bea 
| {\cal M}_{V,coll.}\rangle  &=& -i g_s^2  \left( \frac{\nu}{p_q^+}\right)^\eta \mu^{2\epsilon}
\int \frac{d^D k}{(2\pi)^D} \frac{d^{D-2}l_\perp}{(2\pi)^{D-2}}
\, 
\left[
-2 \frac{2-z}{(1-z)^{1+\eta}} [\lpslash, \kpslash] + 2 [\lpslash, \pjslash]  
\right. 
\nn \\ 
&& \left. 
-\frac{2}{z} \left(\frac{1+z^2}{(1-z)^{1+\eta}} - \epsilon (1-z) \right)(\kpslash - (1-z) \pjslash) \lpslash 
 \right]  
 \frac{z}{k^2 \, (p_j-k)^2} 
  \nn \\
 && 
\times\int d^{D-2}r_\perp
 \frac{e^{i(l_\perp+k_\perp) \cdot r_\perp }}{l_\perp^2}
{\bf T}^a_j \, W_{ab}(r_\perp)
\,
{\bf T}^b_i \, 
|{\cal M}_0(l_\perp,p_{j\perp}-k_{\perp}) \rangle  
\,, \quad 
\eea
where we have let $z \equiv l^+/p_j^+ = 1- k^+/p_j^+$ and implemented the rapidity regulator following Eq.~(\ref{eq:collrap}). $[{a},{b}] \equiv ab - ba$ is the commutator. 

The loop integration over $k^-$ can be performed by contour integral, see~Eq.~(\ref{eq:contour}), which simplifies the virtual matrix element to 
\bea 
| {\cal M}_{V,coll.} \rangle &=&  \frac{g_s^2}{4\pi} \left( \frac{\nu}{p_q^+}\right)^\eta \mu^{2\epsilon}
\int_0^1 d z 
\int \frac{d^{D-2} k_\perp}{(2\pi)^{D-2}} \frac{d^{D-2}l_\perp}{(2\pi)^{D-2}} \, \, 
\Bigg\{
-2 z \frac{2-z}{(1-z)^{1+\eta}} [\lpslash, \kpslash] 
\nn \\ 
&&  \hspace{10.ex}
+ 2 z [\lpslash, \pjslash]  
-2  \left(\frac{1+z^2}{(1-z)^{1+\eta}} 
- \epsilon (1-z) \right)(\kpslash - (1-z) \pjslash) \lpslash 
\Bigg\} 
  \nn \\
 &&  \hspace{-7.ex} \times 
 \int 
 d^{D-2}r_\perp
 \frac{e^{-il_\perp \cdot r_\perp }}{l_\perp^2}
 \frac{ 
e^{-ik_\perp\cdot r_\perp }
}{(k_\perp - (1-z)p_{j\perp})^2} \,
{\bf T}^a_j W_{ab}(r_\perp) 
{\bf T}^b_i \,
|{\cal M}_0(l_\perp,p_{j\perp}-k_\perp) \rangle 
\,.
\eea 
\textcolor{orange}{We note that the rapidity $\eta$-regulator naturally regulates the divergence at $z = 1$, and there is no need for any additional cut-off in the loop integral in our approach. We see this as one advantage of the rapidity regulator we are advocating.} 

Interfering the virtual amplitude with the LO matrix element, one finds 
\bea
&& 
\langle {\cal M}_0 | {\cal M}_{V,coll.} \rangle  + c.c. 
 \nn \\ 
&=& -8 (p_q^+)^2
   \frac{g_s^2}{4\pi} \frac{N_C}{2} 
  \mu^{2\epsilon}
\int_0^1 d z  
  \left[\frac{1+z^2}{(1-z)^{1+\eta}} 
  \left(\frac{\nu}{p_q^+} \right)^{\eta}
  - \epsilon (1-z) \right]
\int \frac{d^{D-2} k_\perp}{(2\pi)^{D-2}} \frac{d^{D-2}l_\perp}{(2\pi)^{D-2}} \, \, 
\nn \\ 
&& \hspace{-6.ex}  \times 
 \int d^{D-2}b_\perp d^{D-2}b_\perp' d^{D-2}r_\perp
 \frac{l_\perp \cdot k_\perp }{l_\perp^2 k_\perp^2}
e^{i(k_\perp+l_\perp) \cdot x_\perp} \, 
e^{i(1-z)p_{j\perp}\cdot x_\perp }
e^{-ip_{j\perp}\cdot z_\perp  }
S_{X_f}^{(3)}(b_\perp,r_\perp,b_\perp')\,, \quad \quad 
\eea 
where we have summed over the spin and averaged over the color. We have done the variable change 
$k_\perp \to k_\perp - (1-z)p_{j\perp}$ and used the color identities in Eq.~(\ref{eq:S3}).

The integral over the loop-momenta $k_\perp$ and $l_\perp$ can be carried out straightforwardly, by using the integral formula Eq.~(\ref{eq:qoverq2-Fourier}) in the Appendix. Plugging the matrix element into the phase space integral in Eq.~(\ref{eq:LO-phase-integral}), and performing the $\eta$ and $\epsilon$ expansions consecutively, we find that the virtual correction to the jet cross section is given by
\bea\label{eq:coll-virtual}
	d\sigma^{(1)}_{V, coll.}	&=& 
	\tau f(\tau)
	\frac{\alpha_s }{2\pi}\frac{N_C}{2}  
	\int 
\frac{d^{D-2} b_{\perp} d^{D-2}b'_{\perp}}{4\pi^{2-\epsilon}}\,
e^{-ip_{J\perp}\cdot z_\perp} 
\Bigg\{  
\left(-\frac{3}{\epsilon} - 1 \right) 
\left(\frac{p_{J\perp}}{\mu}\right)^{-2\epsilon}
S_{X_f}^{(2)}(b_{\perp};b'_\perp)  \nn \\
\, 
&& \hspace{8.ex} +\int \frac{dr_\perp}{\pi} 
\left[ 
		\frac{-2}{\eta}  \left(\frac{\nu}{p_q^+} \right)^\eta 
	 \, \Gamma^2(1-\epsilon) 
	\Bigg( \frac{1}{x_\perp^2}
	(x^2_\perp\mu)^{2\epsilon}
	+ \frac{1}{y_\perp^2}
	(y^2_\perp\mu)^{2\epsilon}	\Bigg)
	 \right.
 \nn \\
&  &\hspace{12.ex} \left.  + 
2 \left(
\frac{e^{ip_{J\perp} \cdot x_\perp} }{x_\perp^{2} }\right)_+ \, 
\int_{0}^{1}d{z}\frac{1+{z}^2}{(1-z)_+}
	 e^{-iz p_{J\perp} \cdot x_\perp} \,  \right] 
S^{(3)}_{X_f}(b_\perp,r_\perp,b_\perp') \Bigg\} 
\,, \quad \quad 
\eea 
where the `$+$'-prescription for a distribution ${\cal D}_+(x)$ has been defined in Eq.~(\ref{eq:plus-pres}).  
We note that the virtual correction is identical to the single hadron production~\cite{Chirilli:2011km,Liu:2019iml}. 

\subsection{The soft virtual correction}\label{sec:soft-virtual}
Now we turn to the virtual correction from the soft contribution, induced by the soft currents discussed in previous sections. The only non-vanishing contribution is given in fig.~\ref{fig:feyn-virt-soft}, which is 
\bea 
| {\cal M}_{V,soft} \rangle 
&=& - \,
g_s^2 \, \mu^{2\epsilon}
 \int \frac{dy}{4\pi}
\int \frac{d^{D-2}l_\perp}{(2\pi)^{D-2}} 
  \frac{d^{D-2}k_\perp}{(2\pi)^{D-2}} 
\frac{4 k\cdot l }{k_\perp^2\, l_\perp^2}  
\left(\frac{l_\perp}{\nu}\right)^{-\eta/2} 
\left(\frac{k_\perp}{\nu}\right)^{-\eta/2} 
e^{-\eta  |y|}
\nn \\ 
&& 
\int 
d^{D-2}r_\perp 
e^{-i  k_\perp \cdot r_\perp} 
e^{-i l_\perp \cdot r_\perp } 
{\bf T}^a_j W_{ab}(r_\perp )  {\bf T}^b_i \, \, 
|{\cal M}_0 (l_\perp,p_{j\perp}-k_{\perp}) \rangle
\,. 
\eea 
Again, the loop integration can be worked out explicitly. When interfering with the LO amplitude, the soft virtual correction to the squared matrix element is found to be 
\bea 
&& 
\langle {\cal M}_0| \, {\cal M}_{V,soft} \rangle  + c.c. \nn \\ 
&=&  - 2 (p_q^+)^2 \,
\alpha_s \,   
\frac{16}{\eta} \left( 
  \frac{i}{(4\pi)^{1-\epsilon}} 
\frac{\Gamma(1-\epsilon-\eta/4)}{\Gamma(1+\eta/4)} 
\right)^2
\left( \frac{x_\perp^2}{4} \right)^{-1+2\epsilon+\eta/2} 
\mu^{2\epsilon} \nu^\eta 
\nn \\ 
&& 
\int d^{D-2}b_\perp   d^{D-2}b'_\perp d^{D-2}r_\perp 
e^{-i p_{j\perp}  \cdot r_\perp}  
{\rm Tr}\left[W^\dagger(b_\perp') t^a W(b_\perp) t^b \right]
W_{ab}(r_\perp ) \,. 
\eea 
Integrating over the phase space in Eq.~(\ref{eq:LO-phase-integral}) and utilizing the color identity in Eq.~(\ref{eq:color-identity}), 
we find the soft correction reads 
\bea\label{eq:soft-virtual}
 d\sigma_{V,soft}
&=& 
\tau f(\tau) \, 
\frac{\alpha_s}{2\pi}  \frac{N_C}{2}
\,
\frac{4}{\eta}
\frac{\Gamma^2(1-\epsilon-\eta/4)}{\Gamma^2(1+\eta/4)}
\int
\frac{ d^{D-2}b_\perp d^{D-2}b_\perp'}{4\pi^{2-\epsilon}} \frac{d^{D-2}r_\perp }{\pi}
e^{-ip_{J\perp} \cdot z_\perp }
S_X^{(3)}(b_\perp,r_\perp,b_\perp') \nn \\
&& \times 
\left(
\frac{1}{x_\perp^2}
\left(\frac{x_\perp^2\nu^2}{c_0^2}\right)^{\eta/2}
(x_\perp^2\mu)^{2\epsilon} 
+
\frac{1}{y_\perp^2}
\left(\frac{y_\perp^2\nu^2}{c_0^2}\right)^{\eta/2}
(y_\perp^2\mu)^{2\epsilon} 
\right) \,,
\eea 
where $c_0 = 2 e^{-\gamma_E}$ with $\gamma_E$ the Euler constant. 
\begin{figure}[htbp]
    \centering
    \subfigure{\includegraphics[width=0.55\textwidth]{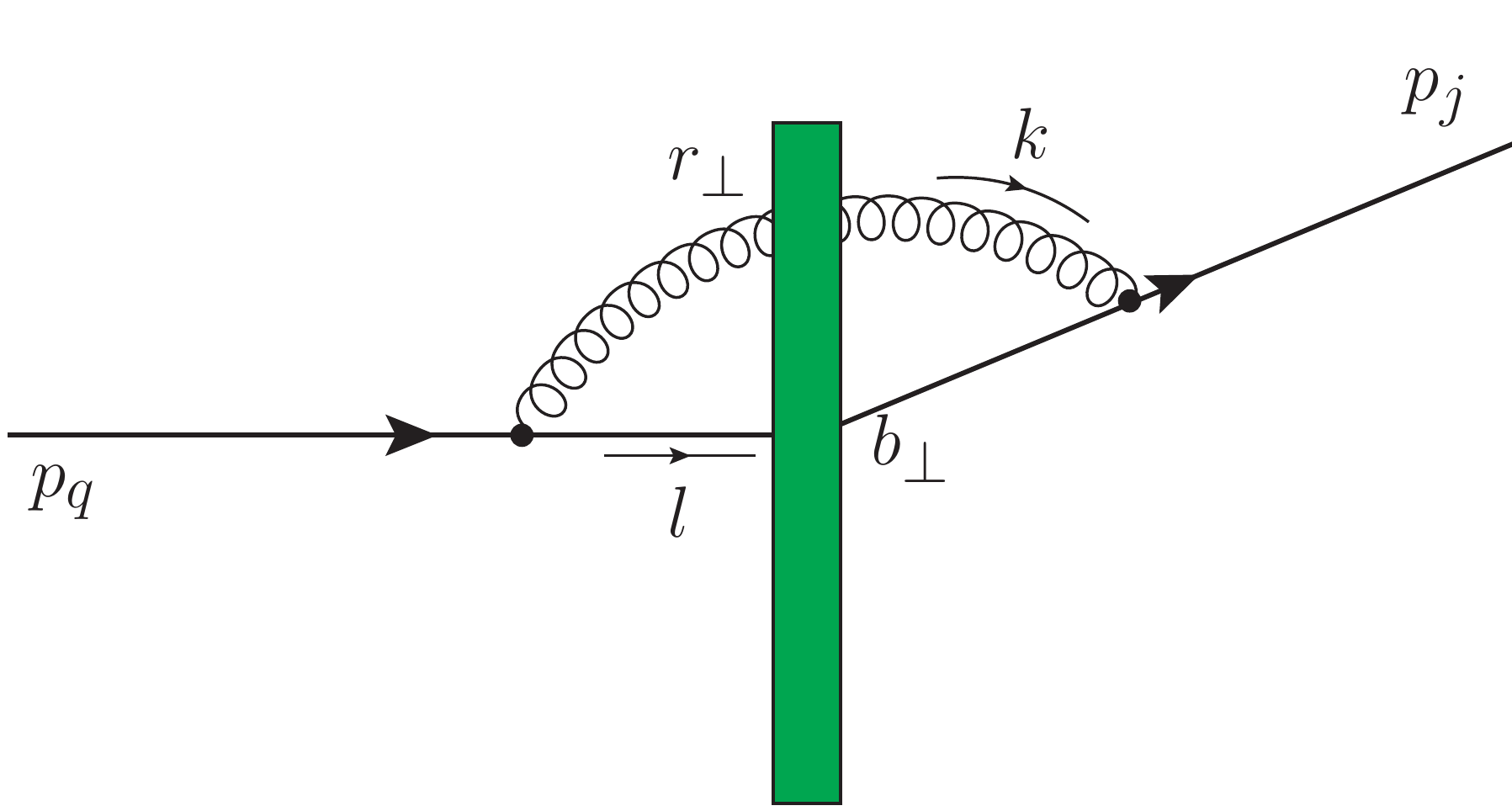}} 
    \caption{The Feynman diagram for the non-vanishing soft virtual correction.}
    \label{fig:feyn-virt-soft}
\end{figure}

\subsection{Real Corrections}
In this section, we go over the real corrections to the inclusive forward jet production in great detail. Unlike the LO and virtual corrections, now we have $2$ partons in the final state and we need to decide when these $2$ particles are clustered into one single jet and when they form distinct jets separately. Therefore, the jet algorithm now imposes non-trivial constraints on the real correction phase space.

We will briefly review the anti-$k_T$ jet algorithm and discuss its restrictions on the kinematics of the $2$-body phase space. Meanwhile, we will study various kinematic limits which are relevant to our NLO calculation. Specifically, we will focus on the collinear, soft and rapidity limits, which will lead to singular behaviors when evaluating the real phase space integrals. 

We will explain the subtraction strategy that handles the singularities in these kinematic limits and present the explicit form of the subtraction terms. 

\subsection{The anti-$k_T$ jet algorithm}\label{subsec:antikt}
Within the anti-$k_T$ jet algorithm, given a list of particles with momenta $\{k_i^\mu \}$,  the distance metrics $\rho_{iB}$ between particle $i$ and the beam $B$, and $\rho_{ij}$ between particles $i$ and $j$ are introduced, in terms of their transverse momenta $k_{i,\perp}$, the azimuthal angles $\phi_i$ and the rapidities $\eta_i$, where~\cite{Cacciari:2008gp}
\bea 
\rho_{iB} = k_{i,\perp}^{-2} \quad
{\text{and}} \quad 
\rho_{ij} = \min( k_{i,\perp}^{-2},k_{j,\perp}^{-2}) \frac{\Delta R_{ij}^2}{R^2} \,.
\eea 
Here $R$ is the jet radius parameter. The geometric distance $\Delta R_{ij}^2$ between particles $i$ and $j$ is given by 
\bea
\Delta R_{ij}^2
= (\phi_i-\phi_j)^2
+ (\eta_i-\eta_j)^2 \,. 
\eea
If $\rho_{ij}$ is the smallest of all the $\rho_{iB}$ and $\rho_{ij}$, particles $i$ and $j$ are clustered to form a pseudo-jet, whereas if $\rho_{iB}$ is the smallest, the $i$-th particle/pseudo-jet is promoted to a jet and removed from the list. The procedure continues until the list is empty and all particles in the list will fall into one of the jets constructed. Additional criteria on the jet transverse momentum $p_{J,\perp}$ or rapidity $\eta_J$ are usually imposed and if the constructed jet passes the criterion thresholds, it is an observed jet, otherwise it is not counted. 

As for the real corrections to the single inclusive jet, the jet clustering procedure will impose kinematic constraints on the phase space of the $2$ real emissions, which can be written as
\bea\label{eq:anti-kt-constrain}
\Theta_{J}  &=& \int d^4p_J \,  \Big\{ 
\Theta(\Delta R_{jk}^2 - R^2)\Big(\delta^{(4)}(p_J - p_j)
+\delta^{(4)}(p_J - p_k) 
\Big) \nn \\ 
&& \hspace{10.ex}
+ \Theta(R^2 - \Delta R_{jk}^2  )\delta^{(4)}(p_J - p_j-p_k)  \Big\}  \,.
\eea 
Such constraints can be understood as follows. When the separation between two particles $p_j$ and $p_k$,  $\Delta R_{jk}$ is greater than $R$, then $j$ and $k$ can either be the signal jet and the jet momentum is given by the momentum of the individual particle. On the other hand, if $\Delta R_{jk} < R$, the $2$ particles are clustered into the single jet and the jet momentum  $p_J$ is the sum of the particle momenta. We should bear in mind that there are additional possible restrictions on the jet momentum $p_J$ (e.g. requirements from experimental cuts) which are suppressed in Eq.~(\ref{eq:anti-kt-constrain}).

\subsection{The phase space measure and the kinematic limits}\label{subsec:kin}
For the real emission, we are dealing with $2$ radiations in the final state. The phase space is given by
\bea 
d\Phi_J
= \frac{1}{2}  
\int_0^1 dx f(x) \, 
\frac{1}{p_q^+}
\int \frac{d^Dp_j}{(2\pi)^{D-1}} \delta(p_j^2) 
\frac{d^Dp_k}{(2\pi)^{D-1}} \delta(p_k^2)
2(2\pi) \delta(p_q^+- p_j^+ - p_k^+) \Theta_J \,, 
\eea 
where in our example $q\to q+g$ we denote the out-going quark momentum as $p_j$ and the gluon momentum $p_k$, respectively. We have included the spin average factor, the flux, the PDF $f(x)$ and the Bjorken-$x$ integration as well as the phase space restriction due to the jet clustering procedure in Eq.~(\ref{eq:anti-kt-constrain}). Here the in-coming quark momentum $p_q^+ = x p_p^+$, with $p_p$ the momentum of the incoming proton in $pA$ collisions. 

If we let $1- \xi = \frac{p_k^+}{p_q^+}$, we can manipulate the phase space to get 
\bea\label{eq: pscoll}
d\Phi_J = 
\frac{1}{8\pi}
\frac{1}{ p_p^+}
\int  
   \frac{dp_J^+}{ p_J^+ } 
\frac{d^{2}p_{J\perp}}{(2\pi)^{2}} 
\int^1 \frac{d\xi }{ \xi(1- \xi) }
\frac{d^{D-2}p_{k\perp}}{(2\pi)^{D-2}} \,
f(x) \, \frac{1}{p_q^+}
\Theta(1-x)\,  \, 
\Big(\Theta_1 + \Theta_2 \Big) \,,   \quad 
\eea 
where $x = \frac{p_j^+ + p_k^+}{p_p^+} = \frac{p_j^+}{p_J^+} \frac{\tau}{\xi}$, $\tau=\frac{p_J^+}{p_p^+}$, and 
\bea \label{eq:clusterTH}
&& \Theta_1 = \Theta(R^2 - \Delta R_{jk}^2  )|_{p_j^+ = p^+_J - p_k^+, p^\mu_{j\perp} = p^\mu_{J\perp} - p^\mu_{k\perp},x = \tau  } \,, \nn \\ 
&& \Theta_2 = \Theta( \Delta R_{jk}^2 - R^2  )|_{p_j^+ = p^+_J, p^\mu_{j\perp} = p^\mu_{J\perp}} \,, 
\eea 
\textcolor{orange}{Here we have chosen to parameterize the phase space using the jet momentum $p_J^+$, $p_{J\perp}$ and the gluon momentum $p_k$. The gluon can be un-resolvable when it is soft or collinear to either the out-going or in-coming quarks. The momentum for the outgoing quark momentum can be generated via $p^+_j = p^+_J - p_k^+$, $p^\mu_{j,\perp} = p^\mu_{J,\perp} - p^\mu_{k,\perp}$ and $p_{j}^- = \frac{p_{j\perp}^2}{p_j^+}$ for $1$ jet and one demands that the constructed $p_j^\mu$ must satisfy $\Theta_1$. While for the $2$-jet case, $p^+_j = p^+_J$, $p^\mu_{j\perp} = p^\mu_{J\perp}$ and $\Theta_2$ should be fulfilled. 
Once we have $p_j^\mu$ and $p_{k}^\mu$, all the other physical variables such as the jet energy $E_J$ and the rapidity $\eta_J$ can be easily constructed in a  straightforward way. } 

For later use, let us study several kinematic limits here. 
\begin{itemize} 
\item $\xi = 1$ rapidity limit. 
In the rapidity limit, if the gluon transverse momentum $p_{k\perp} \ne 0$, the gluon is actually moving backwards with very large $p_{k}^-$ due to the fact that $p_k^- = \frac{p_{k\perp}^2}{(1-\xi) p_q^+} \gg p_k^+$. Hence the backward gluon and the out-going forward quark are highly unlikely to be clustered into $1$ jet and therefore $\Theta_1 \to 0$ while $\Theta_2 \to 1$ in this situation.
More specifically, we have 
\begin{align}\label{eq:etajk}
\Delta \eta_{jk}^2
=(\eta_j-\eta_k)^2
=\left[\frac{1}{2}\ln\left(\frac{p_j^+}{p_j^-}\right)-\frac{1}{2}\ln\left(\frac{p_k^+}{p_k^-}\right)\right]^2=\left[\frac{1}{2}\ln\left(\frac{\xi^2}{(1-\xi)^2}\frac{p_{k\perp}^2}{p_{j\perp}^2}\right)\right]^2\,,
\end{align} 
and we can see that as $\xi = 1$ while $p_{k\perp} \ne 0$, $\Delta \eta_{jk} \gg R$ to prevent $j$ and $k$ to be clustered together, as expected. 
We note that this backward-moving contribution will eventually be subtracted automatically when appropriate rapidity schemes are implemented. On the other hand, if $p_{k\perp} = 0$, we will have $\Theta_1+\Theta_2 = 1$ in the rapidity limit. All in all, as $\xi \to 1$, $\Theta_1+\Theta_2 = 1$. This features will help us to construct the counter term in the following sections.

\item $p_k || p_j$ final-final collinear limit. When $p_k$ and $p_j$ are collinear to each other, they will always be put into one jet to find $\Theta_1 = 1$ and $\Theta_2 = 0$. Also from Eq.~(\ref{eq:etajk}), we find that in this limit since $\Delta \eta_{jk} =0$, we have $\xi p_{k_\perp} = (1-\xi) p_{j\perp}$ or $p_{k\perp} = (1-\xi) p_{J\perp}$.

\item $p_k || p_q$ intial-final collinear limit. In this limit $p_{k\perp} = 0$ and as long as $\xi \ne 1$, which means the gluon is propagating along the beam and can no way be clustered with the out-going quark, then one has $\Theta_2 = 1$ and $\Theta_1 = 0$; meanwhile if $\xi = 1$, $\Theta_1+\Theta_2 = 1$, as one can see from Eq.~(\ref{eq:clusterTH}).

\item The soft gluon $p^\mu_k  \sim p_{J\perp} \ll p_{p}^+$ limit. 
In the soft limit, as the gluon momentum $p_k^\mu \to 0 $, the out-going quark momentum $p_j^+ \to p_q^+ = p_J^+$ and $x \to \tau$. By homogeneous power expansion, the $p_k^+$ component will be completely dropped from the $\delta$-function or the $\Theta$'s in the phase space. As a consequence, in the soft limit, there will be no restrictions on $p_k^+$ or equivalently the gluon rapidity $\eta_k$. Therefore, the phase space will become 
\bea\label{eq:pssoft}
d\Phi_{J,soft} = 
\frac{1}{8\pi}
\frac{1}{ p_p^+}
\, 
   \frac{dp_J^+}{ p_J^+ } 
\frac{d^{2}p_{J\perp}}{(2\pi)^{2}}  \int_{-\infty}^\infty 
d\eta_k \, f(\tau) \, 
\frac{d^{D-2}p_{k\perp}}{(2\pi)^{D-2}}  
\frac{1}{p_q^+}\, 
\Big(\Theta_{1,soft} + \Theta_{2,soft} \Big)  \,,
\eea 
where $x = \tau$, and 
\bea\label{eq:soft-limit-jet} 
&& \Theta_{1,soft} = \Theta(R^2 - \Delta R_{jk}^2  )|_{p_j^+ = p^+_J\,, p^\mu_{j\perp} = p^\mu_{J\perp} - p^\mu_{k\perp} } \,, \nn \\ 
&& \Theta_{2,soft} = \Theta( \Delta R_{jk}^2 - R^2  )|_{p_j^+ = p^+_J\,, p^\mu_{j\perp} = p^\mu_{J\perp}} \,. 
\eea 

Also we note that, if $p_k \to 0$, $\Theta_{1,soft}+\Theta_{2,soft}=1$. And in the rapidity limit when $|\eta_k| \to \infty$, for a jet with finite $\eta_J$, the partons $j$ and $k$ are unlikely to be grouped into one jet and therefore $\Theta_2 \to 1$ and $\Theta_1 \to 0$.

\item Small $R$ limit. In this work, besides the predictions with the full jet algorithm dependence, we will also present analytic results for the jet production in the small-$R$ limit, $R \ll 1$. It is ready to show that in the small-$R$ limit, we have 
\bea 
\Delta R_{jk}^2
= \frac{2 p_{j} \cdot p_{k} }{p_{j\perp} p_{k\perp}} 
= \frac{(\xi p_{k\perp} - (1-\xi)p_{j\perp})^2}{\xi(1-\xi)p_{j\perp} p_{k\perp}}
\equiv \frac{k_\perp^2}{\xi(1-\xi)p_{j\perp} p_{k\perp}}
\,, 
\eea 

up to ${\cal O}(R^2)$ corrections~\cite{Liu:2013hba}. Here we have defined $k_\perp = \xi p_{k\perp}-(1-\xi)p_{j\perp}$ in the last step, and we immediately finds that in the $R\ll 1$ limit
\bea \label{eq: small-r-jet}
&& \Theta_1 \to \Theta_{1,R} = 
\Theta\Big(
\xi^2(1-\xi)^2  R^2 p^2_{J\perp}  - 
k_\perp^2
\Big) |_{p_j = \xi p_J}
\,, \nn \\ 
&& \Theta_2 \to \Theta_{2,R} = \Theta \Big( 
k_\perp^2 -  (1-\xi)^2 R^2 p^2_{J\perp} 
\Big)|_{p_j = p_J}  \,, 
\eea 
where we have used that for $1$ jet,
$p_{j\perp} = \xi p_{J\perp} $ and $p_{k\perp} = (1-\xi) p_{J\perp} $ in the small-$R$ limit, as can be seen from Eq.~(\ref{eq:etajk}), while for
$2$ jets, $p_{j\perp} = p_{J\perp}$ and $p_{k\perp}  =  \frac{1-\xi}{\xi}p_{J\perp}$. Apparently, in the limit $\xi \to 1$, we have
\bea 
\Theta_{1,R} + \Theta_{2,R} =1  \,. 
\eea 

{\color{orange}{\it We note that the $1$ jet phase space area scales as $R^2$ and could be highly suppressed in the small $R$ limit, unless there is $R^{-2}$ compensation from the matrix element, which is the case when $p_k || p_j$ and the matrix element becomes singular.}}

\end{itemize}

\subsection{The collinear contribution}
We now detail the calculation of the collinear real corrections below. 
\begin{figure}[htbp]
    \centering
    \subfigure{\includegraphics[width=0.45\textwidth]{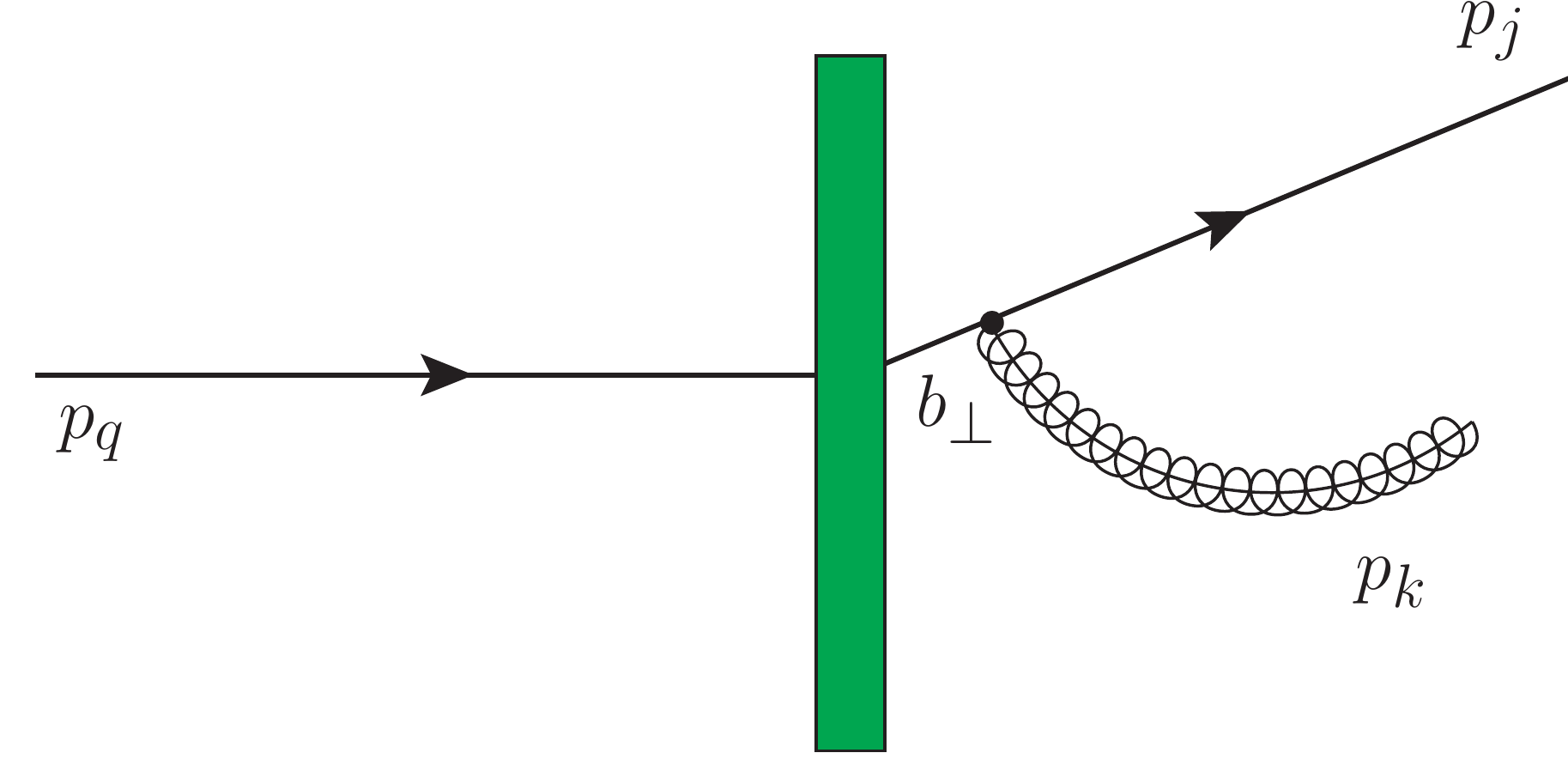}}     
    \subfigure{\includegraphics[width=0.45\textwidth]{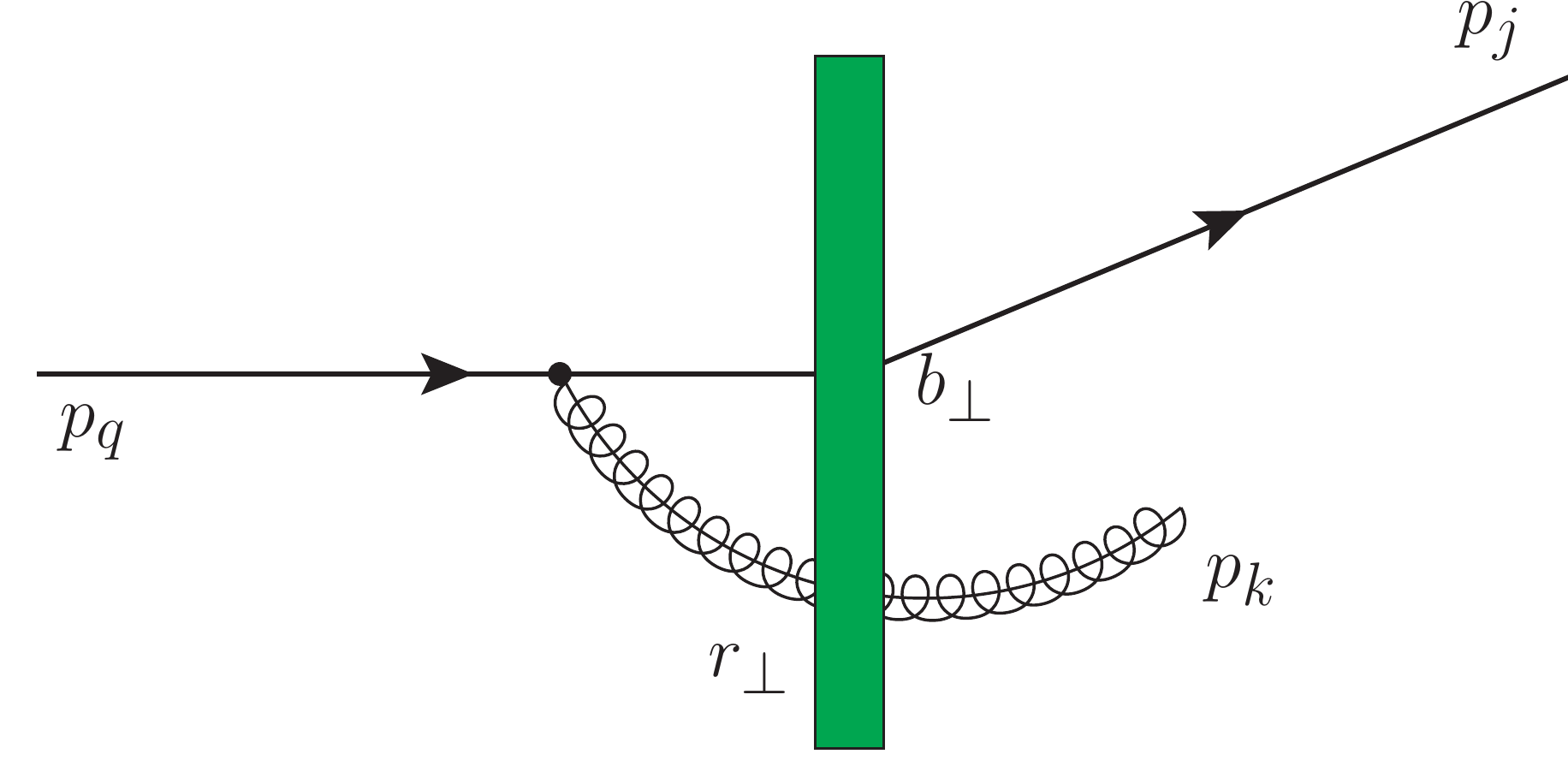}} 
    \caption{Feynman diagrams for the collinear real emissions: (a) final state radiation (left figure); (b) initial state radiation (right figure).}
    \label{fig:feyn-real}
\end{figure}

\subsubsection{The collinear matrix element}
At the NLO, two configurations contribute to the real correction of the quark channel, which we denote as the final state radiation (FSR) if the gluon is emitted after interacting with the shock wave (left figure) and the intial state radiation (ISR) if the gluon is radiated before the interaction (right figure), see fig.~\ref{fig:feyn-real}. The matrix elements are 
\bea 
|{\cal M}_{FSR,coll.}  \rangle 
= 
-g_s \,{\bf T}^a_j
 \,   
\left(
n^\alpha + \frac{\gamma_\perp^\alpha \lpslash}{l^+ }
+ \frac{\pjslash\gamma_\perp^\alpha}{p_j^+}
\right) \, 
 \frac{l^+}{l^2} \, 
 |{\cal M}_0(p_{q\perp},p_{j\perp}+p_{k\perp}) \rangle \,, 
\eea   
for the FSR diagram and 
\bea 
|{\cal M}_{ISR,coll.} \rangle 
&=& - g_s 
\int \frac{d^{D-2}l_\perp}{(2\pi)^{D-2}} 
\left(
2l_\perp^\alpha \frac{l^+}{p_q^+} 
+ \frac{p_k^+}{p_q^+} 
\lpslash \gamma_\perp^\alpha
\right) 
\nn \\
&& \times 
\int 
d^{D-2}r_\perp 
\frac{e^{-i(k_\perp+l_\perp)\cdot r_\perp }}{l_\perp^2}
W_{ab}(r_\perp){\bf T}^b_i \, 
|{\cal M}_0(l_\perp,p_{j\perp}) \rangle 
\,, \quad \quad 
\eea 
for the ISR diagram, respectively. Square the sum of both amplitudes and carry out the phase space integral in Eq.~(\ref{eq: pscoll}), we find that the collinear real correction can be written as the sum of three different pieces 
\bea\label{eq13:NLOr} 
d\sigma_{R,coll.}
=  d\sigma_{fsr} + d\sigma_{isr} + d\sigma_{inter.} \,, 
\eea 
where 
\bea 
&& d\sigma_{fsr} 
=  \frac{ \alpha_s}{2\pi^2}
\frac{N_C}{2} \int d\xi 
\, x f(x) \, \Big(\Theta_1 + \Theta_2 \Big) 
\left( \frac{1+\xi^2}{(1-\xi)^{1+\eta}} - \epsilon(1-\xi) \right) 
\left(\frac{\nu}{p_q^+} \right)^{\eta}
\mu^{2\epsilon }
\nn \\ 
&& \times \int 
\frac{d^{D-2}b_\perp d^{D-2}b_\perp' }{(2\pi)^{D-2}}
d^{D-2}p_{k\perp}
\, 
\, \frac{\Gamma^2(1-\epsilon)
}{(\xi p_{k\perp}-(1-\xi)p_{j\perp})^2}
\, 
e^{-i(p_{k\perp}+p_{j\perp}) \cdot z_\perp}
S_{X_f}^{(2)}(b_\perp,b'_\perp) \,, \quad \quad 
\eea 
which is the contribution from the FSR in which the clustering conditions $\Theta_1$ and $\Theta_2$ are given in Eq.~(\ref{eq:clusterTH}). On the other hand,  
\bea 
&& d\sigma_{isr} = 
\,    \frac{ \alpha_s}{2\pi^2}
\frac{N_C}{2} \int d\xi \, d^{D-2}p_{k\perp} \, 
\, x f(x) \, \Big(\Theta_1 + \Theta_2 \Big) 
\left( \frac{1+\xi^2}{(1-\xi)^{1+\eta}} - \epsilon(1-\xi) \right) 
\left(\frac{\nu}{p_q^+} \right)^{\eta}
\mu^{2\epsilon }
\nn \\ 
&& \times 
\int \, 
\frac{d^{D-2}b_\perp d^{D-2}b_\perp' }{(2\pi)^{D-2}}
\, 
\frac{d^{D-2}r_\perp 
d^{D-2}r_\perp'}{4\pi^{2-2\epsilon}}
\,
 e^{-ip_{j\perp}\cdot z_\perp}
 e^{-i p_{k\perp} \cdot z_\perp'}
\frac{x_\perp \cdot x_\perp'}{x_\perp^{2-2\epsilon} {x_\perp'}^{2-2\epsilon}}
S_{X_f}^{(6)}(b_\perp,r_\perp,b_\perp',r_\perp') \,, 
\quad \quad 
\eea 
represents the ISR contribution where we have integrated over the loop momenta $l_\perp$'s  and let $x_\perp' = b_\perp' - r_\perp'$ and $z_\perp' = r_\perp - r_\perp'$. Here, the $6$-point correlator was introduced previously in Eq.~(\ref{eq:6-point}) and Eq.~(\ref{eq:color-identity}).

The $\sigma_{inter.}$  in Eq.~(\ref{eq13:NLOr}) denotes the billinear term coming from the interference between the ISR and the FSR, which is 
\bea 
&& d \sigma_{inter.} = 
\,    \frac{ \alpha_s}{2\pi^2}
\frac{N_C}{2}
\mu^{2\epsilon }
\int d\xi \, 
\int d^{D-2}p_{k\perp} \,
\frac{d^{D-2}b_\perp d^{D-2}b_\perp' }{(2\pi)^{D-2}}
d^{D-2}r_\perp\, 
\, x f(x) \, \Big(\Theta_1 + \Theta_2 \Big) \nn \\ 
&& \times  
\left( \frac{1+\xi^2}{(1-\xi)^{1+\eta}} - \epsilon(1-\xi) \right)\left(\frac{\nu}{p_q^+} \right)^{\eta}
\nn \\ 
&& \times 
\int 
 \,  \frac{d^{D-2}l_\perp}{(2\pi)^{D-2}}
\frac{2(\xi p_{k\perp}-(1-\xi )p_{j\perp}) \cdot l_\perp }{(\xi p_{k\perp}-(1-\xi )p_{j\perp})^2 l_\perp^2}
e^{il_\perp \cdot y_\perp } 
e^{-ip_{j\perp}\cdot z_\perp }
e^{- i p_{k\perp} \cdot x_\perp}
S_{X_f}^{(3)}(b_\perp,r_\perp,b_\perp') 
 \,.    \quad 
\eea 
We remind that in all contributions, when the matrix elements hit $\Theta_1$ and $\Theta_2$, the Bjorken $x$ and the $p_{j}$ will be replaced by $p_J$ and $p_k$ following the replacement rules in  Eq.~(\ref{eq:clusterTH}) accordingly. We note that unlike the hadron case, for the ISR contribution, the $p_{k\perp}$ integral can not be performed inclusively and thus analytically to reduce the quadrupole structure to dipole, due to the existence of the jet algorithm.

\subsubsection{The idea of subtraction}\label{subsubsect:subtraction}

Now we are at the stage to evaluate the real correction in  Eq.~(\ref{eq13:NLOr}), which contains  collinear and rapidity singularities. For instance, Eq.~(\ref{eq13:NLOr}) involves contributions such as 
\bea\label{eq:pole-example}
d\sigma_{fsr} & = 
&\frac{\alpha_s S_\perp}{2\pi^2}
\frac{N_C}{2} 
\int d\xi \, x f(x)
\left( 
\frac{1+\xi^2}{(1-\xi)^{1+\eta}}
-\epsilon(1-\xi) 
\right)
\left(\frac{\nu}{p_q^+}\right)^\eta\,
\mu^{2\epsilon}
\nn \\
&&
\times 
\int d^{D-2}p_{k\perp} \Bigg\{\Theta_2\,   
+\Theta_1\,  
\Bigg\} 
\frac{\mathcal{F}_F(p_{k\perp}+p_{j\perp};X_f)}{
[\xi p_{k\perp}-(1-\xi)p_{j\perp}]^2
}
\,, 
\eea 
%
%
where the rapidity and the final-final collinear poles arise when $\xi \to 1$ and/or $p_{k\perp} \to 0 $ and when $\xi p_{k\perp} \to (1-\xi) p_{j\perp}$, respectively. 
Here we have applied the Fourier transformation in Eq.~(\ref{eq:dipole-fourier}). 

Different from the forward hadron production, where the integration can be performed analytically and the singularities can thus be extracted in a straightforward way, here in the jet production, the integration is complicated by the existence of the jet clustering algorithm, represented by the $\Theta_1$- and $\Theta_2$-functions. The clustering hinders the analytic computation of the integration and in turn makes the pole isolation less obvious.

In order to deal with this complication, we appeal to the subtraction method and construct counter terms to isolate the poles. The idea can be highlighted by the following simple example. Suppose that we are dealing with the integral $\int_0^1 dx f(x)/x^{1+\epsilon}$, here $f(x)$  is regular but takes a complicated form involving the jet clustering algorithm. The integral can barely be done analytically. Meanwhile the $x\to 0$ singularity also 
discourages the direct numerical integration. 

We can however manipulate the integral to find 
\begin{align}\label{eq:counter1}
\begin{split}
\int_0^1 dx \frac{f(x)}{x^{1+\epsilon}}
=& 
\int_0^1 dx \left(\frac{f(x)}{x^{1+\epsilon}}-\frac{f(0)}{x^{1+\epsilon}}\right)+f(0) \, \int_0^1  \frac{dx}{x^{1+\epsilon}}
\nonumber\\
=& \int_0^1  
\frac{f(x)-f(0)}{x}   - \frac{f(0)}{\epsilon} 
\end{split}
\end{align}
where we subtracted a counter term $x^{-1-\epsilon}\, f(0)$ and then added it back. The counter term is built in such a way that it contains all the infrared, collinear and rapidity (here, denoted as $x\to 0$) singular behaviour of the original integrand $x^{-1-\epsilon}f(x)$. As a result, 
in the first term of the first equation, the combination is everywhere regular in the domain $[0,1]$. We can thus safely set $\epsilon = 0$ to get the second equation. Although $f(x)-f(0)$ remains complicated, since the integral is now finite, we can simply evaluate this term numerically. 

All the $x\to 0 $ singularities are now contained only in the counter term. The good thing about the counter term is that it is constructed as the infrared and collinear limit of a full theory. As a consequence, $f(0)$ is usually significantly simplified when compared to $f(x)$. Especially, due to the infrared-collinear-safety (IRC safe) feature of the jet clustering algorithms, all the jet algorithm dependence will become trivial and reduce to $1$ within the counter term $f(0)$. This eventually allows us to perform the integration over the counter term analytically to extract the poles.

\subsubsection{The collinear contribution}\label{subsubsec:coll-sub}
Back to Eq.~(\ref{eq13:NLOr}), let us see how the subtraction term is constructed. To do so, we need to analyze the infrared behavior of the integrand, which as guaranteed by QCD, only possesses singularities arising from the soft and collinear limits, plus the $\xi \to 1$ rapidity pole due to the power expansion performed in the kinematic constraints. 
Now we take a look at the behaviour of the integrand of $d\sigma_{fsr}$ in those limits one by one: 
\begin{itemize} 
\item 
When approaching the rapidity limit, as $\xi \to 1$, the gluon goes backwards, therefore $\Theta_2 \to 1$ and $\Theta_1 \to 0$ (see prevsious analysis in section~\ref{subsec:kin}). The integrand will behave as
\bea 
\lim_{\xi \to 1}d\sigma_{fsr} 
\propto 
\tau f(\tau) \frac{2}{(1-\xi)^{1+\eta}} 
\frac{{\cal F}_F(p_{k\perp}+p_{J\perp};X_f)}{[p_{k\perp}-(1-\xi)p_{J\perp}]^2}\,,
\eea 
where we have explicitly replaced $p_{j\perp}$ by $p_{J\perp}$ since the matrix element hits $\Theta_2$. 

\item In $d\sigma_{fsr}$, there is only collinear singularity when $\xi p_{k\perp} \to (1-\xi) p_{j\perp}$ in the denominator, which indicates that $p_k||p_j$. The collinear singular behavior only arises in the $1$ jet case, since when the $2$ final state partons are so close to each other, the jet algorithm will always cluster them into one single jet and therefore $\Theta_2 \to 0$. Since the $2$-jet condition will never be satisfied, we have $\Theta_1 \to 1$. Therefore the integrand approaches 
\bea 
\lim_{p_k||p_j}
d\sigma_{fsr} \propto 
\tau f(\tau) \frac{1+\xi^2-\epsilon(1-\xi)^2}{(1-\xi)^{1+\eta}} \frac{{\cal F}_F(p_{J\perp};X_f)}{[p_{k\perp}-(1-\xi)p_{J\perp}]^2}\,,
\eea 
where we have replaced $p_{j\perp}$ by $p_{J\perp}-p_{k\perp}$ since the matrix element now hits $\Theta_1$. 

\item The soft singularity arises when the gluon becomes soft and thus both $\xi \to 1$ and $p_{k\perp} \to 0$. When this happens, $\Theta_1+\Theta_2 \to 1$ (again, see section~\ref{subsec:kin} for details) and $d\sigma_{fsr}$ becomes
\bea 
\lim_{p^0_k \to 0}
d\sigma_{fsr} \propto \tau f(\tau) 
\frac{2}{(1-\xi)^{1+\eta}}\frac{{\cal F}_F(p_{J\perp};X_f)}{[p_{k\perp}-(1-\xi)p_{J\perp}]^2} \,.
\eea 
\end{itemize} 
The demanded subtraction term should satisfy all the singular behaviours, and one option is to construct 
\bea\label{eq:counter}
d\sigma_{fsr}^{c}
=\frac{\alpha_s S_\perp}{2\pi^2}\frac{N_C}{2}
\tau f(\tau)  
\int_{0}^1 d\xi
\frac{1+\xi^2-\epsilon(1-\xi)^2}{(1-\xi)^{1+\eta}}\left(\frac{\nu}{p_q^+}\right)^\eta\, \int
d^{D-2}p_{k\perp}
\frac{\mathcal{F}_F(p_{k\perp}+\xi p_{J\perp};X_f)}{[p_{k\perp}-(1-\xi)p_{J\perp}]^2} \,, \nn \\
\eea
where the superscript $c$ represents the counter-term. This term furnishes the point-wise approximation to the $d\sigma_{fsr}$ in all the singular regions and therefore renders the subtracted combination 
\bea\label{eq:theta1-0}
d\sigma_{fsr}-d\sigma^c_{fsr} &=&\frac{\alpha_s S_\perp}{2\pi^2}
\frac{N_C}{2}
\int_{0}^1d\xi
\frac{1+\xi^2}{1-\xi}
\int  d^2 p_{k\perp}  \Bigg\{\Theta_2 \, xf(x)\frac{\mathcal{F}_F(p_{k\perp}+ p_{J\perp};X_f)}{[\xi p_{k\perp}-(1-\xi)p_{J\perp}]^2} \nn \\
&& 
+\Theta_1 \, \tau f(\tau)
\frac{\mathcal{F}_F(p_{J\perp};X_f)}{[p_{k\perp}-(1-\xi)p_{J\perp}]^2}
-\tau f(\tau)\frac{\mathcal{F}_F(p_{k\perp}+\xi p_{J\perp};X_f)}{[p_{k\perp}-(1-\xi)p_{J\perp}]^2}\Bigg\} \,,
\quad \quad 
\eea 
finite. {\color{orange}{\it The regulators have been removed by setting $\epsilon = 0$ and $\eta = 0$ given that the combination is finite and integrable in $4$-dimension. The integral can be safely evaluated numerically}}. 

One note on the counter term is that the counter term $d\sigma_{fsr}^c$ is fully inclusive over $p_k$, in the sense that {\textcolor{orange}{{\it within the counter term, any jet algorithm and any additional cuts should be placed  on the momentum $p_J$ but never on $p_k$. It is effectively a one jet configuration and the gluon in the counter contribution is essentially un-resolvable, just like the virtual gluon in the loop.}}} The $d\sigma_{fsr}^c$ in Eq.~(\ref{eq:counter}) can be evaluated analytically, which gives
\bea\label{eq:counter-int}
d \sigma_{fsr}^{c,int}&=&\frac{\alpha_s}{2\pi}\frac{N_C}{2}
\tau f(\tau) 
\, \left( 
\frac{2}{\eta \epsilon}
+ \frac{3}{2 \epsilon } + \frac{1}{2}
\right)\left(\frac{\nu}{p_q^+}\right)^\eta
 \nn \\ 
&& \hspace{5.ex}
\times 
\int \frac{d^{D-2} b_{\perp} d^{D-2}b'_{\perp}}{4\pi^2}
\left(\frac{z_\perp \mu}{c_0} \right)^{2\epsilon}\,
e^{-i p_{J\perp}\cdot z_\perp}
S^{(2)}_X(b_\perp,b'_\perp)
\,,
\eea 
where $z_\perp = b_\perp - b_\perp'$. Once we add up Eq.~(\ref{eq:counter-int}) and Eq.~(\ref{eq:theta1-0}), we obtain the contribution from $d\sigma_{fsr}$.

Similar analysis can be applied to the terms which involve the initial-final collinear singularities, which leads to 
%
\bea\label{eq:if-sub}
&&
d\sigma_{isr} -
d \sigma_{isr}^c
=   
\frac{ \alpha_s}{2\pi^2}
\frac{N_C}{2}  
\int^1 d\xi 
 \,    
 \frac{1+\xi^2}{1-\xi} 
\int
d^{2}p_{k\perp} xf(x)
\Big( \Theta_1 +   \Theta_2 - 1 \Big)
\nn \\ 
&& 
\times \int 
\frac{d^2b_\perp d^2b_\perp' }{4\pi^{2}}
\, 
\frac{d^2 r_\perp 
d^2 r_\perp'}{4\pi^{2}}
\,
 e^{-ip_{j\perp}\cdot z_\perp}
 e^{-i p_{k\perp} \cdot z_\perp'}
\frac{x_\perp \cdot x_\perp'}{x_\perp^{2} {x_\perp'}^{2}}
S_{X_f}^{(6)}(b_\perp,r_\perp,b_\perp',r_\perp')\,,  
\eea 
where the $-xf(x)$ term is the subtraction and {\color{orange}{{\it we note that for the subtraction, we always have $p_{j} = p_{J}$ and again within the subtraction term, all kinematic cuts will be applied on the $p_{J}$ only and $p_k$ is un-resolvable just like the loop momentum. These hold for the rest of the section.} }}
To see the combination is finite we notice that, as $\xi \to 1$, $\Theta_1 \to 0$ and $\Theta_2 \to 1$, therefore the rapidity singularity is regulated by the vanishing value of the combination in the bracket in the rapidity limit. There is collinear singularity in $d \sigma_{isr}$, arising when the $p_{k\perp}$ is integrated to $\infty$ and restricts $z_\perp' = 0$. The collinear pole is regulated by the fact that 
\bea\label{eq:insidecone}
\tau f(\tau )\Theta_1 + xf(x) \Theta_2 - xf(x) 
= \Theta(R^2 - \Delta R^2_{jk})
(\tau f(\tau) - xf(x))|_{\text{with correct $p_j$ replacements }}\,, \quad \quad 
\eea 
will restrict the $p_k$ integration to be within the jet cone to remove the singularity. More physically, in the ISR contribution, the pole arises when the emitted gluon is collinear to the in-coming quark. However once the gluon is restricted within the jet cone, the configuration is then forbidden and therefore the integral is regular. Also we note that since it is free of final-final collinear singularities in $d\sigma_{isr}$ as well as in Eq.~(\ref{eq:if-sub}), Eq.~(\ref{eq:if-sub}) will be vanishing as $R \to 0$, since the phase space area allowed by Eq.~(\ref{eq:insidecone}) scales as $R^2$ as $R \to 0$, and there is no $R^{-2}$ compensation from the matrix element. The property will be used to derive the analytical cross section in the small $R$ limit. 

On the other hand, the integrated counter term for the ISR can again be evaluated analytically and we find
\bea \label{eq:if-sin-integrated}
d\sigma_{isr}^{c,int} &=&\frac{\alpha_s}{2\pi}\frac{N_C}{2} 
\left(\frac{\nu}{p^+_q}\right)^\eta\, 
\int_{\tau}^1 d\xi
\left(
\frac{2}{\eta\epsilon} \delta(1-\xi) 
- \frac{1}{\epsilon }\frac{1+\xi^2}{(1-\xi)_+}
+  1-\xi
\right)\, xf(x)
 \nn \\
&& \times  
\int \frac{d^{D-2}b_\perp d^{D-2}b'_\perp}{4\pi^2}
e^{-i   p_{J\perp} \cdot z_\perp}
\left(\frac{z_\perp\mu}{c_0}\right)^{2\epsilon}
S_X^{(2)}(b_\perp,b_\perp')
\,. \quad 
\eea 

Lastly we apply the strategy to the terms which come from the interference between the initial and final state radiations and have the bilinear structure of the dipole. These terms contain no collinear divergence, and all what we need is to subtract the rapidity divergence, which can be achieved by the subtraction as the follow
\bea \label{eq:ibkb}
d\sigma_{inter.}-d\sigma_{inter.}^c &=& - \frac{\alpha_s S_\perp}{2\pi^2} \frac{N_C}{2}
\int_{\tau}^1d\xi
\frac{1+\xi^2}{(1-\xi)}
\int d^{2}p_{k\perp}
\frac{d^{D-2}l_{\perp}}{(2\pi)^{D-4}} \, 
xf(x) \, 
\mathcal{F}_F(l_{\perp};X_f)
\nn \\
&&\hspace{-15.ex} \times \Bigg\{ (\Theta_2 
+\Theta_1 - 1 )\,  \, \mathcal{F}_F(p_{k\perp}+p_{j\perp};X_f)
\frac{2(\xi p_{k\perp}-(1-\xi)p_{j\perp})\cdot(l_{\perp}-p_{j\perp})}{(\xi p_{k\perp}-(1-\xi)p_{J\perp})^2(l_{\perp}-p_{j\perp})^2} \Bigg\} \,.  
\eea 
Here we have introduced the bilinear local counter term 
\bea\label{eq:c-bilinear}
d\sigma_{inter.}^c &=& - \frac{\alpha_s S_\perp}{2\pi^2}\frac{N_C}{2}
\int_{0}^1 d\xi
\frac{1+\xi^2+\frac{(D-4)}{2}(1-\xi)^2}{(1-\xi)^{1+\eta}}\left(\frac{\nu}{p_q^+}\right)^\eta
\frac{d^{D-2}p_{k\perp}}{(2\pi)^{D-4}}
\frac{d^{D-2}l_{\perp}}{(2\pi)^{D-4}} \nn \\
&
&
\times 
xf(x)  \mathcal{F}_F(l_{\perp};X_f)\mathcal{F}_F(p_{k\perp}^\prime;X_f) 
\frac{2(\xi p_{k\perp}^\prime-p_{J\perp})\cdot(l_{\perp}-p_{J\perp})}{(\xi p_{k\perp}^\prime-p_{J\perp})^2(l_{\perp}-p_{J\perp})^2} \,, 
\eea 
which leads to the integrated counter term found to be 
\bea\label{eq:theta3}
d\sigma_{inter.}^{c,int.}&=&  \frac{\alpha_s}{2\pi}\frac{N_C}{2} \tau f(\tau )
\frac{-2}{\eta}\left(\frac{\nu}{p_q^+} \right)^\eta 
\int
\frac{ d^{D-2} b_\perp d^{D-2}b_\perp' }{4\pi^{2-\epsilon}}\frac{dr_\perp}{\pi}\, 
\Gamma^2(1-\epsilon) \nn \\
&& \hspace{20.ex} \times \frac{2x_\perp \cdot y_\perp}{x_\perp^{2}
y_\perp^{2}}\, 
(x_\perp y_\perp \mu)^{2\epsilon}
e^{-ip_{J\perp} \cdot z_\perp } 
S^{(3)}_X(b_{\perp};r_\perp;b'_\perp) \nn  \\
& + & \frac{\alpha_s}{2\pi}\frac{N_C}{2}
\int_{\tau}^1 d\xi 
 \frac{1+\xi^2}{(1-\xi)_+}
 xf(x) 
 \int
 \frac{ d^2 b_\perp d^2b_\perp' }{4\pi^{2}}\frac{d^2r_\perp}{\pi}
 \nn  \\ 
&& \hspace{8.ex}\times 
\, 
\frac{2 x_\perp \cdot y_\perp}{x_\perp^{2}
y_\perp^{2}}\,
\frac{1}{\xi}
e^{-i \frac{1-\xi}{\xi} p_{J\perp} \cdot y_\perp}
e^{-ip_{J\perp} \cdot z_\perp } 
S^{(3)}_X(b_{\perp};r_\perp;b'_\perp) \,. 
\eea 
%
%
%
%
%
%
%
%
%

Add up all contributions, including the collinear virtual corrections in Eq.~(\ref{eq:coll-virtual}), we find that the NLO collinear contribution to the quark channel is given by
\bea
		d\sigma_{coll.} 
		& = & 
		\frac{\alpha_s}{2\pi} \frac{N_C}{2}
		\int \frac{ d^2 b_\perp d^2 b_\perp' }{4\pi^{2}}
			e^{-ip_{J\perp} \cdot z_\perp } 
			\int_\tau^1 d\xi xf(x) 
		\Bigg\{
		-  \frac{1}{ \epsilon } 
P_{qq}^{(1)}(\xi)
\, 
S_X^{(2)}(b_\perp,b_\perp') 
		 \nn \\
&&  \hspace{5.ex}
- \delta(1-\xi)  
\frac{2}{\eta}\left(\frac{\nu}{p_q^+} \right)^\eta
\int
\frac{d^2r_\perp}{\pi}\, 
\left[ \frac{z_\perp^2}{x_\perp^{2} 
y_\perp^{2}} \right]_+
\,
S^{(3)}_X(b_{\perp};r_\perp;b'_\perp)
\nn \\
&&\hspace{5.ex}
+ 
\, 
\left( 
{\cal  H}_{q,2} 
S_X^{(2)}(b_\perp,b_\perp') 
+ \int \frac{dr_\perp}{\pi}
{\cal H}_{q,3}
S_X^{(3)}(b_\perp,r_\perp,b_\perp')   \right)
\bigg\}
\nn \\
&& \hspace{5.ex}
+ (d\sigma_{fsr} - d\sigma^c_{fsr}) + (d\sigma_{isr}-d\sigma^c_{isr})+ (d\sigma_{inter.}-d\sigma^c_{inter.}) \,,  
\eea 
where $P_{qq}^{(1)}(\xi) = \left( \frac{1+\xi^2}{1-\xi} \right)_+$ is the $q\to q$ splitting function. And we have defined
\bea\label{eq:Hq2} 
{\cal H}_{q,2} = 
\left(- P_{qq}^{(1)}(\xi) \ln \frac{z_\perp^2\mu^2}{c_0^2} 
+1-\xi 
+ \left( 
3\ln \frac{z_\perp^2p_{J\perp}^2}{c_0^2}
-\frac{1}{2}
\right)\delta(1-\xi) 
\right) \,, 
\eea
and
\bea\label{eq:Hq3} 
{\cal H}_{q,3} &=& 
2 \frac{1+\xi^2}{(1-\xi)_+} 
\frac{ x_\perp \cdot y_\perp}{x_\perp^{2}
y_\perp^{2}}\,
\frac{1}{\xi}
e^{-i \frac{1-\xi}{\xi} p_{J\perp} \cdot y_\perp} \nn \\ 
&& \hspace{15.ex}
+ 2 \delta(1-\xi) 
\left[ 
\frac{e^{ip_{J\perp} \cdot y_\perp} }{y_\perp^{2} }\right]_+
\int_{0}^{1}d{\xi'}\frac{1+{\xi'}^2}{(1-\xi')_+}
 e^{-i\xi' p_{J\perp} \cdot y_\perp} 
 \,,   
\eea

\subsection{The soft contribution}\label{subsec:nlosoft}
\begin{figure}[htbp]
    \centering
    \subfigure{\includegraphics[width=0.45\textwidth]{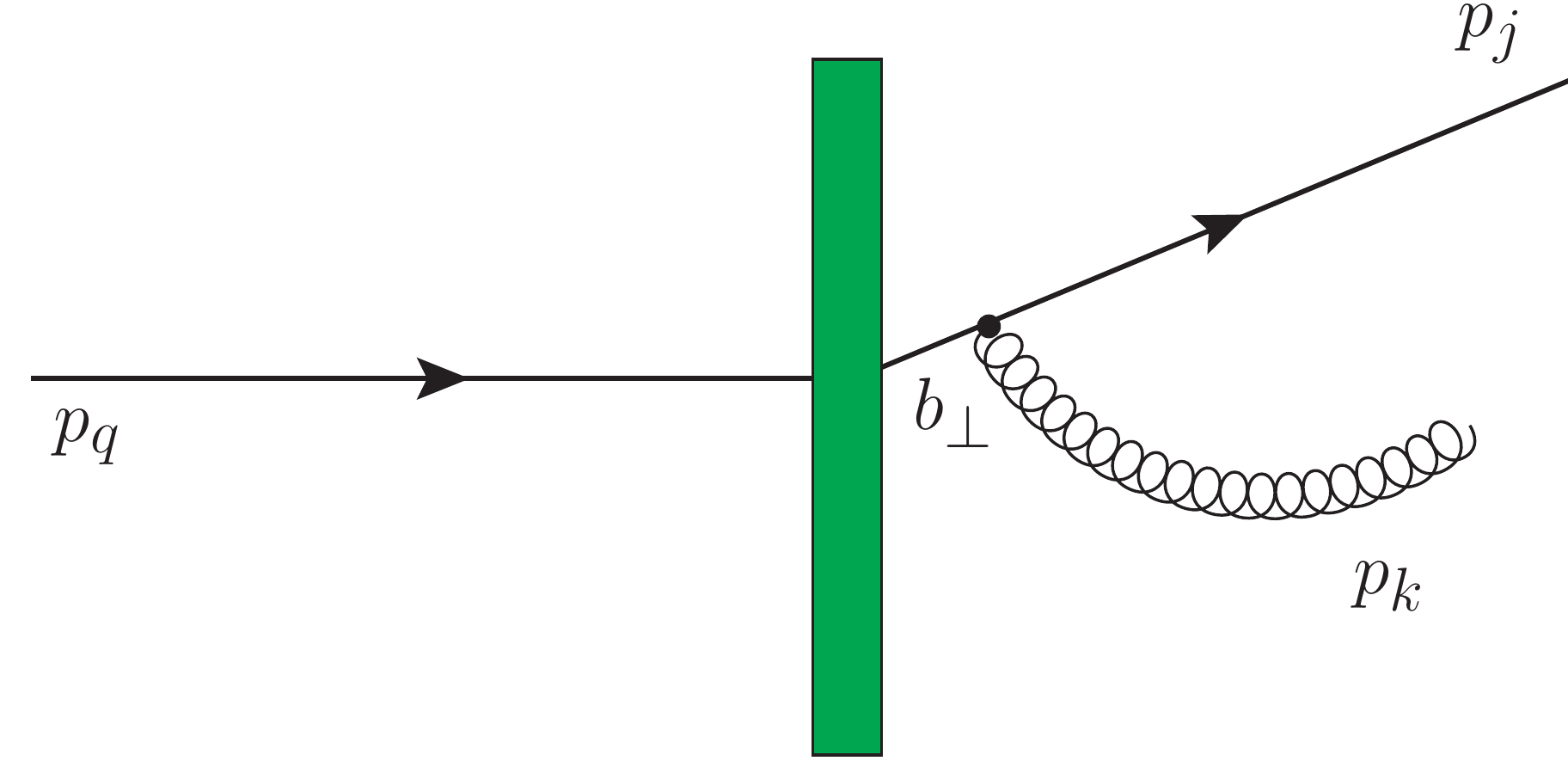}}     
    \subfigure{\includegraphics[width=0.45\textwidth]{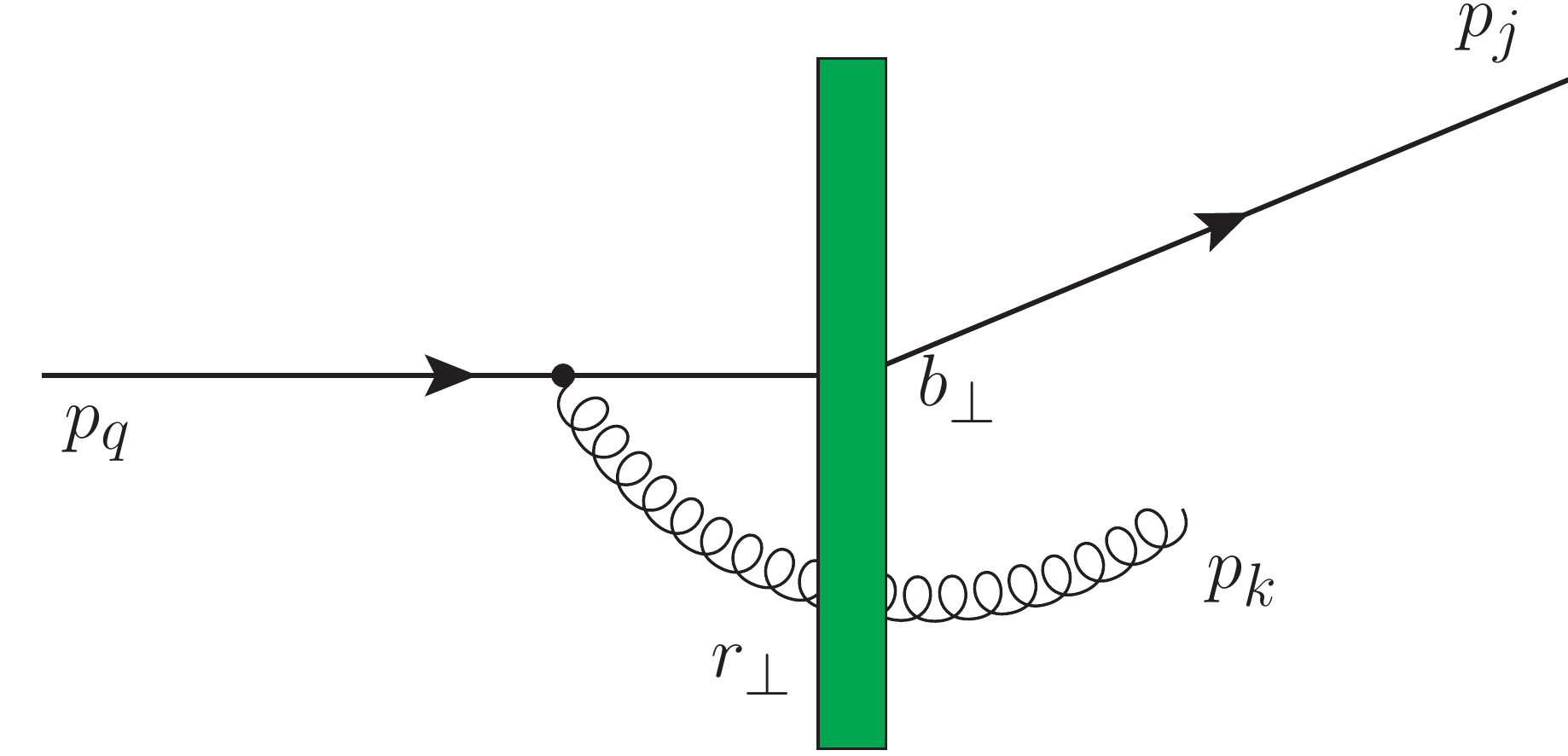}} 
    \caption{Feynman diagrams for the soft real emissions where a soft gluon is emitted: (a) final state radiation; (b) initial state radiation}
    \label{fig:feyn-real-soft}
\end{figure}
Now we consider the real correction to the cross section by emitting a soft gluon, with the momentum scaling as $p_{k} = (p_k^+, p_k^-, p_k^\perp) \sim p_i^+ (\lambda,\lambda,\lambda )$ where $\lambda \sim \frac{p_J^\perp}{\sqrt{\hat{s}}}$. Similar to the collinear radiations, there are $2$ soft contributions to the real amplitudes, one ISR (right figure) and one FSR (left figure), as shown in fig.~\ref{fig:feyn-real-soft}. The amplitudes are 
\bea\label{eq:soft-fsr} 
|{\cal M}_{FSR,soft} \rangle = -g_s  {\bf T}_j^a
\frac{n^\alpha }{p_k^-} \left|\frac{p_{k\perp}}{\nu} e^{|\eta_k|}\right|^{-\eta/2}
|{\cal M}_0(p_{q,\perp},p_{j\perp} + p_{k,\perp}) \rangle \,, 
\eea 
for the FSR which is nothing but the eikonal current acting on the LO matrix element and 
\bea\label{eq:soft-isr} 
|{\cal M}_{ISR,soft}\rangle &=& -g_s 
\int \frac{d^{D-2}l_\perp}{(2\pi)^{D-2}} \frac{2l_\perp^\alpha}{l_\perp^2} \left(\frac{l_\perp}{\nu} e^{|\eta_k|}\right)^{-\eta/2} 
\nn \\
&& \times
\int 
d^{D-2}r_\perp 
e^{-i(p_{k\perp}+l_{\perp})\cdot r_\perp }
W_{ab}(r_\perp) {\bf T}^b_i \, 
|{\cal M}_0(l_\perp,p_{j\perp})\rangle 
\,, 
\eea
for the ISR contribution.

The soft cross section can be obtained by squaring the soft currents and integrating over the soft phase space in Eq.~(\ref{eq:pssoft}), which gives
\bea\label{eq:NLO-softr}
d\sigma_{R,soft}
&=&\frac{\alpha_s S_\perp}{2\pi^2}\frac{N_C}{2} 
\tau f(\tau) \, 
\int_{-\infty}^\infty d \eta_k 2 e^{-\eta |\eta_k|}
\int d^{D-2}p_{k\perp} \, 
\left(\frac{p_{k\perp}}{\nu} \right)^{-\eta/2} \, 
\mathcal{F}_F(p_{k\perp}+p_{j\perp};X_f)
\nn 
\\
&& \left(\Theta_{1,soft} + \Theta_{2,soft} \right) \,
\left[\frac{2}{ p_{k\perp}^2} 
-\int \frac{d^{D-2}l_{\perp}}{(2\pi)^{D-4}}\mathcal{F}_F(l_{\perp}+p_{j\perp};X_f)
\frac{2 p_{k\perp}\cdot l_{\perp}}{ p_{k\perp}^2 l_{\perp}^2} \, \left(\frac{l_\perp}{\nu}\right)^{-\eta/2}\right]  \,\nn \\
&+& 
\frac{\alpha_s}{2\pi^2} \frac{N_C}{2}
\tau f(\tau)
\, 
  \int_{-\infty}^\infty 
d\eta_k  2\, 
\int d^{2}p_{k\perp} 
\, 
\Big(\Theta_{1,soft} + \Theta_{2,soft} -1 \Big) 
\, 
\int 
\frac{d^2b_\perp d^2b_\perp'}{4\pi^2}
 \frac{d^2r_\perp d^2r_\perp'}{4\pi^2} 
\nn \\ 
&& \times 
\frac{x_\perp \cdot x_\perp'}{x^2_\perp\,{x'_\perp}^2}
e^{-ip_{j\perp}\cdot z_\perp} 
e^{-ip_{k\perp}\cdot z'_\perp} 
S_{X_f}^{(6)}(b_\perp,r_\perp,b_\perp',r_\perp')  \,, \quad 
\eea 
where $\Theta_{1,soft}$ and $\Theta_{2,soft}$ are given previsouly in Eq.~(\ref{eq:soft-limit-jet}). The integrand in the last two lines are regular, following the same reason as Eq.~(\ref{eq:if-sub}), although the last two lines are solely manipulated out of the ISR and the combination of the FSR and part of the interference contributions. The last $2$ lines vanish as $R \to 0$. To isolate the poles in the first two lines, we construct the counter term as
\bea \label{eq:NLO-softr-counter}
d\sigma^c_{R,soft} &=&\frac{\alpha_s S_\perp}{2\pi^2}\frac{N_C}{2} 
\tau f(\tau) \, 
\int d \eta_k 2 e^{-\eta |\eta_k|}
\int \frac{d^{D-2}p_{k\perp}}{(2\pi)^{D-4}}
\left(\frac{p_{k\perp}}{\nu}\right)^{-\eta/2} \nn 
\\
&& 
\times\mathcal{F}_F(p_{k\perp}^\prime;X_f)
\left[\frac{2}{ p_{k\perp}^2} 
-\int \frac{d^{D-2}l_{\perp}}{(2\pi)^{D-4}}
\mathcal{F}_F(l_{\perp}+p_{J\perp};X_f)
\frac{2 p_{k\perp}\cdot l_{\perp}}{ p_{k\perp}^2 l_{\perp}^2} \, \left(\frac{l_\perp}{\nu}\right)^{-\eta/2}\right]\,.  \qquad \quad 
\eea 
The subtracted contribution is then free of any singularities and is integrable in $4$-dimension. Therefore we can again remove the regulators to find
\bea\label{eq:NLO-soft-sub}
&& d\sigma_{R,soft} - d\sigma_{R,soft}^c \nn\\
&=&\frac{\alpha_s}{2\pi^2}\frac{N_C}{2} 
\tau f(\tau) \, 2\, 
\int_{-\infty}^\infty d \eta_k 
\int d^{2}p_{k\perp} \, 
\left(\Theta_{1,soft} + \Theta_{2,soft} -1 \right) 
 \, 
\nn 
\\
&&  \,
\Bigg\{ 
S_\perp \mathcal{F}_F(p_{k\perp}+p_{j\perp};X_f) 
\left[\frac{2}{ p_{k\perp}^2} 
-\int d^{D-2}l_{\perp}\mathcal{F}_F(l_{\perp}+p_{j\perp};X_f)
\frac{2 p_{k\perp}\cdot l_{\perp}}{ p_{k\perp}^2 l_{\perp}^2} \right]  \, \nn \\ 
&+&\int 
\frac{d^2b_\perp d^2b_\perp'}{4\pi^2}
 \frac{d^2r_\perp d^2r_\perp'}{4\pi^2}
\frac{x_\perp \cdot x_\perp'}{x^2_\perp\,{x'_\perp}^2}
e^{-ip_{j\perp}\cdot z_\perp} 
e^{-ip_{k\perp}\cdot z'_\perp} 
S_{X_f}^{(6)}(b_\perp,r_\perp,b_\perp',r_\perp') \Bigg\}  \,, 
\eea 
which can be readily evaluated numerically. The integrated counter term is found to be 
\bea 
d\sigma_{R,soft}^{c,int.} &=&
    \frac{\alpha_s}{2\pi} \frac{N_C}{2} \tau f(\tau) \frac{8}{\eta}
    \int \frac{ d^2b_\perp d^2b_\perp' }{4\pi^{2-\epsilon}} 
    e^{-ip_{J\perp}\cdot z_\perp}
    \left\{ 
\frac{\Gamma(-\epsilon-\eta/2)}{\Gamma(1+\eta/2)} 
\left( \frac{z_\perp^2 \nu^2}{c_0^2}\right)^{\eta/2} (z_\perp\mu)^{2\epsilon}  S_X^{(2)}(b_\perp,b_\perp') \right. \nn 
\\ 
& & \hspace{-5.ex}
\left. 
+ 
\frac{\Gamma^2(1-\epsilon-\eta/4)}{\Gamma^2(1+\eta/4)} 
\int \frac{d^2r_\perp}{\pi} 
\frac{x_\perp\cdot y_\perp}{x_{\perp}^{2} y_\perp^{2} }
\left( \frac{x_\perp y_\perp \nu^2}{c_0^2} \right)^{\eta/2}(x_\perp y_\perp \mu^2)^\epsilon
 S_X^{(3)}(b_\perp,r_\perp,b_\perp') \right\} \,.\quad
\eea 
When combined with the soft virtual correction in Eq.~(\ref{eq:soft-virtual}), we arrive at  the contribution  
\bea
d\sigma_{R+V,soft}^{un\text{-}resolv.}
&=
& \frac{\alpha_s}{2\pi}\frac{N_C}{2}  \tau f(\tau)
\int \frac{d^2b_\perp d^2b_\perp'}{4\pi^2} \frac{d^2r_\perp}{\pi}
e^{-ip_{J\perp}\cdot z_\perp }
S_X^{(3)}(b_\perp,r_\perp,b'_\perp)\,
\left\{
\frac{4}{\eta} \left( \frac{\nu}{p_{J\perp}} \right)^\eta 
\left[\frac{z_\perp^2}{x_\perp^2y_\perp^2}\right]_+ \right. \nn \\ 
&& \hspace{-5.ex}
\left. + 
2
\left[ \frac{\ln (x_\perp^2 p_{J\perp}^2/c_0^2)}{x_\perp^2}
+ \frac{\ln (y_\perp^2 p_{J\perp}^2/c_0^2)}{y_\perp^2}
+ \frac{2x_\perp\cdot y_\perp }{x_\perp^2y_\perp^2}
\ln\left(\frac{x_\perp y_\perp p_{J\perp}^2}{c_0^2} \right) \right]_+
\right\} \,. 
\eea
The transverse coordinates are defined as $z_\perp = b_\perp - b_\perp'$, $x_\perp = b_\perp - r_\perp$ and $y_\perp = r_\perp - b_\perp'$.
One may recognize that this unresolved term is nothing but the soft contribution to the single hadron inclusive production and the finite part is the contribution arising from the so-called kinematic constraint~\cite{Watanabe:2015tja,Kang:2019ysm}.
The total soft contribution is therefore given by
\bea
d\sigma_{soft}=
d\sigma_{R+V,soft}^{un\text{-}resolv.}+ (d\sigma_{R,soft} - d\sigma^c_{R,soft}) \,, 
\eea 
where the superscript indicates that the gluons in the virtual and the counter term are unresolvable and the second term can be regarded as the additional kinematic restriction due to the jet clustering.

\subsection{Full NLO Corrections}\label{subsec:fullNLO}
Gluing all pieces from both the collinear and the soft sectors, we find the NLO corrections 
\bea\label{eq: quark-NLO-unren}
		&& d\sigma^{(1)}_{un\text{-}ren.} \nn \\
		& = &
		  \frac{\alpha_s}{2\pi}\frac{N_C}{2}
\int_{\tau}^1 d\xi \, xf(x)\, 
\int \frac{d^2 b_\perp d^2 b'_\perp}{4\pi^2}
e^{-i   p_{J\perp} \cdot z_\perp}
		\Bigg\{ \left( -
\frac{1}{\epsilon}
P_{qq}^{(1)}(\xi)
+ {\cal  H}_{q,2} 
\right) 
\, 
S_{X_f}^{(2)}(b_\perp,b_\perp') 
\nn \\
		&&+ 
\int 
\frac{dr_\perp}{\pi}
S_X^{(3)}(b_\perp,r_\perp,b'_\perp)\,
\left( {\cal H}_{q,3} + 	\delta(1-\xi) 
\left[
\left(\frac{z_\perp^2}{x_\perp^2y_\perp^2}\right)_+
\left( \frac{2}{\eta} + 2 \ln \frac{\nu p_q^+}{p_{J\perp}^2}
\right) 
\right.  \right. \nn \\ 
&& \left. \left. + 
2
\left( \frac{\ln (x_\perp^2 p_{J\perp}^2/c_0^2)}{x_\perp^2}
+ \frac{\ln (y_\perp^2 p_{J\perp}^2/c_0^2)}{y_\perp^2}
+ \frac{2x_\perp\cdot y_\perp }{x_\perp^2y_\perp^2}
\ln\left(\frac{x_\perp y_\perp p_{J\perp}^2}{c_0^2} \right) \right)_+
\right] \right) \Bigg\} \nn \\ 
&&  + (d\sigma_{fsr} - d\sigma^c_{fsr}) + (d\sigma_{isr}-d\sigma^c_{isr})+ (d\sigma_{inter.}-d\sigma^c_{inter.}) 
 + (d\sigma_{R,soft}-d\sigma_{R,soft}^c) 
\,,  \qquad
\eea 
where the collinear $\epsilon$-pole in the first line will be absorbed by the proton PDF. We can immediately realize that the coefficient of the $\eta$-pole is nothing but the BK-kernel to be cancelled by the nucleus dipole distribution. The pole structures explicitly demonstrate the validity of the CGC hybrid scheme when applying to the jet production at the NLO. The finite NLO corrections to the jet production are then found to be
\bea\label{eq: quark-NLO}
		d\sigma^{(1)} & = &
\frac{\alpha_s}{2\pi}\frac{N_C}{2} 
\int \frac{d^2 b_\perp d^2 b'_\perp}{4\pi^2}
\int_{\tau}^1 d\xi \,  
xf(x)
\, 
e^{-i   p_{J\perp} \cdot z_\perp} \nn \\
&& \times 
\left( 
{\cal  H}_{q,2}  \, \, 
S_X^{(2)}(b_\perp,b_\perp') 
+ \int \frac{dr_\perp}{\pi} 
\Big( 
{\cal H}_{q,BK}
+
{\cal H}_{q,3} 
+
{\cal H}_{q,kin.}
\Big) 
S_X^{(3)}(b_\perp,r_\perp,b_\perp')   \right)
\nn \\
&& + (d\sigma_{fsr} - d\sigma^c_{fsr}) + (d\sigma_{isr}-d\sigma^c_{isr})+ (d\sigma_{inter.}-d\sigma^c_{inter.}) 
 + (d\sigma_{R,soft}-d\sigma_{R,soft}^c)
\,,  \quad \quad 
\eea 
where ${\cal H}_{q,2}$ and ${\cal H}_{q,3}$ are given by Eqs.~(\ref{eq:Hq2}) and~(\ref{eq:Hq3}), respectively. The $\eta$-pole induced BK logarithmic term is 
\bea\label{eq:HqBK}
{\cal H}_{q,BK}
= 
2 \left( \ln \frac{\nu p_q^+}{p_{J\perp}^2}
\right) 
\left[\frac{z_\perp^2}{x_\perp^2y_\perp^2}\right]_+\, 
 \delta(1-\xi) \,, 
\eea 
which suggests the scale choice for the rapidity scale $\nu$. By noting that the logarithmic term is proportional to the LO kinematics, it is thus found that $\nu = p_{J\perp}^2/p_q^+ = X_f\, p_A$, where $X_f$ is the momentum fraction carried by the gluon from the nucleus. The scale choice naturally gives rise to the scale for the nucleus distribution required by the CGC framework.  
The kinematic constrain term is given by
\bea\label{eq:Hqkin} 
{\cal H}_{q,kin.}
= 2
\left[ \frac{\ln (x_\perp^2 p_{J\perp}^2/c_0^2)}{x_\perp^2}
+ \frac{\ln (y_\perp^2 p_{J\perp}^2/c_0^2)}{y_\perp^2}
+ \frac{2x_\perp\cdot y_\perp }{x_\perp^2y_\perp^2}
\ln\left(\frac{x_\perp y_\perp p_{J\perp}^2}{c_0^2} \right) \right]_+
\delta(1-\xi) \,. \quad 
\eea 
The final NLO predictions for the inclusive jet production will then be 
\bea 
d\sigma = d\sigma^{(0)} + d\sigma^{(1)} \,, 
\eea 
where the LO cross section $d\sigma^{(0)}$ can be found in Eq.~(\ref{eq:LOsec}) in section~\ref{sec:LOsec}. 
The jet algorithm and other experimental cuts are implemented in the last line of Eq.~(\ref{eq: quark-NLO}), while the rest shares the same kinematics as the LO. 

Once we have $d\sigma$ up to NLO, with full information of momenta $p_j$ and $p_k$, {\color{orange}{\it we can generate any observable distributions by histograms}}. We take the jet energy $E_J$ distribution as an example to highlight how this works: 
\begin{enumerate}
    \item We divided the $E_J$ spectrum into $N$ different bins
    \bea 
    (E_{J,0},E_{J,1})\,, \, (E_{J,1},E_{J,2})\,,  \dots \,, (E_{J,i},E_{J,i+1})\,,\dots \,, (E_{J,N-1},E_{J,N})\,, 
    \eea 
    where the boundary of each bin could follow exactly the experimental setups. 
    
    \item\label{step:kin} We generate the momenta for the quark $p_j^\mu$ and the gluon $p_k^\mu$ in a $1$-jet or $2$-jet event out of the free variables $p_{J}^+$ and $p_{J\perp}$ following Eq.~(\ref{eq:clusterTH}), accordingly. The $1$-jet and the $2$-jet event should satisfy the clustering condition in Eq.~(\ref{eq:clusterTH}), respectively. The event is kept if the condition is fulfilled, otherwise vetoed. 
    
    \item\label{step:fill} The $E_J$ is constructed by $p_j$ and $p_k$
    depending on whether it is an $1$-jet or $2$-jet event. If $E_J \in (E_{J,i},E_{J,i+1})$, the event will be filled into this bin with weight $d\sigma$. {\color{orange}{\it Consistent power counting has to be taken care of here. For instance, if $E_{J,i}\sim \sqrt{s} \gg Q_s \sim \lambda \sqrt{s} $, then in the soft contribution the soft gluon energy $E_k \sim Q_s \sim \lambda  \sqrt{s} $ will never contribute to the non-zoro jet energy bins, since by power counting we will have either $\delta(E_J-E_j-E_k) \approx   \delta(E_J-E_j) $ since $E_j \sim E_J \sim \sqrt{s}$, for one jet or $\delta(E_J-E_j) + \delta(E_J - E_k) \approx \delta(E_J-E_j)+ \delta(E_J)$ for two jets, where we have expanded the $\delta$-functions in terms $\lambda$ and only keep $\lambda^0$ terms. In the former case, we will have a one-jet event with the jet energy given by the energetic quark energy $E_J = E_j$. In the latter case, the gluon jet (contribution associated with $\delta(E_J)$) will never pass the jet energy lower bound $E_J = 0 < E_{J,i}$ and therefore will not contribute to the jet energy spectrum, but only the quark jet does. 
      }}
    
    \item We repeat the steps~\ref{step:kin}) to~\ref{step:fill}) to build up the $E_J$ histogram until we have sufficient statistics. 
    
\end{enumerate}

Obviously, the procedure fits well to other observables and their spectra can be obtained similarly.

\section{An alternative subtraction and the small-$R$ limit}\label{sec:smallR}
 
 The subtraction terms to regulate the real contribution are not uniquely determined. An alternative is to construct the counter term for $d\sigma_{fsr}$ by simply replacing the full anti-$k_T$ jet clustering constraint with its small-$R$ approximation. The other subtraction terms for $d\sigma_{isr}$ and $d\sigma_{inter.}$ as well as the one for the soft contribution remain the same as discussed in the previous sections.  This leads to the new subtraction term for $d\sigma_{fsr}$ such that 
\bea \label{eq:small-R-sub}
d\sigma^c_{r}&=&\frac{\alpha_s S_\perp}{2\pi^2}\frac{N_C}{2}
\int_{0}^1 d\xi
\frac{1+\xi^2+\frac{(D-4)}{2}(1-\xi)^2}{(1-\xi)^{1+\eta}}\left(\frac{\nu}{p^+_q}\right)^\eta\, \nn \\
&& \hspace{8.ex}
\times 
\int
\frac{d^{D-2}p_{k\perp}}{(2\pi)^{D-4}}
\frac{\mathcal{F}_F(p_{k\perp}+ p_{j\perp};X_f)}{[\xi p_{k\perp}-(1-\xi) p_{j\perp}]^2} \left(x f(x) \Theta_{2,R} + 
\tau f(\tau) \Theta_{1,R} 
\right)\,,
\eea 
where $\Theta_{1,R}$ and $\Theta_{2,R}$ are the jet clustering conditions in the small jet radius limit and are given in Eq.~(\ref{eq: small-r-jet}). To see how it can render the subtracted combination finite
\bea\label{eq:small-R-subtracted}
d\sigma_{fsr}-d\sigma^c_{r} &=&\frac{\alpha_s S_\perp}{2\pi^2}
\frac{N_C}{2}
\int_{0}^1d\xi
\frac{1+\xi^2}{1-\xi}
\int  d^2 p_{k\perp}  \nn \\
&& \times 
 \Bigg\{
( \Theta_2 -\Theta_{2,R} )xf(x)
+(\Theta_1 - \Theta_{1,R}) \, 
\tau f(\tau) 
\Bigg\} 
\frac{\mathcal{F}_F(p_{k\perp}+ p_{j\perp};X_f)}{[\xi p_{k\perp}-(1-\xi)p_{j\perp}]^2}
\,, 
\eea 
we first note that in the collinear limit in which $\xi  p_{k\perp}  = (1-\xi) p_{j\perp}$, we have $\Theta_{2} = \Theta_{2,R} = 0$ and $\Theta_{1} = \Theta_{1,R} = 1$. Therefore the divergent behaviour in the collinear limit is removed by $[\Theta_{1}-\Theta_{1,R}] = 0$. As for the rapidity divergence when $\xi \to 1$, we notice that when $\xi = 1$, we will have $\Theta_1+\Theta_{2}  = 1$, $\Theta_{1,R}+\Theta_{2,R}  = 1$ and $x = \tau$, and hence the integrand vanishes and the integral is again finite in this limit.
 
 We work out the part proportional to $\Theta_2$ and $\Theta_1$ of this new integrated counter term respectively, and find  
 \bea\label{eq:small-R-int-counter2}
d\sigma_{r,\Theta_2}^{c,int.} 
&=&\frac{\alpha_s}{2\pi}\frac{N_C}{2} 
\int_\tau^1 \frac{d\xi}{\xi^2} \int  \frac{d^{D-2}b_\perp d^{D-2}b_\perp'}{(2\pi)^{D-2}}\Bigg\{
\delta(1-\xi)
  \frac{2}{\epsilon \eta}   
\left(\frac{\nu}{p^+_q}\right)^\eta\, 
\, (4\pi)^{ - \epsilon } 
\Gamma (1 -\epsilon ) \, 
  (z_\perp \mu)^{2 \epsilon } 
 \nn \\ 
& + &
\left( 
- \frac{1}{\epsilon} 
\frac{1+\xi^2}{(1-\xi)_+} 
+ (1-\xi)
\right) 
(4\pi)^{ - \epsilon } 
\Gamma (1-\epsilon )
\left( 
  z_\perp \mu 
\right)^{2 \epsilon } 
 \nn \\ 
&-&
 \left[\frac{1}{\epsilon^2}\delta(1-\xi)-\frac{1}{\epsilon}\frac{1+\xi^2}{(1-\xi)_+}+(1-\xi)+2(1+\xi^2)\left(\frac{\log(1-\xi)}{1-\xi}\right)_+\right] \\ 
&& 
\hspace{8.ex}
\times 
\, 
\left( 
 \frac{ p_{J\perp} R}{\xi \mu}\right)^{-2\epsilon}  
 \frac{ \pi^{-\epsilon}}{\Gamma(1-\epsilon)}
\Bigg\} \, xf(x) \, 
S^{(2)}_{X_f}(b_\perp,b_\perp') 
e^{- \frac{i}{\xi} p_{J\perp} \cdot z_\perp }
\,, 
\eea 
for the $\Theta_2$ term, and 
\bea \label{eq:small-R-int-counter1}
d\sigma^{c,int.}_{r,\Theta_1}&=&\frac{\alpha_s }{2\pi^2}\frac{N_C}{2}\int\frac{d^{D-2}b_\perp d^{D-2}b_\perp'}{(2\pi)^{D-2}}
\left[ \frac{1}{\epsilon^2}+ \frac{3}{2\epsilon}+\left(\frac{13}{2}-\frac{2\pi^2}{3}\right) \right]e^{-i p_{J\perp}\cdot z_\perp } \nn \\
&&\times \left(\frac{p_{J\perp} R}{\mu}\right)^{-2\epsilon}\frac{\pi^{-\epsilon}}{\Gamma(1-\epsilon)} \tau f(\tau)S^{(2)}_{X_f}(b_\perp,b_\perp')\,,
\eea 
for the $\Theta_1$ term. Adding up these two contributions, we find 
\bea\label{eq:small-R-int-counter}
d\sigma_{r}^{c,int.} 
&=&\frac{\alpha_s}{2\pi}\frac{N_C}{2} 
\int_\tau^1 \frac{d\xi}{\xi^2} \int  \frac{d^{D-2}b_\perp d^{D-2}b_\perp'}{(2\pi)^{D-2}}\Bigg\{
\delta(1-\xi)
  \frac{2}{\epsilon \eta}   
\left(\frac{\nu}{p^+_q}\right)^\eta\, 
\, (4\pi)^{ - \epsilon } 
\Gamma (1 -\epsilon ) \, 
  (z_\perp \mu)^{2 \epsilon } 
 \nn \\ 
& + &
\left( 
- \frac{1}{\epsilon} 
\frac{1+\xi^2}{(1-\xi)_+} 
+ (1-\xi)
\right) 
(4\pi)^{ - \epsilon } 
\Gamma (1-\epsilon )
\left( 
  z_\perp \mu 
\right)^{2 \epsilon } 
 \nn \\ 
&+&
\left(   
\frac{1}{\epsilon} P_{qq}^{(1)}(\xi) 
- 2 (1+\xi^2) \left(\frac{\ln (1-\xi) }{1-\xi} \right)_+
-  (1-\xi)   
+ \left(
\frac{13}{2} - \frac{2}{3}\pi^2 
\right)\delta(1-\xi) 
\right)  \nn \\ 
&& 
\hspace{8.ex}
\times 
\, 
\left( 
 \frac{ p_{J\perp} R}{\xi \mu}\right)^{-2\epsilon}  
 \frac{ \pi^{-\epsilon}}{\Gamma(1-\epsilon)}
\Bigg\} \, xf(x) \, 
S^{(2)}_{X_f}(b_\perp,b_\perp') 
e^{- \frac{i}{\xi} p_{J\perp} \cdot z_\perp }
\,, 
\eea 
and therefore the $d\sigma_{fsr}$ contribution is given by 
\bea 
d\sigma_{fsr} = d\sigma_{r}^{c,int.} + (d\sigma_{fsr} - d\sigma^c_r) \,,
\eea 
where the second combination can be evaluated numerically. 
\begin{figure}[htbp]
	\begin{center}
		\includegraphics*[width=0.55\textwidth]{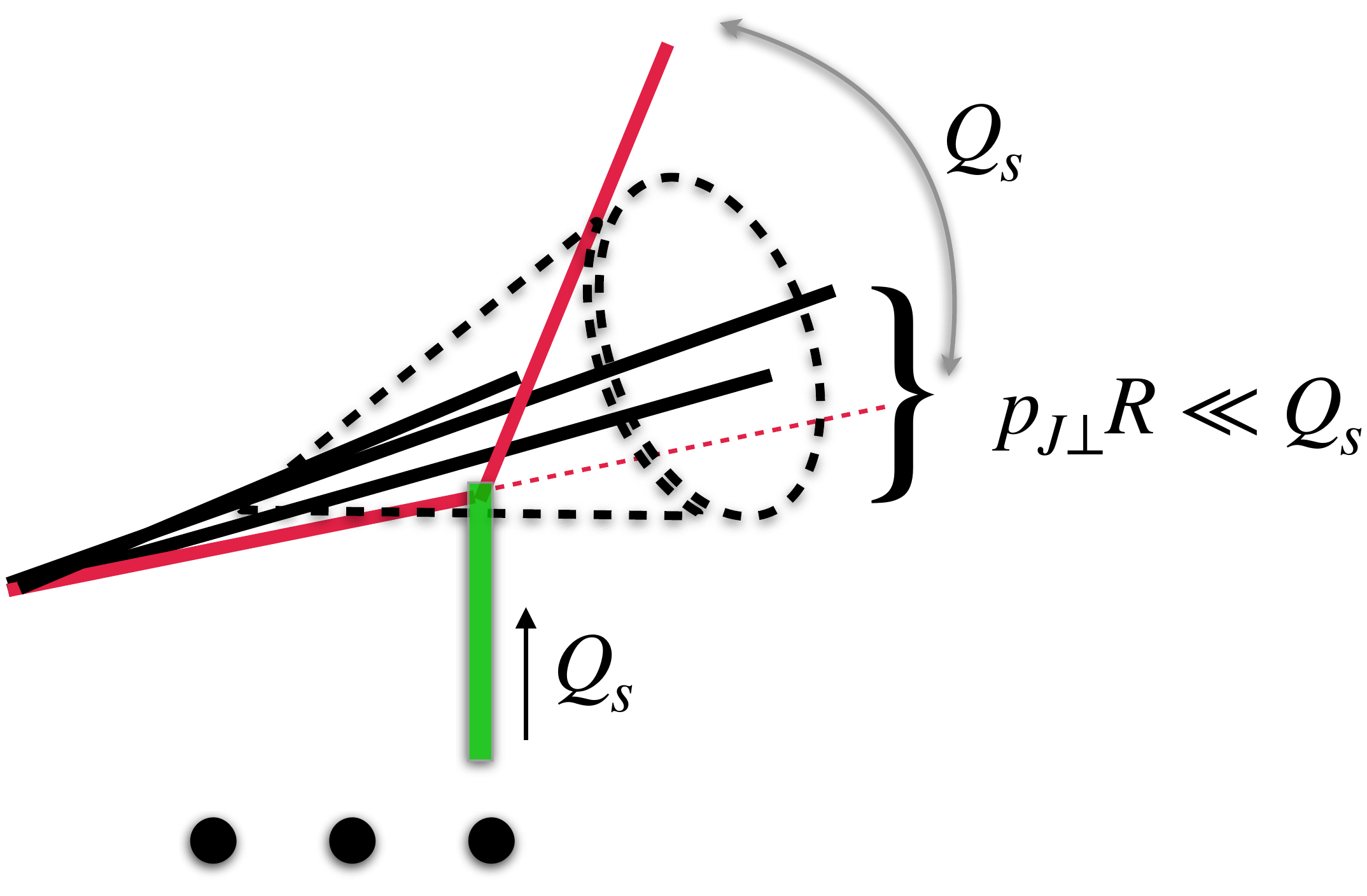}
		\caption{A parton (in red) being knocked out of the small-$R$ jet by interacting with the shock wave with typical momentum transfer ${\cal O}(Q_s)$.}
		\label{fig:siJF}
	\end{center}
\end{figure}

There are several interesting aspects to notice
\begin{itemize}
    \item The $d\sigma_{r}^{c,int.}$ is composed of $2$ parts. 
    \begin{enumerate}
        \item 
    The first part, as the first $2$ lines of Eq.~(\ref{eq:small-R-int-counter}), is identical to the NLO calculations of producing a parton in the single hadron production. 
    
    \item The jet clustering information are encoded in the last $2$ lines, which reproduces exactly the un-renormalized semi-inclusive jet function (siJF) derived in~\cite{Kang:2016mcy,Dai:2016hzf}, which consists of the $1$-jet configuration in which both partons are clustered into a single jet
    \bea 
    J_{\Theta_1} = \frac{\alpha_s}{2\pi} \frac{N_C}{2} \left[ \frac{1}{\epsilon^2}
    +\frac{3}{2\epsilon} 
    +\left(\frac{13}{2} - \frac{2\pi^2}{e} \right)
    \right]
    \left( \frac{p_{J\perp}R}{\mu} \right)^{-2\epsilon}
 \frac{1}{\Gamma(1-\epsilon)} \delta(1-\xi) \,,
    \eea 
    and the $2$-jet case which is
    \bea 
    J_{\Theta_2} &=&-
    \frac{\alpha_s}{2\pi}\frac{N_C}{2}
    \left[\frac{1}{\epsilon^2}\delta(1-\xi)-\frac{1}{\epsilon}\frac{1+\xi^2}{(1-\xi)_+}+(1-\xi)+2(1+\xi^2)\left(\frac{\log(1-\xi)}{1-\xi}\right)_+\right]  \nn \\
&& 
\hspace{8.ex}
\times 
\, 
\left( 
 \frac{ p_{J\perp} R}{\xi \mu}\right)^{-2\epsilon}  
 \frac{ 1}{\Gamma(1-\epsilon)}
 \,. 
    \eea 
    \end{enumerate}
    
    The $\epsilon$-poles in these $2$ parts cancel exactly with each other when the corresponding  virtual corrections are added into the first part. The cancellation of the $\epsilon$-poles is guaranteed by the Kinoshita-Lee-Nauenberg (KLN) theorem~\cite{Kinoshita:1962ur,Lee:1964is}, since unlike the hadron production, the jets are inclusive over the final states. The cancellation of the poles serves as a strong check of our calculation.

    \item The combination $d\sigma_{fsr} - d\sigma^c_r$ is of order ${\cal O}(R^2)$ and is negligible when $R^2 \ll 1$. Hence when $R$ is small, we can approximate
    \bea
    d\sigma_{fsr} = d\sigma^c_{r,int} \,, 
    \eea 
    which is practically a very good approximation since usually $R$ is chosen around $0.4$. It is also true for $d\sigma_{isr} - d\sigma^c_{isr}$, $d\sigma_{inter.} - d\sigma^c_{inter.}$ and $d\sigma_{R,soft}-d\sigma_{R,soft}^c$, and therefore they are all negligible in the small-$R$ limit.
    
    \item As a consequence, in the small-$R$ limit, we can derive the cross section fully analytically. With some manipulations, we find that the cross section can be written as 
    \bea\label{eq:small-r-xsec-fact}
	&& 	d\sigma^{(1)}_{R} = 
		\int_{\tau}^1  d\xi \,   \frac{d\zeta}{\zeta^2} \, 
		 xf(x)
		\Bigg\{  
	d\hat{\sigma}^{(1)}_{q\to q}(\xi, p_J/\zeta)
	\delta(1-\zeta)  
	\nn \\ 
 && 
 +
 d\hat{\sigma}^{(0)}_{q\to q}(p_J/\zeta) \delta(1-\xi)  
 \times
\frac{\alpha_s }{2\pi}\frac{N_C}{2} 
\, 
\left(   - P^{(1)}_{qq}(\zeta ) \ln \frac{p^2_{J\perp}R^2}{\zeta^2 \mu^2}
- 2 (1+\zeta^2) \left(\frac{\ln (1-\zeta) }{1-\zeta} \right)_+
\right.
\nn \\
&& 
\left. 
\hspace{25.ex}
-  (1-\zeta)   
+ \left(
\frac{13}{2} - \frac{2}{3}\pi^2 
\right)\delta(1-\zeta) 
\right)  
\Bigg\}  \,, 
\eea 
where $x = \tau/\xi\zeta$, while $d\hat{\sigma}^{(0)}_{q\to q}(p_{J\perp}/\zeta)$ and $d\hat{\sigma}^{(1)}_{q\to q}(\xi,p_{J\perp}/\zeta)$ are the LO and NLO cross section to produce a quark with momentum $p_{j\perp} = p_{J\perp}/\zeta$, respectively. Their explicit results are given by
\bea
d\hat{\sigma}_{q\to q}^{(0)}(\xi,p_j)
 =  
 \int \frac{d^{2}b_\perp d^{2}b_\perp'}{(2\pi)^{2}}  \,
 e^{-i p_{j\perp} \cdot z_\perp} 
 S_{X_f}^{(2)}(b_\perp,b_\perp')
 \,, 
 \eea 
 and 
\bea\label{eq:dsigqqsmallr}
d\hat{\sigma}^{(1)}_{q\to q}(p_j) & = &
\frac{\alpha_s}{2\pi}\frac{N_C}{2} 
\int \frac{d^2 b_\perp d^2 b'_\perp}{4\pi^2}
\, 
e^{-i   p_{j\perp} \cdot z_\perp} 
\Bigg\{ {\cal  H}^{q\to q}_2  \, \, 
S_{X_f}^{(2)}(b_\perp,b_\perp') 
\nn \\
&&  \hspace{3.ex}  
+ \int \frac{d^2 r_\perp}{\pi}
\Big( 
{\cal H}_{BK}^{q\to q}
+
{\cal H}^{q\to q}_3 
+
{\cal H}^{q\to q}_{kin.}
\Big) 
S_{X_f}^{(3)}(b_\perp,r_\perp,b_\perp')   \Bigg\} \,,
\eea 
where ${\cal H}_{3}^{q\to q}$, ${\cal H}^{q\to q}_{BK}$ and ${\cal H}^{q\to q}_{kin.}$ can be obtained by replacing $p_{J\perp}$ with $p_{j\perp}$ in Eqs.~(\ref{eq:Hq3}), (\ref{eq:HqBK}) and~(\ref{eq:Hqkin}), respectively. The term ${\cal H}^{q\to q}_2$ is given by
\bea 
{\cal H}_2^{q\to q} =  \left(
- P_{qq}^{(1)}(\xi) \ln \frac{z_\perp^2\mu^2}{c_0^2}
+ 1-\xi + 
\left(\frac{3}{2}\ln \frac{z_\perp^2p_{j\perp}^2}{c_0^2}
- \frac{1}{2} \right)\delta(1-\xi)  \right)
\left( 
1+ \frac{1}{\xi^2}
e^{-i\frac{1-\xi}{\xi}p_{j\perp}\cdot z_\perp}
\right) \,. \nn \\
\eea 
We note that $d\hat{\sigma}_{q\to q}$ is nothing but the partonic cross section for the single hadron production in $pA$ collisions in the CGC formalism.

Obviously, at NLO, we can work out one of the $\xi$ and $\zeta$ integrations, to find the more familiar form of Eq.~(\ref{eq:small-r-xsec})  
    \bea\label{eq:small-r-xsec}
	&& 	d\sigma^{(1)}_{R} = 
		\int_{\tau}^1  d\xi \,   
		 xf(x)
		\Bigg\{  
	d\hat{\sigma}^{(1)}_{q\to q}(\xi, p_J)
	\nn \\ 
 && 
 +
 d\hat{\sigma}^{(0)}_{q\to q}(p_J/\xi)  \times
\frac{\alpha_s  }{2\pi \xi^2 }\frac{N_C}{2} 
\, 
\left(   - P^{(1)}_{qq}(\xi ) \ln \frac{p^2_{J\perp}R^2}{\xi^2 \mu^2}
- 2 (1+\xi^2) \left(\frac{\ln (1-\xi) }{1-\xi} \right)_+
\right.
\nn \\
&& 
\left. 
\hspace{25.ex}
-  (1-\xi)   
+ \left(
\frac{13}{2} - \frac{2}{3}\pi^2 
\right)\delta(1-\xi) 
\right)  
\Bigg\}  \,, 
\eea 
with $x = \tau/\xi$.

\item 
{\color{orange}{\it It is very interesting to point out that the structure of the foward jet production in Eq.~(\ref{eq:small-r-xsec-fact}) is exactly the same as the central jet production using collinear factorization~\cite{Kang:2016mcy}, i.e., the cross section is factorized into a cross section that produces a parton and a part encoding the jet formation in the vacuum.}} One difference between the CGC formalism Eq.~(\ref{eq:small-r-xsec}) and the collinear factorization lies in the cross section $d\sigma_{q\to q}$ due to the different mechanisms to produce a parton. While the form of the term inside the bracket in Eq.~(\ref{eq:small-r-xsec}) {\color{orange}{\it is identical to}} the NLO quark jet function, obtained before within the collinear factorization~\cite{Kang:2016mcy}, with an additional $\xi^{-2}$ in the argument of the first logarithm. The difference is due to the fact that in the central region one looks at the high transverse momentum of order $E_J$ much larger than the angular separation of the splitting partons, while in the forward scattering, the jet transverse momentum is about the same order of the splitting, which introduces the additional $\xi$ after evaluating the phase space integration. 
Apparently the small radius jet is {\color{orange}{\it totally ignorant of the interaction with the CGC shock wave.}}

This feature can be understood by noting that when $R \ll 1$, $p_{J\perp} R \ll p_{J\perp} \sim Q_s$ where $p_{J\perp} R$ sets the typical scale of the parton transverse momentum with respect to the jet axis inside the jet. As sketched in fig.~\ref{fig:siJF}, any parton will be knocked out of the jet if interacts with the shock wave, since it will require an additional transverse momentum $p_\perp \sim Q_s \gg p_{J\perp} R$, and therefore does not contribute to the jet function. As a consequence, any parton inside the jet can not experience the shock wave and the jet function remains the same in the CGC formalism as the collinear counter part.

Another way to understand the feature is that the typical time scale for the jet formation is of order ${\cal O}(1/p_{J\perp }R)$ which occurs much later than the semi-hard interaction that happens within the time period of order ${\cal O}(1/Q_s)$ for $R \ll 1$, and therefore the formation of the jet feels no shock waves.

\item 
Following the previous argument, using the power counting proposed in~\cite{Kang:2019ysm}, we can derive that in the small-$R$ limit, to all orders, the jet cross section can be written as the factorized form that~\footnote{We note that when we derive the factorization theorem, we default to the energetic jet assumption in which $E_J \gg Q_s \sim p_{J\perp}$. } 
\bea\label{eq:small-r-factorization} 
&& d\sigma_{R}
= \int d\xi
\frac{d\zeta}{\zeta^2}  xf(x)
d\hat{\sigma}_{q\to q}(\xi,p_{J}/\zeta) \, J_q(\zeta)  \,, 
\eea 
where $x=\tau/\xi\zeta$ and $J_q(\zeta)$ is the quark siJF~\cite{Kang:2016mcy,Dai:2016hzf} in the large $N_C$ limit, which is 
\bea\label{eq:siJF} 
J_q(\zeta) = 
\delta(1-\zeta)  & - &
\frac{\alpha_s }{2\pi}\frac{N_C}{2}
\, 
\left[    \Big( P^{(1)}_{qq}(\zeta) +P^{(1)}_{gq}(\zeta) \Big)\ln \frac{p^2_{J\perp}R^2}{\zeta^2\mu^2}
+ 2 (1+\zeta^2) \left(\frac{\ln (1-\zeta) }{1-\zeta} \right)_+ \right. \nn \\
&& \left. + (1-\zeta)   
- \left(
\frac{13}{2} - \frac{2}{3}\pi^2 
\right)\delta(1-\zeta) 
+ 2P^{(1)}_{gq}(\zeta) \ln(1-\zeta) + \zeta
\right] \,, \quad 
\eea 
where we have included the contribution from the gluon as the signal jet and $P^{(1)}_{gq}(\zeta) = \frac{1+(1-\zeta)^2}{\zeta}$. 
Again compare with the jet function in the central region with large transverse momentum, there is an additional $\zeta^{-2}$ in the first logarithm. 
The factorization is illustrated diagrammatically in fig.~\ref{fig:small-R-fact}. The small-$R$ approximation for the gluon channel can be found in the Appendix~\ref{sec:small-R-g}. 

\item {\color{orange}{\it We note that the same argument holds for any other jet substructure observable $v$, such as the jet mass, the angularities, the soft drop, as long as $v \ll p_{J,\perp}$, which indicates that
for any jet substructure studies in the small-$x$ region, the cross section can always be factorized into the product of a $d\hat\sigma$ that produces partons, similar to the one in  Eq.~(\ref{eq:small-r-factorization}), and a jet function for the observable $v$, occurs also in the collinear factorization.}} Given that for many interesting jet substructures, the corresponding jet functions have already be obtained up to next-to-next-to-next-to-leading order (N${}^3$LO), see for instance~\cite{Becher:2006qw,Becher:2010pd,Ellis:2010rwa,Kang:2018jwa,Bruser:2018rad,Kang:2018vgn,Kang:2019prh,Li:2021zcf,Liu:2021xzi,Chen:2020adz,Chen:2022jhb}, this factorization feature can thus be used to dramatically simplify and realize future calculations of the jet substructure distributions in the small-$x$ regime. Last but not the least, the factorized form lays down the foundation for combining the existing parton shower with the CGC calculations, in which one manage to calculate the fully differential cross section within the CGC framework just like what we have presented in this work, and then shower the physical states in vacuum due to the factorization we pointed out.

%
%
%

\end{itemize}

\begin{figure}[htbp]
	\begin{center}
	\includegraphics*[width=0.9\textwidth]{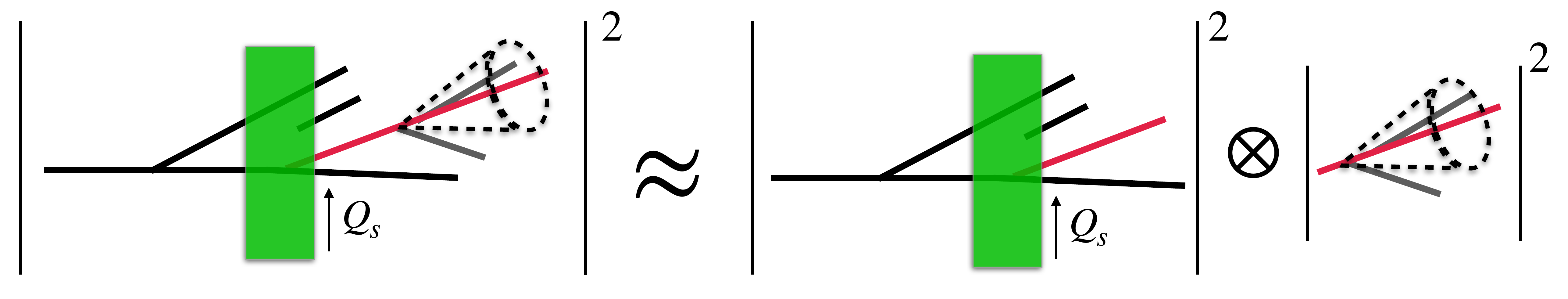}
		\caption{The factorization of the jet cross section in the small-$R$ limit}
		\label{fig:small-R-fact}
	\end{center}
\end{figure}

\section{Threshold resummation}\label{sec:threshold}
Just like the hadron production, there exist large threshold logarithms in the jet production cross section, represented by $\left(\frac{\ln^n(1-\xi)}{1-\xi}\right)_+$. When the jet $p_{J\perp}$ becomes large, we will have
$\tau = p_{J}^+/p_p^+ = 
p_{J\perp} e^{\eta_J} /p_p^+
$ quickly approaches $x$ and thus $\xi \to 1$. This means that the radiations outside the jet cone are restricted to be soft. 
In the threshold limit, the $\left(\frac{\ln^n(1-\xi)}{1-\xi}\right)_+$ becomes dominantly large. 
The threshold logarithms are the most obviously seen by applying the Mellin transformation $\int_0^1 \, d\xi \,  \xi^{N-1}  f(\xi)  $ to the small-$R$ limit cross section $d\hat{\sigma}_{q\to q}$ in Eq.~(\ref{eq:small-r-xsec}) 
\bea\label{eq:threshxsec} 
&& d\hat{\sigma}^{(1)}_{q\to q,thr.} = 
\langle {\cal M}_0 | 
\frac{\alpha_s}{\pi} 
\left(
{\bf T}_i^2 + {\bf T}_j^2 
\right) 
\ln {\bar N}
\ln \frac{\mu^2}{p_{J\perp}^2}  \nn \\
&-& \frac{\alpha_s}{\pi} 
\int \frac{d^2r_\perp}{\pi} \Bigg[
-2 \ln {\bar N}
\left( \frac{x_\perp \cdot y_\perp}{x_\perp^2 y_\perp^2} \right)_+
+ \ln\frac{X_f}{X_A} 
\left(\frac{z^2_\perp}{x_\perp^2y_\perp^2} \right)_+
 \Bigg] 
{\bf T}_j^{a'}  W_{a'a}(r_\perp)
{\bf T}_i^{a} |{\cal M}_0 \rangle \,,  \quad
\eea 
and 
\bea\label{eq:threshjet} 
J^{(1)}_{q,thr.}(\xi) =  
-\frac{\alpha_s}{\pi} {\bf T}_j^2 \, 
\left( - \ln{\bar N} \ln \frac{p_{J\perp}^2R^2}{\mu^2}
+ \ln^2{\bar N}  
+ \frac{\pi^2}{2} 
- 
 \frac{13}{4}  
\right) \,, \quad 
\eea 
where we have applied the Mellin transformation to Eqs.~(\ref{eq:small-r-xsec}) and (\ref{eq:dsigqqsmallr}) in the threshold limit $\xi \to 1$, following the threshold Mellin transform rules in Eq.~(\ref{eq:MellinThr}).
Here, ${\bar N} = N e^{\gamma_E}$. The threshold $\xi \to 1$ limit maps to the large ${\bar N}$ limit after taking the Mellin moment, and we have only kept the threshold contribution, i.e., the non-vanishing terms as ${\bar N}\to \infty$.  
\textcolor{orange}{Here to clearly identify the types of the logarithms, we unfold back the color constant $N_C$ in the NLO cross section to the color operators ${\bf T}$ and the Wilson line $W_{a'a}$ to keep track of the color structures. We have assigned $i$ to the incoming parton and $j$ to the out-going quark.}
 
 The logarithms can be resummed to all orders. We will leave the details to another work~\cite{Xie:future1}. Here let us emphasize several points associated with the threshold logarithms 
 \begin{itemize}
     \item Some of the logarithmic terms in Eqs.~(\ref{eq:threshxsec}) and~(\ref{eq:threshjet}) can be reduced by the scale choice $\mu = p_{J\perp}$ and $X_f = X_A$ in $d{\hat \sigma}_{q\to q}$. The scale choice  characterizes the scale where the semi-hard interaction happens. Note that $X_f = X_A$ is the typical scale choice for the CGC framework.
     The scale evolution associated with $X_f$ gives the BK equation. This set of scale choice is good for eliminating large logarithms when being away from the threshold. 
     
     \item However when $p_{J\perp}$ is approaching the threshold, the remaining logarithms in $1-\xi$ (or in $\bar N$ in Mellin space) become overwhelmingly large and draw the cross section negative~\cite{Liu:2020mpy}, see also fig.~\ref{fig:rthrdp} in the numerical analyses Section~\ref{sec:numerics}. These logarithms should be resummed to all orders to secure the perturbative predictive power.
     
     \item Thanks to the factorization theorem~\cite{Xie:future1}, the remaining logarithms in the jet function and $d\hat{\sigma}_{q\to q,thr.}$ can be treated independently. The $\ln{\bar N} \ln R^2$ and $\ln^2{\bar N}$ terms within the jet function are Sudakov logarithms and have been studied extensively before~\cite{Sterman:1986aj,Catani:1989ne,deFlorian:2007fv,Liu:2012sz,Liu:2013hba,Kumar:2013hia,Dai:2017dpc,Liu:2017pbb}. Their resummation can be achieved by re-factorizing the siJF into a exclusive jet function and a collinear-soft function~\cite{Liu:2017pbb,Xie:future1}. The procedure is standard in the language of SCET and we will leave the  derivation to future work~\cite{Xie:future1}. At the leading logarithmic accuracy, the resummed jet function is 
     \bea 
     J_{thr.,resum} = \exp\left[ 
     -\frac{\alpha_s}{\pi} {\bf T}_j^2 \, 
\left( - \ln{\bar N} \ln \frac{p_{J\perp}^2R^2}{\mu^2}
+ \ln^2{\bar N}  \right)
      \right] \,,
     \eea 
     which is the direct exponentiation of the Sudakov logarithms. 
     
    \item It is more involved to resum the logarithmic term from the interference of the ISR and FSR
    \bea\label{eq:threshlog} 
    \langle {\cal M}_0 | 
    2 \, \frac{\alpha_s}{\pi} \, 
 \ln{\bar N}
\int 
\frac{d^2r_\perp}{\pi}  
\left( \frac{x_\perp \cdot y_\perp}{x_\perp^2 y_\perp^2} \right)_+ {\bf T}_i^a W_{aa'}(r_\perp) {\bf T}_{j}^{a'}  | {\cal M}_0 \rangle \,, 
    \eea 
     {\color{orange}{\it which calls for the threshold resummation technique different from the standard ones in the collinear factorization~\cite{Becher:2006nr,Becher:2006mr,Liu:2017pbb}. One way to see this is to notice that this term shares the same color structure ${\bf T}_i^a W_{aa'}(r_\perp) {\bf T}_{j}^{a'}$ as the non-linear BK evolution, see Eq.~(\ref{eq:BK-T-solve}).}} Indeed, as we have seen in Section~\ref{subsec:colorcharge}, at higher orders, extra Wilson lines $W$'s arise, which suggests the need for a non-linear evolution to resum this contribution.  The standard Sudakov resummation techniques only turn the color charge ${\bf T}$ into a resummed form, see for instance~\cite{Catani:2013vaa,Becher:2014oda,Hinderer:2018nkb,Liu:2017pbb},  while will leave the CGC Wilson line $W$ untouched.
     
     \item A systematic method to resum the logs proportional to ${\bf T}_i^a W_{aa'}(r_\perp) {\bf T}_{j}^{a'}$ in Eq.~(\ref{eq:threshxsec}) has been developed recently in~\cite{Liu:2020mpy}, which will be detailed in~\cite{Xie:future1}  for the jet production. However, we note that for the leading log resummation it is sufficient to consider the $n$ independent soft gluon emissions at N$^{(n)}$LO, with strong ordering $q^-_1 \gg q^-_2 \gg \dots \gg q^-_m$, and $p^-_1 \gg p^-_2 \gg \dots p^-_{n-m}$, where $q_i$ and $p_i$ are the ISR and FSR momenta, respectively. It can be shown that the matrix could be written as   
     \bea\label{eq:norderxsec}  
&&  \frac{1}{n!} \left|\sum_{m=0}^n \Graph{0.23}{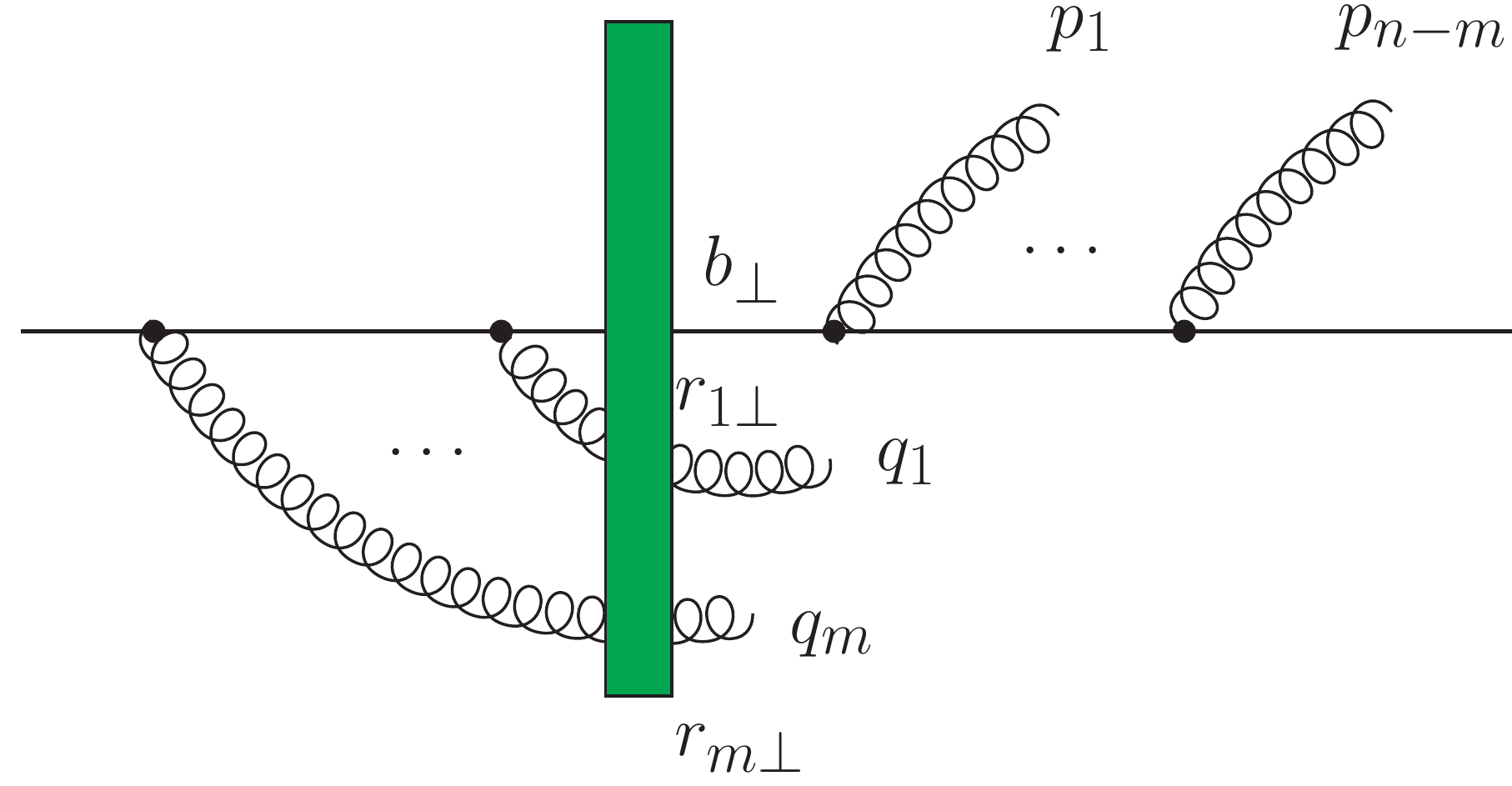} \right|^2
\nn \\ 
&=& \frac{1}{n!}
\langle {\cal M}_0 |  \left\{ 
 - \frac{\alpha_s}{\pi} 
\int \frac{d^2r_\perp}{\pi} \Bigg[
-2 \ln {\bar N}
\left( \frac{x_\perp \cdot y_\perp}{x_\perp^2 y_\perp^2} \right)_+  \right. \nn \\ 
&&  \hspace{20.ex}
\left. 
+ \ln\frac{X_f}{X_A} 
\left(\frac{z^2_\perp}{x_\perp^2y_\perp^2} \right)_+
 \Bigg] 
{\bf T}_j^{a'}  W_{a'a}(r_\perp)
{\bf T}_i^{a} \right\}^n |{\cal M}_0 \rangle  
 \,, 
\eea 
where integration over the phase space is implicitly understood and we have used the property of the color charge operator in Eq.~(\ref{eq:colorforresum}).  When sum up all order contributions, the above equation leads to the exponentiation of the logarithms, which reads
\bea\label{eq:xsecLL} 
&& \sum_{n=0}^\infty \frac{1}{n!} \left|\sum_{m=0}^n \Graph{0.23}{real-soft-multi-resum.pdf} \right|^2
\nn \\ 
&=&
\langle {\cal M}_0 | 
\exp\left\{ - \frac{\alpha_s}{\pi} 
\int \frac{d^2r_\perp}{\pi} \Bigg[
-2 \ln {\bar N}
\left( \frac{x_\perp \cdot y_\perp}{x_\perp^2 y_\perp^2} \right)_+  \right. 
\nn \\ 
&& 
\hspace{20.ex}
\left. 
+ \ln\frac{X_f}{X_A} 
\left(\frac{z^2_\perp}{x_\perp^2y_\perp^2} \right)_+
 \Bigg]
{\bf T}_j^{a'}  W_{a'a}(r_\perp)
{\bf T}_i^{a}  \right\}  | {\cal M}_0 \rangle \,. 
\eea 
One realizes that the above equation reproduce what was found in~\cite{Liu:2020mpy} 
based on factorization and the rapidity renormalization group equations. 
The resummed formula suggests that when approaching the threshold, instead of choosing the CGC scale $X_f = X_A$, one should use a dynamic scale such that $X_f$ minimizes the exponent~\cite{Liu:2020mpy}.

 \end{itemize}
We briefly highlight how we derive the resummation form by going to the soft strong ordering limit. We start with NLO. At NLO, we can have for the matrix element  
\bea 
\Graph{0.43}{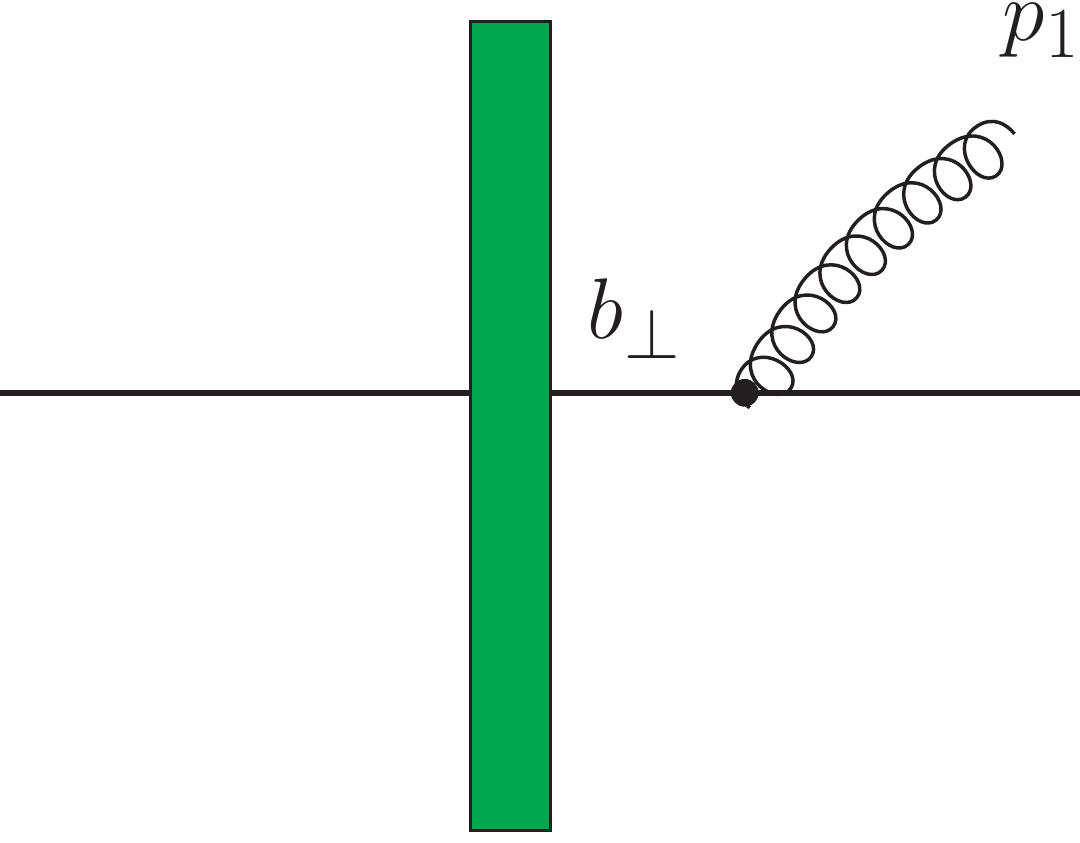} 
= - g_s {\bf T}_j^b \frac{n_\beta}{p_1^-} |{\cal M}_0 \rangle \,
\equiv  {\bfcal J}_\beta^{b,(1)}(p_1) |{\cal M}_0 \rangle \,, 
\eea 
which is the eikonal approximation, presented by the soft eikonal current ${\bfcal J}$ acting on $|{\cal M}_0 \rangle$. We also have the ISR contribution, which is  
\bea 
&& 
\Graph{0.43}{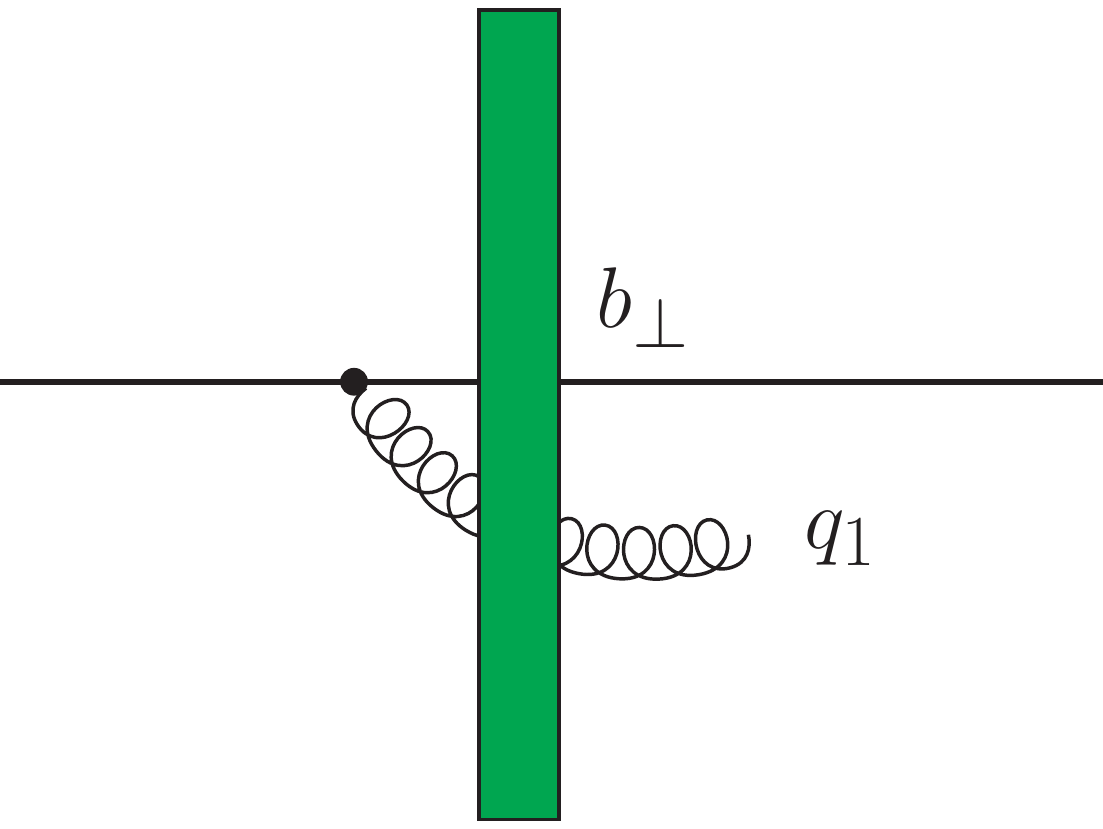} 
  \equiv {\bfcal K}_\beta^{b,(1)}(q_1) |{\cal M}_0 \rangle 
\nn \\ 
&=& i g_s {\bf T}_i^a 
\int \frac{d l^- dl_\perp}{(2\pi)^{D-1}}
\frac{n^\alpha}{-l^-} e^{il_\perp \cdot b_\perp} 
d_{\alpha\beta}(l)  \frac{(- q_1^+ )}{-l^2}  \int d^2r_\perp e^{-i(q_{1\perp} + l_\perp)\cdot r_\perp} W_{ba}(r_\perp) |{\cal M}_0 \rangle \nn \\  
&=& i g_s {\bf T}_i^a 
\int\frac{d l^- dl_\perp}{(2\pi)^{D-1}}
\frac{2e^{il_\perp \cdot b_\perp} }{- l^- l^+} 
\left( 
-\frac{n_\beta}{2} l^+ 
+   l_\beta  
\right)  \frac{(- q_1^+ )}{l^2} \int d^2r_\perp e^{-i(q_{1\perp} + l_\perp)\cdot r_\perp} W_{ba}(r_\perp) |{\cal M}_0 \rangle \nn \\ 
&=& i g_s {\bf T}_i^a 
\int \frac{d l^- dl_\perp}{(2\pi)^{D-1}}
\frac{e^{il_\perp \cdot b_\perp}  }{- l^- l^+} 
\frac{ 2 l_{\perp \beta} }{l^2}  (- q_1^+ ) \int d^2r_\perp e^{-i(q_{1\perp} + l_\perp)\cdot r_\perp} W_{ba}(r_\perp) |{\cal M}_0 \rangle \nn \\ 
&=& - g_s  
\int \frac{  dl_\perp}{(2\pi)^{D-2}}
\frac{ 2 l_{\perp \beta} }{l_\perp^2 }
  e^{il_\perp \cdot b_\perp} 
  \int d^2r_\perp e^{-i(q_{1\perp} + l_\perp)\cdot r_\perp} W_{ba}(r_\perp) 
  {\bf T}_i^a
  |{\cal M}_0 \rangle  \,,  
  \eea 
where we introduced $d_{\alpha\beta}(l) \equiv -g_{\alpha \beta} + \frac{{\bar n}_\alpha l_\beta + {\bar n}_{\beta}l_\alpha}{\bar{n}\cdot l}$ and have used the light-cone gauge condition ${\bar n}\cdot \epsilon=0$. We  performed the contour integration on $l^-$ similar to Eq.~(\ref{eq:contour}) to get the final result. Here we have suppressed the rapidity regulators. 

The LL approximation to the NLO cross section will then be 
\bea 
d\hat{\sigma}^{(1)}_{LL} &=& 
\int d \Phi_{p_1} \langle {\cal M}_0 | 
\left( {\bfcal J}^{ (1) \dagger }_\beta(p_1) + {\bfcal K}^{ (1) \dagger}_\beta
(p_1)
\right) \cdot
\left( {\bfcal J}^{ (1)}_\beta(p_1) + {\bfcal K}^{ (1)}_\beta(p_1)
\right)
|{\cal M}_0 \rangle \nn \\
& \equiv  &
 \langle {\cal M}_0 |  d \hat{\boldsymbol \sigma}^{(1)} |{\cal M}_0 \rangle \,. 
\eea 

At NNLO, to produce LLs, we only need to consider $2$ independent soft gluon emissions with momenta $p_1$ and $p_2$. We require $p_1^- \gg p_2^-$. For the double FSR, the eiknoal current in this strongly order limit is well known, which gives
\bea 
\Graph{0.4}{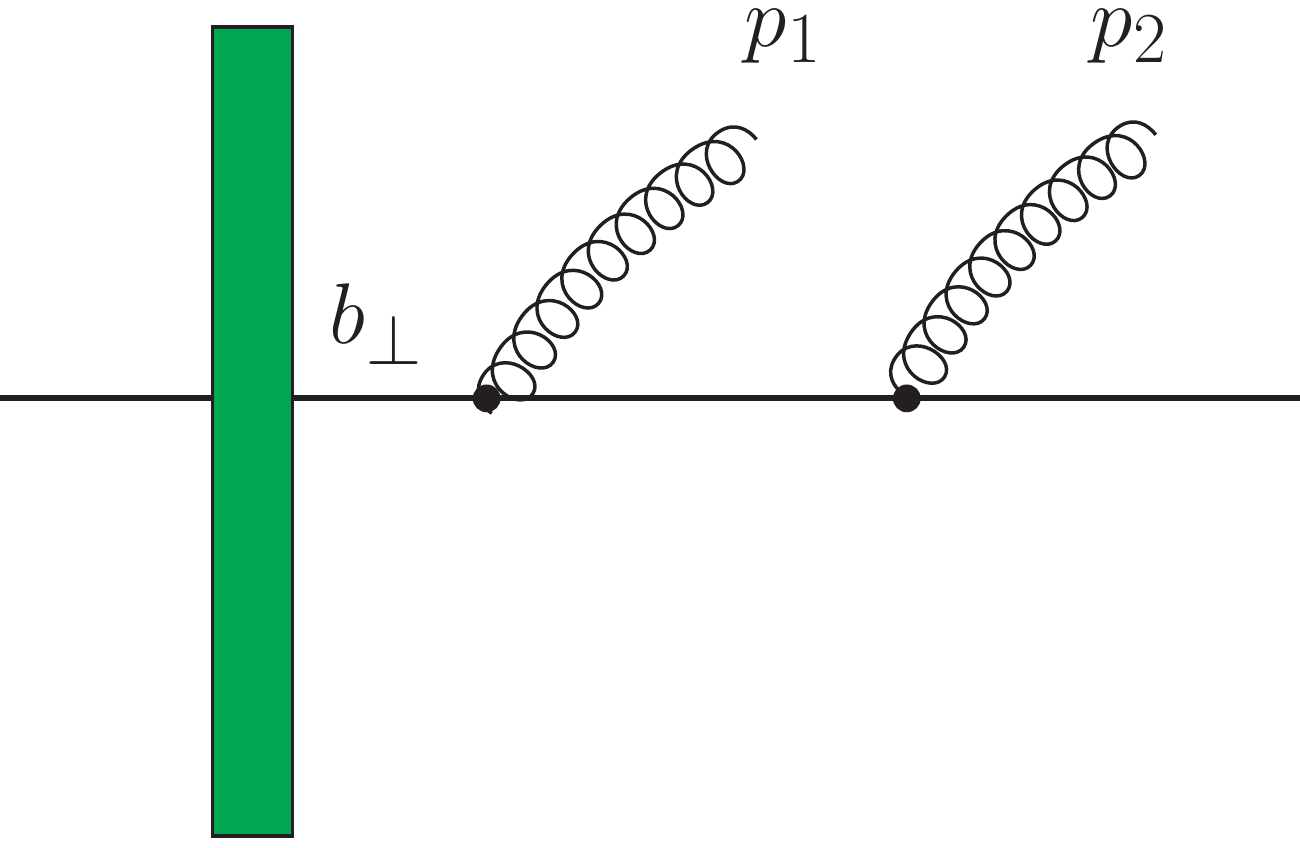} 
=  {\bfcal J}_\beta^{b_1,(1)}(p_1)  {\bfcal J}_\alpha^{b_2,(1)}(p_2)  |{\cal M}_0 \rangle \,, 
\eea 
with $p_1^- \gg p_2^- $. 
And for double ISR, the matrix element reads 
\bea 
&& \Graph{0.4}{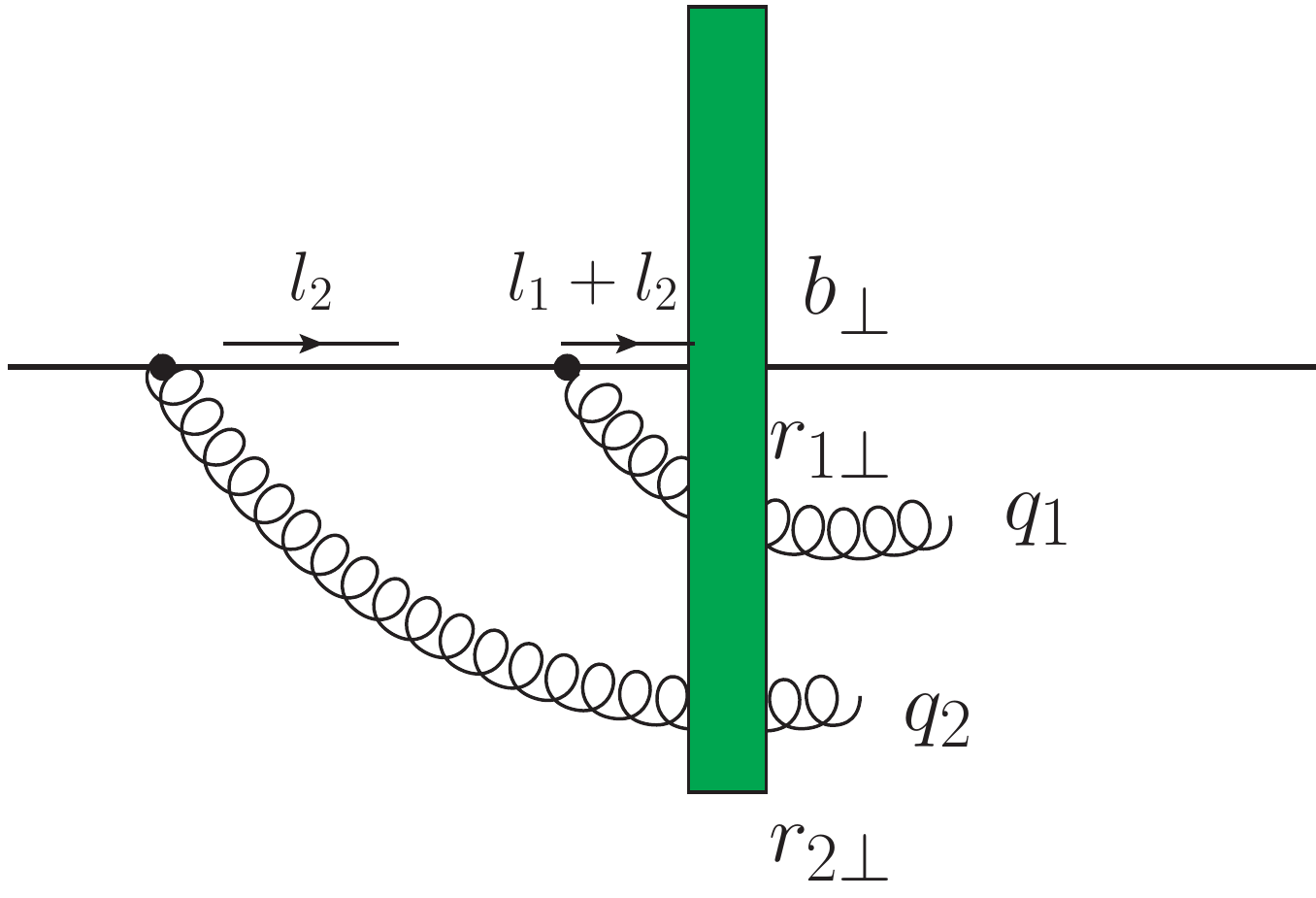} \nn \\
&= & i g_s {\bf T}_i^{a_2} 
\int \frac{d l_2^- dl_{2\perp}}{(2\pi)^{D-1}}
\frac{n^\rho}{-l_2^-} e^{il_{2\perp} \cdot b_\perp} 
d_{\rho\alpha}(l_2)  \frac{(- q_2^+ )}{-l_2^2}  \int d^2r_{2\perp} e^{-i(q_{2\perp} + l_{2\perp})\cdot r_{2\perp}} W_{b_2a_2}(r_{2\perp}) 
|{\cal M}_0 \rangle \nn \\ 
&& \times i g_s {\bf T}_i^{a_1} 
\int \frac{d l_1^- dl_{1\perp}}{(2\pi)^{D-1}}
\frac{n^\sigma}{-(l_1^-+l_2^-)} e^{il_{1\perp} \cdot b_\perp} 
d_{\sigma \beta}(l_1)  \frac{(- q_1^+ )}{-l_1^2}  \int d^2r_{1\perp} e^{-i(q_{1\perp} + l_{1\perp})\cdot r_{1\perp}} W_{b_1a_1}(r_{1\perp}) \nn \\ 
&\approx& i g_s {\bf T}_i^{a_1} 
\int \frac{d l_1^- dl_{1\perp}}{(2\pi)^{D-1}}
\frac{n^\sigma}{-l_1^-} e^{il_{1\perp} \cdot b_\perp} 
d_{\sigma\beta}(l_1)  \frac{(- q_1^+ )}{-l_1^2}  \int d^2r_{1\perp} e^{-i(q_{1\perp} + l_{1\perp})\cdot r_{1\perp}} W_{b_1a_1}(r_{1\perp}) \nn \\ 
&& \times i g_s {\bf T}_i^{a_2} 
\int \frac{d l_2^- dl_{2\perp}}{(2\pi)^{D-1}}
\frac{n^\rho}{-l_2^-} e^{il_{2\perp} \cdot b_\perp} 
d_{\rho\alpha}(l_2)  \frac{(- q_2^+ )}{-l_2^2}  \int d^2r_{2\perp} e^{-i(q_{2\perp} + l_{2\perp})\cdot r_{2\perp}} W_{b_2a_2}(r_{2\perp}) 
|{\cal M}_0 \rangle \nn \\ 
&=& - g_s  
\int \frac{  dl_{1\perp}}{(2\pi)^{D-2}}
\frac{ 2 l_{1\perp \beta} }{l_{1\perp}^2 }
  e^{il_{1\perp} \cdot b_\perp} 
  \int d^2r_{1\perp} e^{-i(q_{1\perp} + l_{1\perp})\cdot r_{1\perp}} W_{ba}(r_{1\perp}) 
  {\bf T}_i^a
   \nn \\  
&\times & - g_s  
\int \frac{  dl_{2\perp}}{(2\pi)^{D-2}}
\frac{ 2 l_{2\perp \alpha} }{l_{2\perp}^2 }
  e^{il_{2\perp} \cdot b_\perp} 
  \int d^2r_{2\perp} e^{-i(q_{2\perp} + l_{2\perp})\cdot r_{2\perp}} W_{ba}(r_{2\perp}) 
  {\bf T}_i^a
  |{\cal M}_0 \rangle  \nn \\ 
&=&
{{\bfcal K}}_\beta^{b_1,(1)}(q_1)
{{\bfcal K}}_\alpha^{b_2,(1)}(q_2)
  |{\cal M}_0 \rangle \,,
\eea 
in the strongly ordered limit, 
where to get the approximation we have used the strongly ordered limit that $l_2^- \ll l_1^-$. 

It is easy to get the matrx element with one ISR and one FSR which gives  
\bea 
\Graph{0.4}{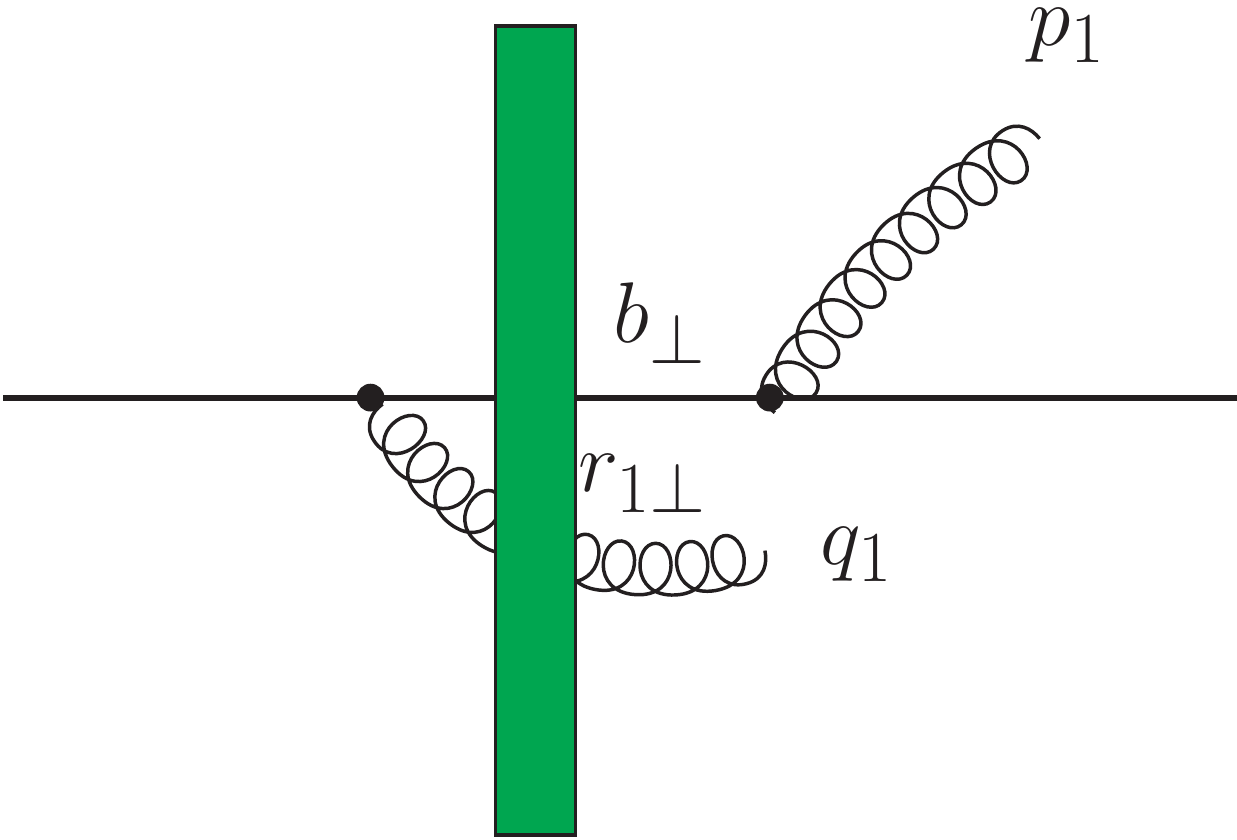}
= \left( 
{\bfcal J}^{b_1,(1)}_\beta (p_1)
 {\bfcal K}^{b_2,(1)}_\alpha(q_1) +  
 {\bfcal K}^{b_1,(1)}_\beta(q_1) 
 {\bfcal J}^{b_2,(1)}_\alpha(p_1)  
 \right) |{\cal M}_0\rangle 
 \,. \quad \quad 
\eea 
Here since there is no strong ordering requirements in $q_1$ and $p_1$, we have splitted the contribution to 2 terms with $p_1 > q_1$ and $q_1 > p_1$, respectively.

Therefore, one could find the LL approximation to the NNLO matrix element be
\bea 
|{\cal M}_2 \rangle= \left( {\bfcal J}^{b_1,(1)}_\beta + {\bfcal K}^{b_1,(1)}_\beta \right) \left({\bfcal J}^{b_2,(1)}_\alpha   + {\bfcal K}^{b_2,(1)}_\alpha   \right) 
| {\cal M}_0 \rangle  \,, 
\eea 
and thus 
\bea 
d\hat{\sigma}^{(2)}_{LL} = 
\langle {\cal M}_0 | \frac{1}{2!} \left(d {\boldsymbol \sigma}^{(1)} 
\right)^2
|{\cal M}_0 \rangle \,.
\eea 
Here the factor $1/2!$ accounts for symmetry factor of $2$ indistinguishable gluons in the final state.
The calculation is extendable to arbitrary $n$-th order to find
\bea 
d\hat{\sigma}^{(n)}_{LL} = 
\langle {\cal M}_0 | \frac{1}{n!} \left(d \hat{\boldsymbol \sigma}^{(1)} 
\right)^n
|{\cal M}_0 \rangle \,,
\eea 
which is exactly Eq.~(\ref{eq:norderxsec}). Once one sums up all orders, 
one realizes the LL resummation 
\bea 
d\hat{\sigma}_{LL} = 
\langle {\cal M}_0 | 
\sum_{n=0}^\infty 
\frac{1}{n!} \left(
d \hat{\boldsymbol \sigma}^{(1)} 
\right)^n |{\cal M}_0 \rangle  
=  \langle {\cal M}_0 |e^{d \hat{\boldsymbol \sigma}^{(1)} }  |   {\cal M}_0   \rangle 
\,, 
\eea 
which is exactly Eq.~(\ref{eq:xsecLL}) when going to the Mellin space.

\section{Numerical Analyses}~\label{sec:numerics}

In this section, we perform numerical analyses. For the time being, we pay attention to the numerical validation of the theoretical framework developed previously, while we leave the comparison of the phenomenological predictions with the LHC inclusive jet data to another work~\cite{Xie:future1}. 
\begin{figure}[htbp] 
	\begin{center}
		\includegraphics*[width=0.45\textwidth]{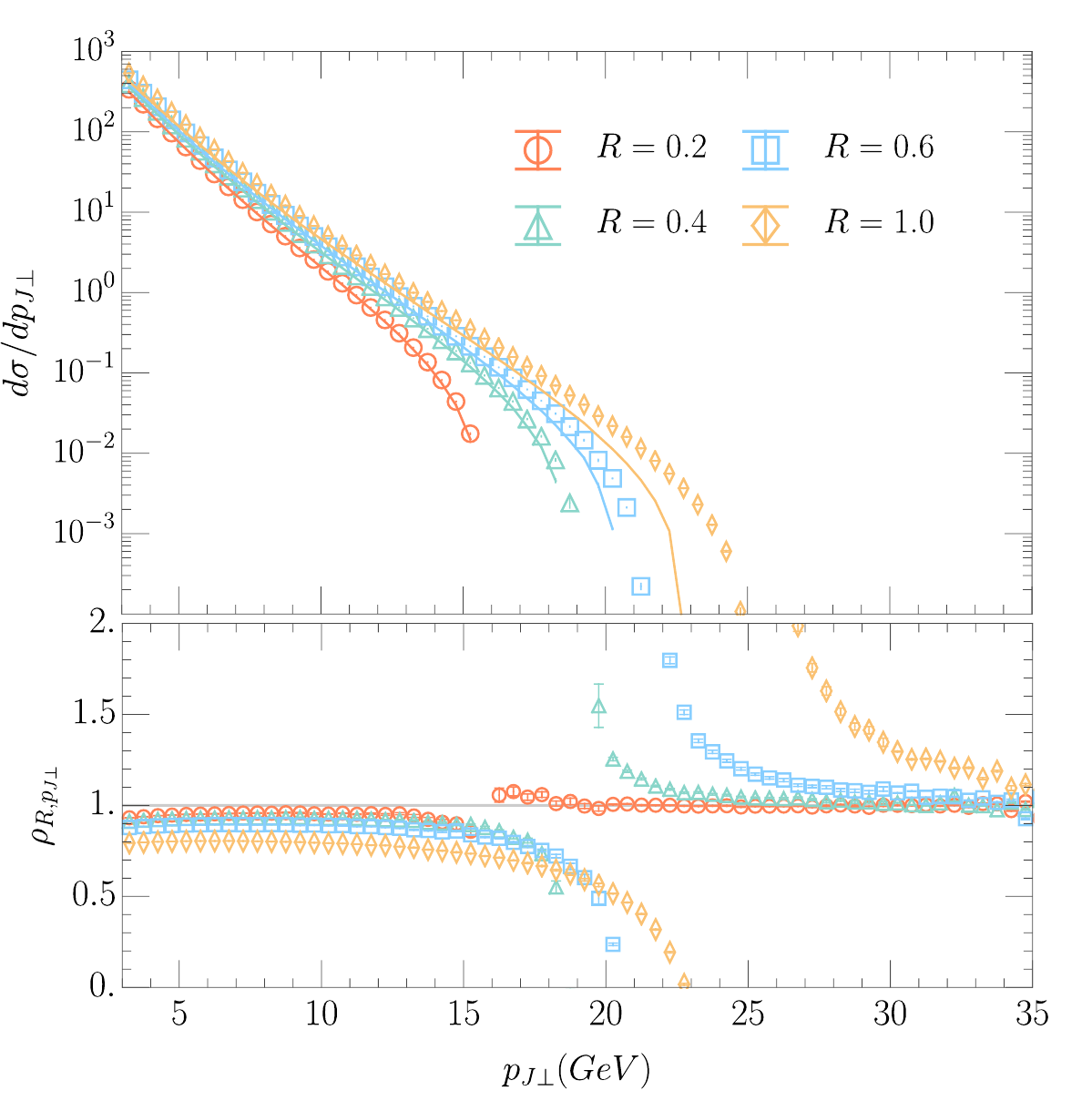}
		\includegraphics*[width=0.45\textwidth]{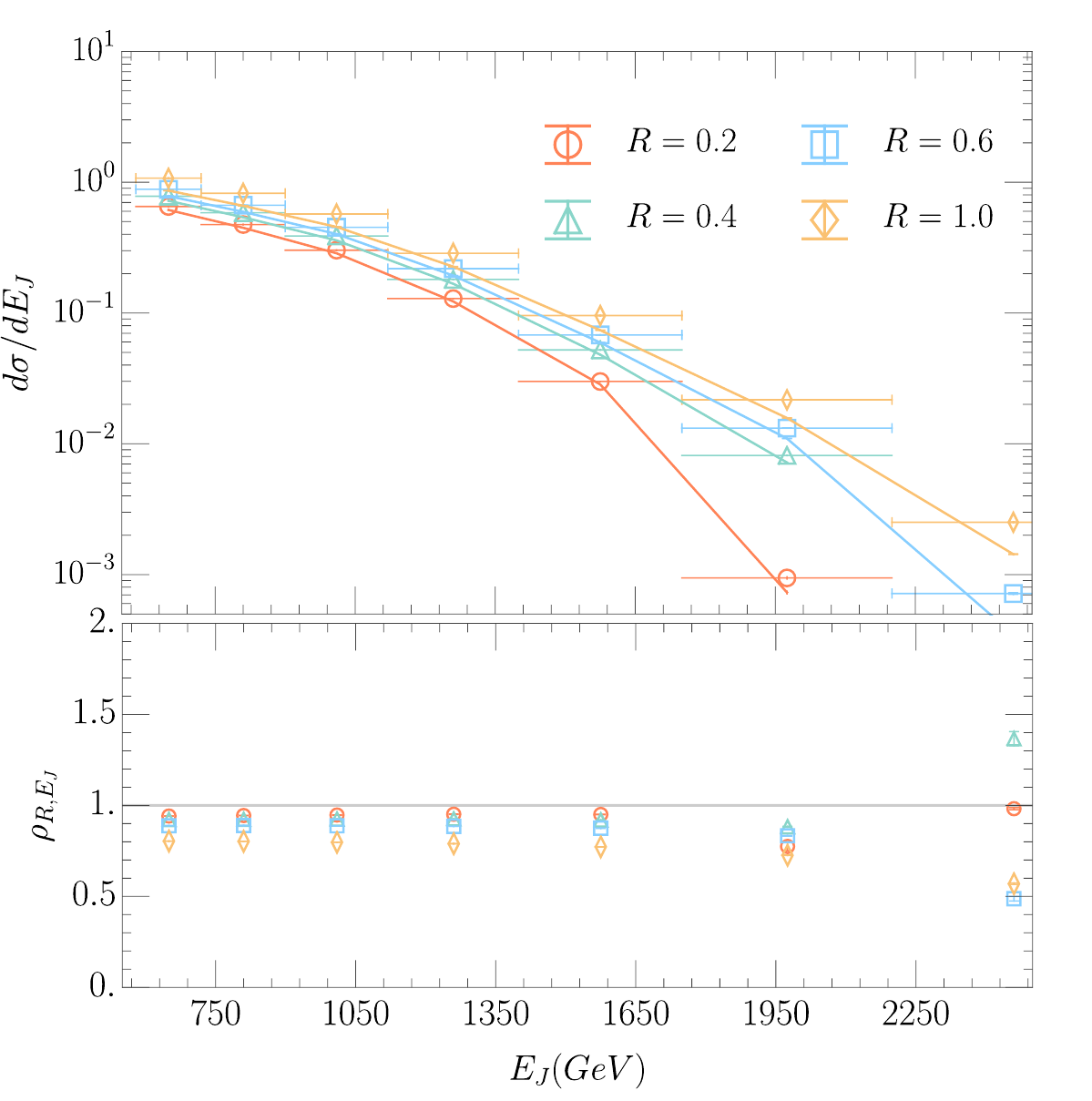}
		\caption{Comparison between the NLO calculations with full jet algorithm dependence and the small-$R$ limits. The open points represent the NLO calculations in quark channel with the full jet algorithm dependence, using different shapes to represent different jet radius $R$. The lower panels are the ratio of these two results defined in Eq.~\eqref{eq:ratio_full-R}. The solid lines are the corresponding results in the small-$R$ limit. Left panel shows the inclusive jet $p_{J\perp}$ spectra. Right panel for the jet energy $E_J$ distribution. These are for jets constructed via the anti-$k_T$ algorithm with $R=0.5$ in p+Pb collisions at center-of-mass energy $\sqrt{s} = 5.02$ TeV per nucleon pair and rapidity $5.2 < \eta < 6.6$.}
		\label{fig:FullvsSmallR}
	\end{center}
\end{figure}
 
For the numerical calculation, we use the NLO \textsc{MSTW2008} PDF sets~\cite{Martin:2009iq} for the proton PDFs. 
We applied the LL BK evolution with $\alpha_s$ running~\cite{Kovchegov:2006vj,Balitsky:2006wa,Fujii:2013gxa} to the nucleus dipole distributions. As for the initial condition, the MV-like model ``$\gamma_{1119}$'' are used and we stick to the parameter settings in~\cite{Fujii:2013gxa}. As for the $S_\perp$, we follow the choice in \cite{Watanabe:2015tja}. In the calculation, we implement the kinematic cuts following strictly the CMS experimental set-ups in~\cite{CMS:2018yhi}, in which the proton and the Pb nucleus are colliding at the center of mass energy $\sqrt{s} = 5.02\, {\rm~TeV}$ per each nucleon pair. The jets are constructed by the anti-$k_T$ algorithm with the radius parameter $R = 0.5$ and required to have a minimum transverse momentum $p_{J\perp} > 3~{\rm GeV}$. We select the jets at very forward pseudo-rapidities $5.2 <\eta_J < 6.6$ in the laboratory frame, the boost between the laboratory frame and center-of-mass frame
is $\delta \eta_J =0.465$. Throughout the analyses, we set the  factorization scale $\mu = p_{J\perp}$ and the rapidity scale $X_f = X_A$ where $X_A$ is the
momentum fraction carried by the gluon from the nucleus. 

In fig.~\ref{fig:FullvsSmallR}, we first display the comparison between the NLO single inclusive jet cross section with full jet algorithm dependence and its analytical small-$R$ approximation. We stick to the quark channel for this study. The NLO distributions are obtained by the histogram procedure described in Section~\ref{subsec:fullNLO}. The error bars reflect the numerical uncertainties. 
We have performed the comparisons for $R = 0.2$, $0.4$, $0.6$ and $1.0$. 

In the left top panel of fig.~\ref{fig:FullvsSmallR}, we plot the inclusive jet $p_{T\perp}$ distributions from the full NLO predictions (in open dots) and the small-$R$ approximations (in solid lines). From the plot, we can observe that for larger values of $R$, jet spectrum spans to larger transverse momentum. This is expected since with a larger jet radius, one clusters more particles into one jet and thus more effortlessly generates larger jet transverse momentum to pass the $3\, {\rm GeV}$ threshold than the small jet radius case. We also observe that just like the single hadron production in $pA$ collisions, the inclusive jet $p_{J\perp}$ spectrum manifests the negative cross section problem for large $p_{J\perp}$. The issue can be resolved by the threshold resummation~\cite{Liu:2020mpy,Xie:future1} as sketched in Section~\ref{sec:threshold}. As we go to the smaller values of the jet radius $R$, the better the small-$R$ approximation becomes. This can be demonstrated from the ratio 
\bea
\label{eq:ratio_full-R}
\rho_{R,p_{J\perp}} \equiv \frac{d\sigma_{\rm full}/dp_{J,\perp}}{d\sigma_{\text{small $R$}}/dp_{J,\perp}}\,,
\eea
where the numerator and denominator represent the jet cross section with the full jet algorithm and its small-$R$ limit, respectively. This ratio is shown in the lower part of the left panel, which shows strongly that as $R$ becomes smaller, the ratio $\rho_{R,p_{J\perp}}$ approaches 1. The comparison serves as a strong validation of the analytic small-$R$ approximation and the factorization we derived. 
From the ratio, we can see that for small $p_{J\perp}$, the small-$R$ approximation makes up $\ge 90\%$ of the complete NLO result for narrow jets with $R =0.2$, $0.4$ and $0.6$, while the percentage drops to $80\%$ for fat jets with $R = 1.0$. The approximation is even better for large $p_{J\perp}$. The small-$R$ approximation could break down when strong cancellation kicks in between different contributions, and therefore one should be careful when invoking the approximation for phenomenology studies. In the present case, it happens to be when the cross section becomes negative. 

We plot the comparison using the inclusive jet $E_J$ spectrum in the right panel of fig.~\ref{fig:FullvsSmallR}. The division of the energy bins follows the CMS~\cite{CMS:2018yhi}. The plot demonstrates the feasibility of our NLO calculation for various different observables and experimental set-ups. 
Similar behaviours are observed in the $E_J$ distribution as the $p_{J\perp}$ spectrum. The NLO cross section eventually turns negative and the threshold resummation comes to rescue~\cite{Xie:future1}. The small-$R$ approximation did a better job in the $E_J$ prediction than the $p_{J\perp}$ case, which can be understood that it is mostly the low $p_{J\perp}$ jet events that fill up the $E_J$ spectrum. This also explains why the NLO $E_J$ spectrum can stay positive for a wider range. 
\begin{figure}[htbp]
	\begin{center}
		\includegraphics*[width=0.65\textwidth]{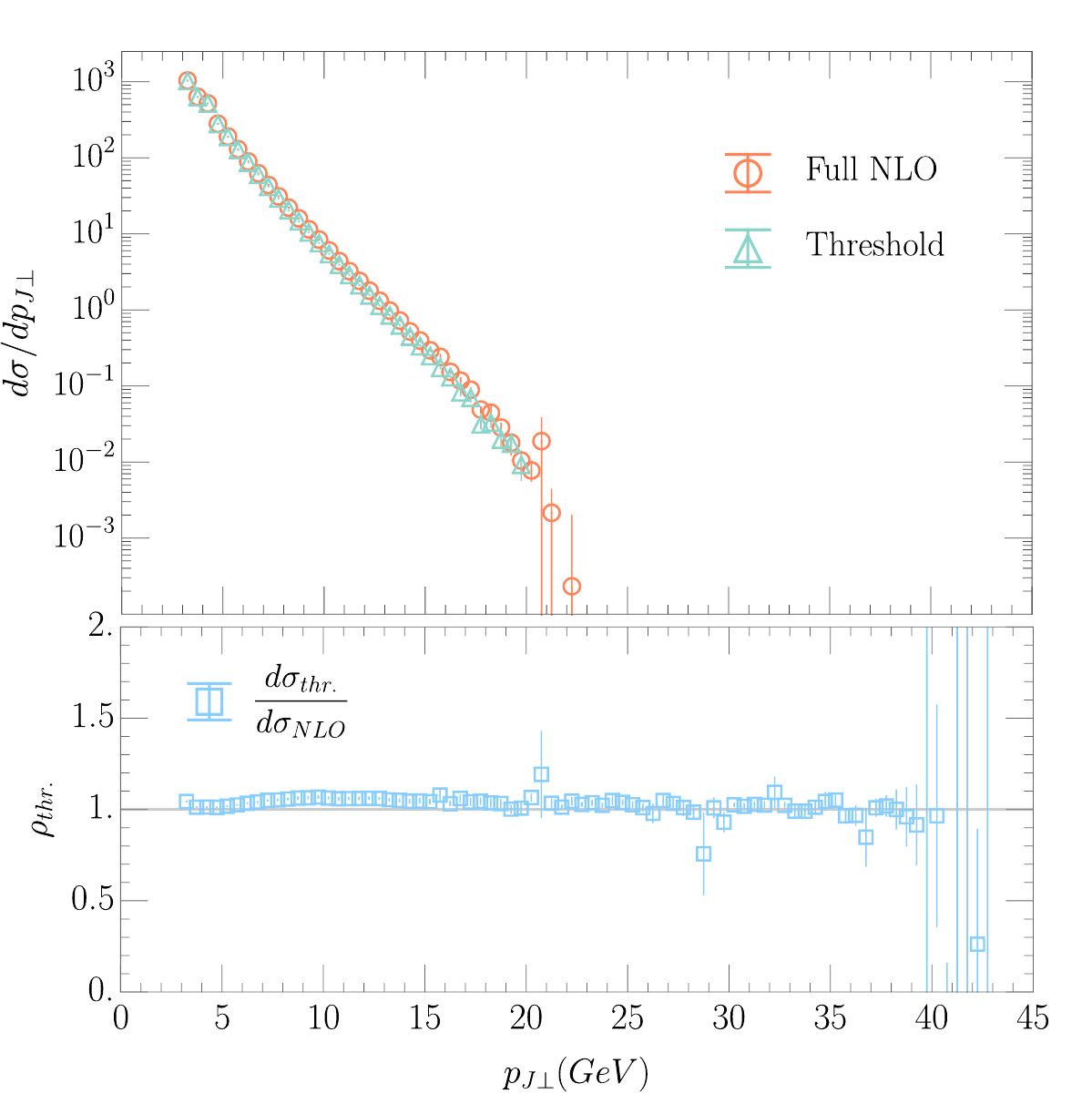}
		\caption{This figure shows the comparison of the threshold contribution with the $\text{NLO}$ result. The error band represents the numerical uncertainty.}
		\label{fig:rthrdp}
	\end{center}
\end{figure}

Now we turn to study the threshold limit in fig.~\ref{fig:rthrdp}. Here we include all partonic channels which contain threshold logs (the $q\to q$ and $g \to g$ channels). The upper panel displays the jet $p_{J\perp}$ spectrum predicted by the complete NLO calculation (in red dots) and its threshold approximation (in green triangles). The NLO cross section becomes negative when $p_{J\perp} > 20\, {\rm GeV}$. To investigate the size of the threshold contributions, we plot the ratio of the threshold limit to the full NLO prediction $\rho_{thr.} =\frac{ d\sigma_{thr.}/dp_{J\perp}}{d\sigma_{{\rm NLO}}/dp_{J\perp}}$ as a function of the jet transverse momentum $p_{J\perp}$. To manifest the effect of the threshold logarithms $\left(\frac{\ln^i(1-\xi)}{1-\xi}\right)_+$, we have removed from the ratio the common $\delta(1 - \xi)$ terms shared by both $d\sigma_{thr.}$ and $d\sigma_{\rm NLO}$. We see that when we crank up $p_{J\perp}$, the threshold logarithms
are overwhelmingly dominant and the ratio $\rho_{thr.}$ approaches $1$.

\section{Conclusions}\label{sec:conclusion}
In this work, we applied the recently proposed computational techniques~\cite{Kang:2019ysm,Liu:2020mpy} to derive the complete NLO corrections to the
single inclusive jet production in proton-nucleus ($pA$) collisions at forward rapidities, using the CGC effective theory. Within the framework, the cross section is factorized into the product of the proton PDFs, the small-$x$ nucleus multi-pole distributions and the perturbatively calculable partonic cross sections. Our result clearly shows that the CGC factorization is valid at the NLO level for jet production.

The root of the factorization is the consistent and homogeneous power expansion in the power counting parameter $\lambda \sim {\cal O}\left(\sqrt{-t/s} \right)$. The power expansion relaxes the phase space kinematic constraint which in turn is compensated by the existence of the soft contribution. Our explicit NLO calculation showed that the soft mode is required in order to generate the kinematic restrictions due to the jet clustering in addition to the constraint in the hadron production case~\cite{Watanabe:2015tja}.   

Our ${\cal O}(\alpha_s)$ calculation of the jet cross sections is fully differential over the final state
physical kinematics. The calculation is thus not limited to the jet spectra predictions but is able to predict any distribution of infra-red safe quantities. The full NLO jet cross section is given in Eq.~(\ref{eq: quark-NLO}) and Appendix~\ref{sec:allchannels}. To perform this computation, we have discussed a subtraction method to single out the singular contributions from the phase space integration. The singular terms have been evaluated analytically. The remaining non-singular part contains the jet algorithm and the experimental cuts. Since the remaining integration is finite and integrable in $4$-dimension, it was obtained numerically. 

We further investigated the small jet radius limit of jet production. Given that in practice the jet radius $R$ is usually chosen as ${\cal O}(0.4)$, the small-$R$ limit in general is believed to be a very good approximation to the full $R$ result and has been applied in several small-$x$ jet studies. In this work, we are able to carry out the fully analytic small-$R$ jet cross section in Eq.~(\ref{eq:small-r-xsec}) and  compare it with the full jet algorithm dependent cross section to validate the approximation. We showed that in the small jet radius limit, the single inclusive
jet cross section can be further factorized into the same short-distance cross section as the single inclusive hadron
production, with only the fragmentation functions replaced by the semi-inclusive jet functions (siJFs). The siJFs  share the same form as the ones that show up
in the jet production in the central region within the collinear factorization with slightly different $\zeta$ dependence in the logarithm (see Eq.~(\ref{eq:siJF})) due to the different jet transverse momentum range. We
argued that this factorization feature of the small-$R$ jet cross section holds for generic jet processes and jet substructure observables in the
CGC framework. 
We carefully examined numerically when the small-$R$ limit reliably approximates the full result. The obtained analytic result serves as a guide to appropriately using the small-$R$ approximation in the forward scatterings, meanwhile the observed factorization can facilitate high order 
and resummation calculations of the jet cross sections and open up the opportunities to realize the computations of event shapes in other forward scatterings. Furthermore, the observed factorized form lays down the foundation for combining the CGC fixed order calculations with the existing parton shower techniques. 

Like the forward hadron production, the obtained NLO jet spectrum also
exhibits the negative cross section feature when the jet transverse momentum becomes large. Following the suggestions in~\cite{Liu:2020mpy}, we resolved the negative cross section problem by the threshold resummation, where additional Sudakov logarithms arising in the jet production have also been resummed. Different from~\cite{Liu:2020mpy}, in this manuscript, we accomplish the leading logarithmic threshold resummation
using a different approach by considering the strongly ordered independent emissions to all orders. The achieved resummation  agrees with the one obtained by the renormalizaton group equation~\cite{Liu:2020mpy,Xie:future1}, which suggests the plausibility of both approaches.

We look forward to comparing the NLO and resummation predictions against the LHC $pA$ jet production data~\cite{Xie:future1} in the future. Meanwhile, we expect the computational set-ups developed in the present work and in~\cite{Kang:2019ysm,Liu:2020mpy} lay down the theoretical foundation of the perturbative predictions for the CGC phenomenology involving jet clustering procedures and any other jet substructure and event shape observables.

\acknowledgments
We thank Farid Salazar for suggestive comments on our manuscript. This work is supported by the National Natural Science Foundation of China under Grant No.~12175016 (H.L., K.X. and X.L.), China Postdoctoral Science Foudation under Grant No.~212400211 (H.L.) and the National Science Foundation under grant No. PHY-1945471 (Z.B.).

\newpage 

\appendix 

\section{Useful Formulae}
\subsection{Contour Integral} 
The integral of the form 
\bea
I = \int \frac{d^Dk}{(2\pi)^D} \frac{1}{(k^2+i0^+)((p_j-k)^2+i0^+)}  \,,
\eea 
is frequently 
encountered in this work. The $k$-loop integration can be performed by contour, which gives 
\bea\label{eq:contour} 
&&\int \frac{d^Dk}{(2\pi)^D} \frac{1}{(k^2+i0^+)((p_j-k)^2+i0^+)}  \nn \\
& = & - 
\frac{1}{2} \int \frac{dk^+ dk^-}{(2\pi)^2}
\frac{d^{D-2}k_\perp}{(2\pi)^{D-2}}
\frac{(k^+(p_j^+-k^+))^{-1}}{(k^- - \frac{k_\perp^2- i0^+}{k^+} )
(k^- - (p_j^-   - \frac{(p_{j\perp}-k_\perp)^2-i0^+}{p_j^+ - k^+}))} 
\nn \\
& = & - (2\pi i )
\frac{1}{2} p_j^+ 
\int_0^1 \frac{dz}{(2\pi)^2} \int 
\frac{d^{D-2}k_\perp}{(2\pi)^{D-2}}
\frac{(k^+(p_j^+-k^+))^{-1}}{
\frac{p_{j\perp}^2}{p_j^+}
- \frac{(p_{j\perp}-k_\perp)^2}{p_j^+ - k^+} - \frac{k_\perp^2}{k^+} 
} \nn \\ 
& = &  
\frac{i}{4\pi}  
\int_0^1 d z \int 
\frac{d^{D-2}k_\perp}{(2\pi)^{D-2}}
\frac{1}{(k_\perp - (1-z)p_{j\perp})^2} \,,
\eea 
We first did the $k^-$-integral. The integral is only non-vanishing when $0 < k^+ < p_j^+$ otherwise all the poles occur at the same side of the complex $k^-$-plain and one can choose the contour in the other side of the plain to enclose no poles. 
To get the 3rd line, we have performed the $k^-$-contour integral around the upper plane which picks the pole $k^- = p_j^-   - \frac{(p_{j\perp}-k_\perp)^2-i0^+}{p_j^+ - k^+}$.  and we let $p_j^- = p_\perp^2/p_j^+$ to reach the final result. 

\subsection{Fourier Transformation} 
We list some integral formulae for deriving the NLO results. 
\bea\label{eq:q2-Fourier} 
\int d^{D-2} q_\perp \frac{e^{-iq_\perp \cdot r_\perp} }{q_\perp^{2+2\alpha}} 
&=& \Omega_{D-3}  
\int d^{D-2} q_\perp  \, q_\perp^{D-3} \, 
\int_0^\pi d \phi  \, 
 \, \sin^{-2\epsilon}\phi \, 
\frac{e^{-iq_\perp r_\perp \cos \phi }}{q_\perp^{2+2\alpha }} \nn \\
&=& \int dq_\perp q_\perp^{D-5-2\alpha} \,  \Omega_{D-3} \, 
\sqrt{\pi} 
\frac{\Gamma\left(\frac{1}{2} - \epsilon \right) }{\Gamma(1-\epsilon)} 
{}_0F_1\left(1-\epsilon,-\frac{1}{4}q_\perp^2 r_\perp^2 \right) \nn \\
&=&  
\pi^{1-\epsilon} \, 
\frac{\Gamma(-\epsilon-\alpha)  }{\Gamma(1+\alpha )}
\left( \frac{r_\perp}{2} \right)^{2\epsilon+2\alpha} \,.
\eea 
Here $\Omega_{D-3} = \frac{2\pi^{(D-3)/2}}{\Gamma((D-3)/2)}$ the $(D-3)$-dimensional solid angle, 
and from the result we can easily deduce that 
\bea\label{eq:qoverq2-Fourier} 
\int \mathrm{d}^{D-2} q_\perp
\frac{e^{-i q_\perp \cdot r_\perp } }{q_\perp^2} q_\perp^\alpha 
=  i \partial_{r_{\perp\alpha}} \int \mathrm{d}^{D-2} q_\perp
\frac{e^{-i q_\perp \cdot r_\perp } }{q_\perp^2} 
= -i \, 2^{1-2\epsilon} \pi^{1-\epsilon} \Gamma(1-\epsilon) \frac{r_\perp^\alpha}{r_\perp^2} r_\perp^{2\epsilon} \,.  
\eea 

\subsection{Mellin Transformation}
The Mellin transformation or the $N$th-moment of a function is given by
\bea 
M_N(f(\xi)) = \int_0^1 d\xi \xi^{N-1}f(\xi) \,.
\eea 
If we apply the Mellin transform to the convolution of $2$ functions 
\bea 
f(\xi)\otimes g(\xi) 
= \int_0^1 \frac{d\xi'}{\xi'} f(\xi') g\left(\frac{\xi}{\xi'} \right)  \,, 
\eea 
we will have 
\bea 
M_N(f(\xi)\otimes g(\xi) ) 
= M_N(f(\xi)) M_N(g(\xi)) \,. 
\eea 

In the resummation part of this work, the transformation we used are 
\bea 
&& M_N(\delta(1-\xi)) = 1 \,,  \nn \\ 
&& M_N(1) = \frac{1}{N} \,,  \nn \\ 
&& M_N(1-\xi) = \frac{1}{N(N-1)} \,, \nn \\ 
&& M_N\left( \frac{1}{(1-\xi)_+} \right)
= - H_{N-1} \,,  \nn \\ 
&& M_N\left( \left[\frac{\ln(1-\xi)}{1-\xi}\right]_+ \right) = \sum_{i=1}^{N-1} \frac{H_i}{i} \,, 
\eea 
where $H_n$ is the harmonic number, defined as 
\bea 
H_n = \sum_{i=1}^n \frac{1}{i} \,. 
\eea 
The threshold limit $\xi \to 1$ will map onto the large $N$ limit, which leads to  the limit
\bea\label{eq:MellinThr} 
&& M_N(\delta(1-\xi)) \to  1 \,,  \nn \\ 
&& M_N(1) \to 0  \,,  \nn \\ 
&& M_N(1-\xi) \to  0  \,, \nn \\ 
&& M_N\left( \frac{1}{(1-\xi)_+} \right)
\to - \ln {\bar N} \,,  \nn \\ 
&& M_N\left( \left[\frac{\ln(1-\xi)}{1-\xi}\right]_+ \right) \to \frac{1}{2} \ln^2{\bar N} + \frac{\pi^2}{12} \,,
\eea 
where ${\bar N} = N e^{\gamma_E}$.

\subsection{Color identities}\label{subsec:color-id}
We provide several color identities used in the CGC multi-point correlators. 

Through the Fiertz identity, one can derive that  
\bea\label{eq:WWWab}
&& 
{\rm Tr}[ W^\dagger(b'_\perp) t^a W^\dagger(b_\perp')t^b] W_{ab}(r_\perp) \nn \\
&=& \frac{1}{2}
{\rm Tr}[W(b_\perp)W^\dagger(r_\perp)]
{\rm Tr}[W(r_\perp)W^\dagger(b'_\perp)]
- \frac{1}{2N_C}
{\rm Tr}[W(b_\perp)W^\dagger(b_\perp')] \,, 
\eea 
which relates the $3$-point functions to dipole distributions. 

Also one needs the following relation that 
\bea\label{eq:color-identity} 
&& \frac{2}{N_C^2}{\rm Tr}\left[ t^{b'} 
W^\dagger(b_\perp)
W(b_\perp) t^b
\right]
W^\dagger_{b'a}(r_\perp')
W_{ab}(r_\perp) \nn \\ 
&=& \left( \frac{1}{N_C^2}
{\rm Tr}[W(b_\perp)W^\dagger(b_\perp')W(r_\perp')W^\dagger(r_\perp)]
{\rm Tr}[W(r_\perp)W^\dagger(r'_\perp)] - \frac{1}{N^3_C}  {\rm Tr}[W(b_\perp) W^\dagger(b'_\perp)] \right) \,. \nn \\ 
\eea

\section{Contributions from the other channels}\label{sec:allchannels}
In this section, we list all the remaining channels that contribute to the inclusive jet production, which are the $g\to g$,$q\to g$ and $g\to q$ sub-processes. 
\subsection{$g\to g$ channel}
The LO $g\to g$ is given by 
\bea\label{eq:LOsecg} 
d\sigma^{(0)}_{gg}
 = \tau G(\tau) 
 \int \frac{d^{D-2}b_\perp d^{D-2}b_\perp'}{(2\pi)^{D-2}}  \, 
 e^{-i p_{J\perp} \cdot z_\perp} 
 S_{X_f}^{(2)}(b_\perp,b_\perp')S_{X_f}^{(2)}(b_\perp,b_\perp') \,, 
 \eea 
whose form in the momentum space is
\bea 
d\sigma^{(0)}_{gg}
 = \tau G(\tau) S_\perp {\cal F}_A(p_{J\perp};X_f) \,,
 \eea 
where $\tau = p^+_J/p_p^+$, $G(x)$ denotes the gluon PDF of the proton and $\mathcal{F}_A(k_\perp;X_f)$ represents the momentum space nucleus dipole distribution in the adjoint representation, whose definition is
\bea 
(2\pi)^{D-2}
S_\perp {\cal F}_A(k_\perp,X_f)
= \int d b_\perp db_\perp' 
e^{-i k_\perp \cdot z_\perp}
S^{(2)}_{X_f}(b_\perp,b_\perp')S^{(2)}_{X_f}(b_\perp,b_\perp')  \,, \quad 
\eea 
where $z_\perp = b_\perp - b'_\perp$. The relation between the dipole in the adjoint representation ${\cal F}_A$ and the fundamental representation ${\cal F}_F$ is given by 
\begin{align}
	\mathcal{F}_A(k_\perp;X_f)=\int d^{D-2} k_{1\perp}\mathcal{F}_F(k_\perp;X_f)\mathcal{F}_F(k_\perp-k_{1\perp};X_f).
\end{align}

The NLO correction of the $g\to g$ channel to the jet cross section is found to be 
\bea\label{eq: gluon-NLO}
		d\sigma^{(1)}_{gg} && = 
\frac{\alpha_s}{2\pi}N_C
\int \frac{d b_\perp d b'_\perp}{4\pi^2}
\int_{\tau}^1 d\xi \,  
xG(x)
\, 
e^{-i   p_{J\perp} \cdot z_\perp} \nn \\
&& \times 
\left( 
{\cal  H}_{g,2}  \, \, 
S_X^{(2)}(b_\perp,b_\perp') S_X^{(2)}(b_\perp,b_\perp') + {\cal  H}_{q\bar{q},2}  \, \, 
S_X^{(3)}(b_\perp,r_\perp, b_\perp')  \right. \nn \\
&& \left. + \int \frac{dr_\perp}{\pi} 
\Big( 
{\cal H}_{g,BK}
+
{\cal H}_{g,3} 
+
{\cal H}_{g,kin.}
\Big) 
S_X^{(2)}(b_\perp,b_\perp') S_X^{(2)}(b_\perp,r_\perp)S_X^{(2)}(r_\perp,b_\perp')   \right)
\nn \\
&& + (d\sigma_{g,fsr} - d\sigma^c_{g,fsr}) + (d\sigma_{g,isr}-d\sigma^c_{g,isr})+ (d\sigma_{g,inter.}-d\sigma^c_{g,inter.}) 
 + (d\sigma_{gR,soft}-d\sigma_{gR,soft}^c),
\,\nn \\  
\eea 
where $x=\tau/\xi$ and $\xi=\frac{p_{jg}^+}{p_g^+}$. Here $p_g$ is the momentum of the initial state gluon, $p_{jg}$ denotes the momentum of one of the final state gluons that will become the signal jet in the 2-jets case.

The hard factor ${\cal H}_{g,2}$, ${\cal H}_{q\bar{q},2}$ and ${\cal H}_{g,3}$ are found as the following,
\bea\label{eq:Hg2} 
{\cal H}_{g,2} &=& 
\left\{- P_{gg}^{(1)}(\xi) \ln \frac{z_\perp^2\mu^2}{c_0^2} 
+\left[ \frac{2 N_f T_R}{3 N_C}\ln \frac{z_\perp^2\mu^2}{c_0^2}\right.\right.\nn \\
&&\left.\left.+\left(\frac{11}{3}-\frac{4 N_f T_R}{3 N_C}\right)\ln \frac{z_\perp^2p_{J\perp}^2}{c_0^2}-\frac{2N_f T_R}{3N_C}
\right]\delta(1-\xi) 
\right\} \,, 
\eea
\bea\label{eq:Hqqbar2} 
{\cal H}_{q\bar{q},2} = 
\frac{2 N_f T_R}{N_C}\delta(1-\xi) 
\int \frac{dr_\perp}{\pi}\left[ 
\frac{e^{ip_{J\perp} \cdot y_\perp} }{y_\perp^{2} }\right]_+
\int_{0}^{1}d{\xi'}[\xi'^2+(1-\xi')^2]
 e^{-i\xi' p_{J\perp} \cdot y_\perp} \,, 
\eea
\bea\label{eq:Hg3} 
{\cal H}_{g,3} = &&4 \frac{[1-\xi(1-\xi)]^2}{(1-\xi)_+} 
\frac{ x_\perp \cdot y_\perp}{x_\perp^{2}
y_\perp^{2}}\,
\frac{1}{\xi^2}
e^{-i \frac{1-\xi}{\xi} p_{J\perp} \cdot y_\perp}\nn \\
&&+ 4 \delta(1-\xi) 
\left[ 
\frac{e^{ip_{J\perp} \cdot y_\perp} }{y_\perp^{2} }\right]_+
\int_{0}^{1}d{\xi'}[\frac{\xi'}{(1-\xi')_+}-\frac{1}{2}\xi'(1-\xi')]
 e^{-i\xi' p_{J\perp} \cdot y_\perp} 
 \,,  \quad \quad 
\eea 
where $x_\perp = b_\perp - r_\perp$, $y_\perp = r_\perp - b_\perp'$ and 
\bea\label{eq:splitgg}
P^{(1)}_{gg}(\xi)=2\left[\frac{\xi}{(1-\xi)^+}+\frac{1-\xi}{\xi}+\xi(1-\xi)\right]+\left(\frac{11}{6}-\frac{2 N_f T_R}{3 N_C}\right)\delta(1-\xi),
\eea
in Eq.~\ref{eq:splitgg}, $T_R=\frac{1}{2}$, $N_f$ denotes the number of flavors in the quark loop, which is taken to be $3$ in our calculation.

The $\nu$-related logarithmic term is
\bea \label{eq:HgBK}
{\cal H}_{g,BK}
= 
2 \left( \ln \frac{\nu p_g^+}{p_{J\perp}^2}
\right) 
\left[\frac{z_\perp^2}{x_\perp^2y_\perp^2}\right]_+\, 
 \delta(1-\xi) \,, 
\eea 
which is also proportional to the LO kinematics, similar to the $q \to q$ channel case.

The kinematic constraint term is given by
\bea \label{eq:Hgkin}
{\cal H}_{g,kin.}
= 2
\left[ \frac{\ln (x_\perp^2 p_{J\perp}^2/c_0^2)}{x_\perp^2}
+ \frac{\ln (y_\perp^2 p_{J\perp}^2/c_0^2)}{y_\perp^2}
+ \frac{2x_\perp\cdot y_\perp }{x_\perp^2y_\perp^2}
\ln\left(\frac{x_\perp y_\perp p_{J\perp}^2}{c_0^2} \right) \right]_+
\delta(1-\xi) \,. \quad 
\eea 

Finally, the terms in the last line of Eq.~\ref{eq: gluon-NLO} are
\bea\label{eq:theta1-0g}
d\sigma_{g,fsr}-d\sigma^c_{g,fsr} &=&\frac{\alpha_s S_\perp}{2\pi^2}
N_C
\int_{0}^1d\xi
\frac{[1-\xi(1-\xi)]^2)}{\xi(1-\xi)}
\int  d^2 p_{k\perp}  \Bigg\{2 \times \Theta_2 \, xG(x)\frac{\mathcal{F}_A(p_{k\perp}+ p_{J\perp};X_f)}{[\xi p_{k\perp}-(1-\xi)p_{J\perp}]^2} \nn \\
&& 
+\Theta_1 \, \tau G(\tau)
\frac{\mathcal{F}_A(p_{J\perp};X_f)}{[p_{k\perp}-(1-\xi)p_{J\perp}]^2}
-\tau G(\tau)\frac{\mathcal{F}_A(p_{k\perp}+\xi p_{J\perp};X_f)}{[p_{k\perp}-(1-\xi)p_{J\perp}]^2}\Bigg\} \,,
\quad \quad 
\eea 
where the factor $2$ before $\Theta_2$ accounts for the contribution from filling the second jet to the histogram if it passes the jet threshold, where we have used the fact that the filling is symmetric between the $2$ gluon jets. 

The interference term is given by
 \bea \label{eq:ibkbg}
d\sigma_{g,inter.}-d\sigma_{g,inter.}^c &=& - \frac{\alpha_s S_\perp}{2\pi^2} N_C
\int_{\tau}^1d\xi
 \frac{[1-\xi(1-\xi)]^2)}{\xi(1-\xi)}\\ 
&&\hspace{-5.ex}\int d^{2}p_{k\perp}
\frac{d^{D-2}l_{\perp}}{(2\pi)^{D-4}}\frac{d^{D-2}l_{\perp}^\prime}{(2\pi)^{D-4}} \, 
xG(x) \, 
\mathcal{F}_F(l_{\perp}';X_f)\mathcal{F}_F(l_{\perp}-l_{\perp}';X_f)
\nn \\
&&\hspace{-5.ex} \Bigg\{  (2\Theta_2 
+\Theta_1 - 2 )\,  \, \mathcal{F}_F(p_{k\perp}+p_{j\perp}-l_{\perp}';X_f)
\frac{2(\xi p_{k\perp}-(1-\xi)p_{j\perp})\cdot(l_{\perp}-p_{j\perp})}{(\xi p_{k\perp}-(1-\xi)p_{J\perp})^2(l_{\perp}-p_{j\perp})^2} \Bigg\} \,\nn ,\\\quad  
\eea 
where a factor $2$ is again included in the $2$-jet case to account for filling both of the jets in the final state. In this case, the counter contribution will also be filled twice. The same situation also happens to the ISR contribution which gives 

\bea\label{eq:if-subg}
&&
d\sigma_{g,isr} -
d \sigma_{g,isr}^c
=   
\frac{ \alpha_s}{2\pi^2}
N_C 
\int^1 d\xi 
 \,    
 \frac{[1-\xi(1-\xi)]^2)}{\xi(1-\xi)}
\int
d^{2}p_{k\perp} xG(x)
\Big( \Theta_1 +   2\Theta_2 - 2 \Big)
\nn \\ 
&& 
\times \int 
\frac{db_\perp db_\perp' }{4\pi^{2}}
\, 
\frac{d r_\perp 
d r_\perp'}{4\pi^{2}}
\,
 e^{-ip_{j\perp}\cdot z_\perp}
 e^{-i p_{k\perp} \cdot z_\perp'}
\frac{x_\perp \cdot x_\perp'}{x_\perp^{2} {x_\perp'}^{2}}
S_{X_f}^{(8)}(b_\perp,r_\perp,b_\perp',r_\perp')\,.
\eea 

The soft contribution gives 
\bea\label{eq:NLO-soft-subg}
&& d\sigma_{gR,soft} - d\sigma_{gR,soft}^c \nn\\
&=&\frac{\alpha_s}{\pi^2}\frac{N_C}{2} 
\tau G(\tau) \, 2\, 
\int_{-\infty}^\infty d \eta_k 
\int d^{2}p_{k\perp}\frac{d^{D-2}l_{\perp}^\prime}{(2\pi)^{D-4}} \, 
\left(\Theta_{1,soft} + \Theta_{2,soft} -1 \right) 
 \, 
\nn 
\\
&&  \,
\Bigg\{ 
S_\perp \mathcal{F}_F(l_{\perp}';X_f) \mathcal{F}_F(p_{k\perp}+p_{j\perp}-l_{\perp}';X_f) 
\left[\frac{2}{ p_{k\perp}^2} 
-\int d^{D-2}l_{\perp}\mathcal{F}_F(l_{\perp}+p_{j\perp}-l_{\perp}';X_f)
\frac{2 p_{k\perp}\cdot l_{\perp}}{ p_{k\perp}^2 l_{\perp}^2} \right]  \, \nn \\ 
&+&\int 
\frac{db_\perp db_\perp'}{4\pi^2}
 \frac{dr_\perp dr_\perp'}{4\pi^2}
\frac{x_\perp \cdot x_\perp'}{x^2_\perp\,{x'_\perp}^2}
e^{-ip_{j\perp}\cdot z_\perp} 
e^{-ip_{k\perp}\cdot z'_\perp} 
S_{X_f}^{(8)}(b_\perp,r_\perp,b_\perp',r_\perp') \Bigg\}  \,, 
\eea 
where $x_\perp' = b_\perp' - r_\perp'$ and $z_\perp' = r_\perp - r_\perp'$ and
\bea\label{eq:8-point} 
S_{X_f}^{(8)}(b_\perp,r_\perp,b_\perp',r_\perp') = 
\frac{1}{N_C^3}
{\rm Tr}[W(b_\perp)W^\dagger(b_\perp')W(r_\perp')W^\dagger(r_\perp)]
{\rm Tr}[W(r_\perp)W^\dagger(r'_\perp)]{\rm Tr}[W(b_\perp)W^\dagger(b'_\perp)] .\,\nn \\
\eea 
Here $p_k$  and $\eta_k$ denote the momentum and the rapidity of the final state soft gluon, respectively. We note that in this work, we assume all the signal jets are energetic with $E_J \gg Q_s \sim p_{J\perp}$, therefore the soft gluon itself could not form a jet that passes the jet criteria and therefore the soft gluon jet will not contribute to the jet spectrum and the $\Theta_2$ term will only be filled once to the histogram. However, if the condition is relieved, the soft jet could contribute to the  $p_{J\perp}$ spectrum.

The $8$-point correlator is defined as 
\bea 
S^{(8)}_{X_f}(b_\perp,r_\perp,b_\perp',r_\perp')
= \frac{1}{N_C^3}{\rm Tr}\left[f_{cde}W_{bd}^\dagger(r_\perp)W_{ae}(r_\perp')f_{cfg}W^\dagger_{bf}(b_\perp)W_{ag}(b_\perp')\right]  . 
\eea 
Similar to $S_{X_f}^{(6)}$, in the large $N_C$ limit, the $S_{X_f}^{(8)}$ can be written as products of the $4$-point function and the dipoles such that \bea 
S_{X_f}^{(8)} \approx 
S_{X_f}^{(4)}(b_\perp,b_\perp',r_\perp',r_\perp) S_{X_f}^{(2)}(r_\perp,r_\perp')S_{X_f}^{(2)}(b_\perp,b_\perp') \,,
\eea 
which is originated from the color identity that 
\bea\label{eq:color-identity-g} 
&& \frac{1}{N_C^3}{\rm Tr}\left[f_{cde}W_{bd}^\dagger(r_\perp)W_{ae}(r_\perp')f_{cfg}W^\dagger_{bf}(b_\perp)W_{ag}(b_\perp')\right]  .\nn \\ 
&=& \left( \frac{1}{N_C^3}
{\rm Tr}[W(b_\perp)W^\dagger(b_\perp')W(r_\perp')W^\dagger(r_\perp)]
{\rm Tr}[W(r_\perp)W^\dagger(r'_\perp)]{\rm Tr}[W(b_\perp)W^\dagger(b'_\perp)] \right.\nn \\
&&\left. - \frac{1}{N^3_C}  {\rm Tr}[W(b_\perp) W^\dagger(r'_\perp)W(r_\perp) W^\dagger(b_\perp)W(b'_\perp) W^\dagger(r_\perp)W(r'_\perp) W^\dagger(b'_\perp)] \right) \,.
\eea 

We can see that the $g\to g$ channel is very similar to the result of the $q \to q$ channel except for the splitting function and the color structure. In QCD, for the collinear limit, the only difference between cross sections of different channels is with respect to the splitting functions ~\cite{Catani:1998nv} . And the shock-wave introduced in CGC, which is the origin of the difference in dipole structure, will not change this property. For the soft limit, according to the conclusion of the soft theorem~\cite{Weinberg:1965nx}, the only difference between these two channels is the color factor.

%

\subsection{$g\to q$ channel}
For the $g\to q +\bar{q}$ channel case, we will suppose that the final state quark becomes the jet in 2-jets case, and then we can get the NLO correction of this channel to the jet cross section as
\bea\label{eq: gluonquark-NLO}
		d\sigma^{(1)}_{gq} & = &
\frac{\alpha_s N_f T_R}{\pi}
\int \frac{d b_\perp d b'_\perp}{4\pi^2}
\int_{\tau}^1 d\xi \,  
xG(x)
\, 
e^{-i   p_{J\perp} \cdot z_\perp} \nn \\
&& \hspace{-7.ex}\times 
\left( 
{\cal  H}_{gq,12}  \, \, 
S_X^{(2)}(b_\perp,b_\perp') + {\cal  H}_{gq,22}  \, \, 
S_X^{(2)}(b_\perp,b_\perp') S_X^{(2)}(b_\perp,b_\perp') + \int \frac{dr_\perp}{\pi} 
 {\cal H}_{gq,3} S_X^{(3)}(b_\perp,r_\perp,b_\perp')   \right)
\nn \\
&& \hspace{-7.ex}+ (d\sigma_{gq,fsr} - d\sigma^c_{gq,fsr}) + (d\sigma_{gq,isr}-d\sigma^c_{gq,isr})+ (d\sigma_{gq,inter.}-d\sigma^c_{gq,inter.}) 
\,,  
\eea 
where $\xi=\frac{p_{jq}^+}{p_g^+}$, $p_g$ is the momentum of the initial state gluon, $p_{jq}$ denotes the momentum of the final state quark.

The hard factors are
\bea\label{eq:Hgq12} 
{\cal H}_{gq,12} = 
\left(- P_{qg}^{(1)}(\xi) \ln \frac{z_\perp^2\mu^2}{c_0^2} -2\xi^2+2\xi \right)
 \,, 
\eea
\bea\label{eq:Hgq22} 
{\cal H}_{gq,22} = 
\left(-\frac{1}{3}\ln \frac{z_\perp^2\mu^2}{c_0^2}+\frac{1}{6}\right) \delta(1-\xi)
 \,, 
\eea
\bea\label{eq:Hgq3} 
{\cal H}_{gq,3} = &&P_{qg}^{(1)}(\xi) 
\frac{ z_\perp \cdot y_\perp}{z_\perp^{2}
y_\perp^{2}}\,
\frac{2}{\xi}
e^{-i \frac{p_{J\perp}}{\xi}  \cdot y_\perp},
\eea 
where $P^{(1)}_{qg}(\xi)=[\xi^2+(1-\xi)^2]$.

The results of terms in the last line of Eq.~\ref{eq: gluonquark-NLO} are
\bea\label{eq:theta1-0gq}
d\sigma_{gq,fsr}-d\sigma^c_{gq,fsr} &=&\frac{\alpha_s N_f T_R}{2\pi^2}
\int_{0}^1d\xi
P_{qg}^{(1)}(\xi) 
\int  d^2 p_{k\perp}  \Bigg\{2 \Theta_2 \, xG(x)\frac{\mathcal{F}_A(p_{k\perp}+ p_{J\perp};X_f)}{[\xi p_{k\perp}-(1-\xi)p_{J\perp}]^2} \nn \\
&& 
\hspace{-5.ex}+\Theta_1 \, \tau G(\tau)
\frac{\mathcal{F}_A(p_{J\perp};X_f)}{[p_{k\perp}-(1-\xi)p_{J\perp}]^2}
-\tau G(\tau)\frac{\mathcal{F}_A(p_{k\perp}+\xi p_{J\perp};X_f)}{[p_{k\perp}-(1-\xi)p_{J\perp}]^2}\Bigg\} \,,
\quad \quad 
\eea 
where the factor of $2$ before $\Theta_2$ accounts for the contribution from the ${\bar q}$ jet, which at NLO is identical to the $q$ jet case.

The interference term is found to be 
\bea \label{eq:ibkbgq}
d\sigma_{gq,inter.}-d\sigma_{gq,inter.}^c &=& - \frac{\alpha_s N_f T_R}{2\pi^2}
\int_{\tau}^1d\xi
 P_{qg}^{(1)}(\xi) 
\int d^{2}p_{k\perp}
\frac{d^{D-2}l_{\perp}}{(2\pi)^{D-4}} \, 
xG(x) \, 
\mathcal{F}_F(l_{\perp};X_f)
\nn \\
&&\hspace{-5.ex} \Bigg\{ (2 \Theta_2 
+\Theta_1 - 2 )\,  \, \mathcal{F}_F(p_{k\perp}+p_{J\perp}-l_{\perp};X_f)
\frac{2(\xi p_{k\perp}-(1-\xi)p_{j\perp})\cdot(l_{\perp}-p_{j\perp})}{(\xi p_{k\perp}-(1-\xi)p_{J\perp})^2(l_{\perp}-p_{j\perp})^2} \Bigg\}, \,\nn \\ \quad  
\eea 
and the contribution from the ISR gives
\bea\label{eq:if-subgq}
&&
d\sigma_{gq,isr} -
d \sigma_{gq,isr}^c
=   
\frac{\alpha_s N_f T_R}{2\pi^2}
\int^1 d\xi 
 \,    
 P_{qg}^{(1)}(\xi) 
\int
d^{2}p_{k\perp} xG(x)
\Big( \Theta_1 +  2 \Theta_2 - 2 \Big)
\nn \\ 
&& 
\times \int 
\frac{db_\perp db_\perp' }{4\pi^{2}}
\, 
\frac{d r_\perp 
d r_\perp'}{4\pi^{2}}
\,
 e^{-ip_{j\perp}\cdot z_\perp}
 e^{-i p_{k\perp} \cdot z_\perp'}
\frac{x_\perp \cdot x_\perp'}{x_\perp^{2} {x_\perp'}^{2}}
S_{X_f}^{(2)}(b_\perp,b_\perp') S_{X_f}^{(2)}(r_\perp,r_\perp') \,.  
\eea 
In both cases, the contribution from the ${\bar q}$ jet is included through the factor of $2$ before $\Theta_2$ and the counter terms.



\subsection{$q\to g$ channel}
For the $q\to g$ channel which means the gluon becomes the jet in 2-jets case of $q\to  q + g$ process, the NLO correction to the jet cross section is found to be 
\bea\label{eq: quarkgluon-NLO}
		d\sigma^{(1)}_{qg} & = &
\frac{\alpha_sN_C}{2\pi}
\int \frac{d b_\perp d b'_\perp}{4\pi^2}
\int_{\tau}^1 d\xi \,  
xf(x)
\, 
e^{-i   p_{J\perp} \cdot z_\perp} 
{\cal  H}_{qg,22}  \, \, 
S_X^{(2)}(b_\perp,b_\perp') S_X^{(2)}(b_\perp,b_\perp')  
\nn \\
&& + d\sigma_{qg,fsr} + (d\sigma_{qg,isr}-d\sigma^c_{qg,isr})+ d\sigma_{qg,inter.}
\,,  
\eea 
where $\xi=\frac{p_{jg}^+}{p_q^+}$ and $p_q$ is the momentum of the initial state quark, $p_{jg}$ denotes the momentum of the final state gluon.

The hard factor is
\bea\label{eq:Hqg22} 
{\cal H}_{qg,22} = 
\left(- P_{gq}^{(1)}(\xi) \ln \frac{z_\perp^2\mu^2}{c_0^2} +\xi \right)
 \,, 
\eea
where $\mathcal{P}_{gq}(\xi)=\frac{1}{\xi}[1+(1-\xi)^2]$.

The results of terms in the last line of Eq.~\ref{eq: quarkgluon-NLO} are
\bea\label{eq:theta1-0qg}
d\sigma_{qg,fsr} &=&\frac{\alpha_sN_C}{2\pi^2}
\int_{\tau}^1d\xi
P_{gq}^{(1)}(\xi) 
\int  d^2 p_{k\perp}  \Theta_2 \, xf(x)\frac{\mathcal{F}_F(p_{k\perp}+ p_{J\perp};X_f)}{[\xi p_{k\perp}-(1-\xi)p_{J\perp}]^2}  \,.\nn \\
\quad \quad 
\eea 
\bea \label{eq:ibkbqg}
d\sigma_{qg,inter.} &=& - \frac{\alpha_sN_C}{2\pi^2}
\int_{\tau}^1d\xi
 P_{gq}^{(1)}(\xi) 
\int d^{2}p_{k\perp}
\frac{d^{D-2}l_{\perp}}{(2\pi)^{D-4}} \, 
xf(x) \, 
\Theta_2 \mathcal{F}_F(p_{k\perp}+p_{J\perp};X_f)
\nn \\
&&\hspace{-5.ex} 
 \,  \, \mathcal{F}_F(p_{k\perp}+p_{J\perp} - l_{\perp};X_f)
\frac{2(\xi p_{k\perp}-(1-\xi)p_{J\perp})\cdot(l_{\perp}-p_{J\perp})}{(\xi p_{k\perp}-(1-\xi)p_{J\perp})^2(l_{\perp}-p_{J\perp})^2} \,,\nn \\ \quad  
\eea 
and
\bea\label{eq:if-subqg}
&&
d\sigma_{qg,isr} -
d \sigma_{qg,isr}^c
=   
\frac{\alpha_sN_C}{2\pi^2}
\int^1 d\xi 
 \,    
 P_{gq}^{(1)}(\xi) 
\int
d^{2}p_{k\perp} xf(x)
\Big( \Theta_2 - 1 \Big)
\nn \\ 
&& 
\times \int 
\frac{db_\perp db_\perp' }{4\pi^{2}}
\, 
\frac{d r_\perp 
d r_\perp'}{4\pi^{2}}
\,
 e^{-ip_{j\perp}\cdot z_\perp}
 e^{-i p_{k\perp} \cdot z_\perp'}
\frac{x_\perp \cdot x_\perp'}{x_\perp^{2} {x_\perp'}^{2}}
S_{X_f}^{(6)}(r_\perp,b_\perp,r_\perp',b_\perp')  \,.  
\eea 
Here $p_k$ denotes the momentum of the final state quark. Under this definition $\Theta_2$  has the same form as the $q \to q$ channel cases in terms of $p_k$ ,$\xi$ and $p_J$.


\section{The small-$R$ limit for other channels channel}~\label{sec:small-R-ohter}
In this section, we list all the remaining channels that contribute to the result of jet production in the small-$R$ limit, which are the $g\to g$,$q\to g$ and $g\to q$ sub-processes. 

\subsection{$g\to g$ channel}~\label{sec:small-R-g}
In the small-$R$ limit, by performing subtractions similar to the method in Section~\ref{sec:smallR} and omitting ${\cal O}(R^2)$ contributions, we can get the factorized form for the $g\to g$ channel as 
\bea\label{eq:small-r-factorization-g} 
&& d\sigma_{g,R}
= \int d\xi
\frac{d\zeta}{\zeta^2}  xG(x)
d\hat{\sigma}_{g\to g}(\xi,p_{J}/\zeta) \, J_g(\zeta)  \,, 
\eea 
where $x=\tau/\xi\zeta$ and $J_g(\zeta)$ is the gluon siJF~\cite{Kang:2016mcy,Dai:2016hzf} in the large $N_C$ limit, which is 
\bea \label{eq:siJF-g}
J_g(\zeta) = 
\delta(1-\zeta)  & - &
\frac{\alpha_s }{2\pi}N_C
\, 
\left\{    \Big( P^{(1)}_{gg}(\zeta) +\frac{2N_f T_R}{N_c} P_{qg}(\zeta) \Big)\ln \frac{p^2_{J\perp}R^2}{\zeta^2\mu^2}
+ 4 \frac{(1-\zeta+\zeta^2)^2}{\zeta} \left(\frac{\ln (1-\zeta) }{1-\zeta} \right)_+ \right. \nn \\
&& \left.    
- \left(
\frac{67}{9} - \frac{2}{3}\pi^2-\frac{N_f T_R}{N_c}\frac{23}{9} 
\right)\delta(1-\zeta) 
+ \frac{4 N_f T_R}{N_c}[P^{(1)}_{qg}(\zeta) \ln(1-\zeta) + \zeta(1-\zeta)]
\right\},\nn \\
\,
\eea 
and we have included the contribution from the $g\to q \bar{q}$ channel,  $P^{(1)}_{qg}(\zeta)=[\zeta^2+(1-\zeta)^2]$.

$d\hat{\sigma}_{g\to g}(p_{J\perp}/\zeta)$ is the cross section to produce a gluon with momentum $p_{j\perp} = p_{J\perp}/\zeta$. The explicit results at LO and NLO are given by
\bea
d\hat{\sigma}_{g\to g}^{(0)}(\xi,p_j)
 =  
 \int \frac{d^{2}b_\perp d^{2}b_\perp'}{(2\pi)^{2}}  \,
 e^{-i p_{j\perp} \cdot z_\perp} 
 S_{X_f}^{(2)}(b_\perp,b_\perp')S_{X_f}^{(2)}(b_\perp,b_\perp')
 \,, 
 \eea 
 and 
\bea\label{eq: gluon-NLO-small-r}
		d\hat{\sigma}^{(1)}_{g\to g}(p_j) & = &
\frac{\alpha_s}{2\pi}N_C
\int \frac{d b_\perp d b'_\perp}{4\pi^2}
\int_{\tau}^1 d\xi \,  
xG(x)
\, 
e^{-i   p_{j\perp} \cdot z_\perp} \nn \\
&& \times 
\left( 
{\cal  H}^{g\to g}_{2}  \, \, 
S_X^{(2)}(b_\perp,b_\perp') S_X^{(2)}(b_\perp,b_\perp') + {\cal  H}^{g\to q\bar{q}}_{2}  \, \, 
S_X^{(3)}(b_\perp,r_\perp, b_\perp')  \right. \nn \\
&& \left. + \int \frac{dr_\perp}{\pi} 
\Big( 
{\cal H}^{g\to g}_{BK}
+
{\cal H}^{g\to g}_{3} 
+
{\cal H}^{g\to g}_{kin.}
\Big) 
S_X^{(2)}(b_\perp,b_\perp') S_X^{(2)}(b_\perp,r_\perp)S_X^{(2)}(r_\perp,b_\perp')   \right),
\nn \\
\eea 
where ${\cal H}_{2}^{g\to q\bar{q}}$, ${\cal H}_{3}^{g\to g}$, ${\cal H}^{g\to g}_{BK}$ and ${\cal H}^{g\to g}_{kin.}$ can be obtained by replacing $p_{J\perp}$ with $p_{j\perp}$ in Eq.~(\ref{eq:Hqqbar2}), Eq.~(\ref{eq:Hg3}),  Eq.~(\ref{eq:HgBK}) and Eq.~(\ref{eq:Hgkin}), respectively. The ${\cal H}^{g\to g}_2$ is given by
\bea\label{eq:Hg2-small-r} 
{\cal H}^{g\to g}_{2}& =& 
\left\{- P_{gg}^{(1)}(\xi) \ln \frac{z_\perp^2\mu^2}{c_0^2} 
+\left[\left(\frac{11}{6}-\frac{2 N_f T_R}{3 N_C}\right)\ln \frac{z_\perp^2p_{J\perp}^2}{c_0^2} \right.\right.
\nn \\&&\left.\left.-\frac{N_f T_R}{3N_C}\right]\delta(1-\xi) 
\right\}\left( 
1+ \frac{1}{\xi^2}
e^{-i\frac{1-\xi}{\xi}p_{j\perp}\cdot z_\perp}
\right) \,. 
\eea
We note that $d\hat{\sigma}_{g\to g}$ is the gluonic cross section for the single hadron production. 

\subsection{$g\to q(\bar{q})$ channel}~\label{sec:small-R-gq}
The contribution of the $g\to q$ channel to the small-$R$ limit result is found to be 
\bea\label{eq:small-r-factorization-gq} 
&& d\sigma_{gq,R}
= 2\int d\xi
\frac{d\zeta}{\zeta^2}  xG(x)
d\hat{\sigma}_{g\to q}(\xi,p_{J}/\zeta) \, J_q(\zeta)  \,, 
\eea 
where $x=\tau/\xi\zeta$ and $J_q(\zeta)$ is the quark siJF~\cite{Kang:2016mcy,Dai:2016hzf} in the large $N_C$ limit whose form is given by Eq.~(\ref{eq:siJF}). The factor 2 comes from summing up the contributions from the $g \to q$ channel and $g \to \bar{q}$ channel, which are the same at this order. Here, $d\hat{\sigma}_{g\to q}(p_{J\perp}/\zeta)$ is the cross section to produce a quark(anti-quark) with momentum $p_{j\perp} = p_{J\perp}/\zeta$ when the initial state parton is a gluon . The explicit results at NLO is given by
\bea\label{eq: gluonquark-NLO-small-r}
		d\sigma^{(1)}_{g \to q}(p_j) & = &
\frac{\alpha_s N_f T_R}{2\pi}
\int \frac{d b_\perp d b'_\perp}{4\pi^2}
\int_{\tau}^1 d\xi \,  
xG(x)
\, 
e^{-i   p_{j\perp} \cdot z_\perp} \nn \\
&& \hspace{-7.ex}\times 
\left( 
{\cal  H}^{g \to q}_{12}  \, \, 
S_X^{(2)}(b_\perp,b_\perp') + {\cal  H}^{g \to q}_{22}  \, \, 
S_X^{(2)}(b_\perp,b_\perp') S_X^{(2)}(b_\perp,b_\perp') + \int \frac{dr_\perp}{\pi} 
 {\cal H}^{g \to q}_{3} S_X^{(3)}(b_\perp,r_\perp,b_\perp')   \right),\nn
 \\
\eea 

where ${\cal H}_{12}^{g\to q}$ and ${\cal H}^{g\to q}_{3}$  can be obtained by replacing $p_{J\perp}$ with $p_{j\perp}$ in  Eq.~(\ref{eq:Hgq12}) and Eq.~(\ref{eq:Hgq3}) respectively. The ${\cal H}^{g\to q}_{22}$ is given by

\bea\label{eq:Hgtoq22} 
{\cal H}^{g\to q}_{22} = 
\left(- P_{qg}^{(1)}(\xi) \ln \frac{z_\perp^2\mu^2}{c_0^2} -2\xi^2+2\xi \right)\frac{1}{\xi^2}
e^{-i\frac{1-\xi}{\xi}p_{j\perp}\cdot z_\perp}
 \,, 
\eea

\subsection{$q\to g$ channel}~\label{sec:small-R-qg}
The contribution of the $q\to g$ channel to the small-$R$ limit result is found to be 
\bea\label{eq:small-r-factorization-qg} 
&& d\sigma_{qg,R}
= \int d\xi
\frac{d\zeta}{\zeta^2}  xf(x)
d\hat{\sigma}_{q\to g}(\xi,p_{J}/\zeta) \, J_g(\zeta)  \,, 
\eea 
where $x=\tau/\xi\zeta$ and $J_g(\zeta)$ is the gluon siJF~\cite{Kang:2016mcy,Dai:2016hzf} in the large $N_C$ limit whose form is given by Eq.~(\ref{eq:siJF-g}).

$d\hat{\sigma}_{q\to g}(p_{J\perp}/\zeta)$ is the cross section to produce a gluon with momentum $p_{j\perp} = p_{J\perp}/\zeta$ when the initial state parton is a qurak. The explicit results at NLO is given by

\bea\label{eq: quarkgluon-NLO-small-r}
		d\sigma^{(1)}_{qg} & = &
\frac{\alpha_sN_C}{2\pi}
\int \frac{d b_\perp d b'_\perp}{4\pi^2}
\int_{\tau}^1 d\xi \,  
xf(x)
\, 
e^{-i   p_{j\perp} \cdot z_\perp} \nn \\
&& \hspace{-7.ex}\times 
\left( 
{\cal  H}^{q \to g}_{12}  \, \, 
S_X^{(2)}(b_\perp,b_\perp')+ {\cal  H}^{q \to g}_{22}  \, \, 
S_X^{(2)}(b_\perp,b_\perp')S_X^{(2)}(b_\perp,b_\perp')   + \int \frac{dr_\perp}{\pi} 
 {\cal H}^{q \to g}_{3} S_X^{(3)}(b_\perp,r_\perp,b_\perp')   \right),\nn
 \\
\eea 

where ${\cal H}_{22}^{q\to g}$ can be obtained by replacing $p_{J\perp}$ with $p_{j\perp}$ in  Eq.~(\ref{eq:Hqg22}). The ${\cal H}^{q\to g}_{12}$ and ${\cal H}^{q\to g}_{3}$ are given by
\bea\label{eq:Hqtog12} 
{\cal H}^{q \to g}_{12} = 
\left(- P_{gq}^{(1)}(\xi) \ln \frac{z_\perp^2\mu^2}{c_0^2} +\xi \right)\frac{1}{\xi^2}
e^{-i\frac{1-\xi}{\xi}p_{j\perp}\cdot z_\perp}
 \,, 
\eea
and
\bea\label{eq:Hqtog3} 
{\cal H}^{q \to g}_{3} = &&P_{gq}^{(1)}(\xi) 
\frac{ z_\perp \cdot y_\perp}{z_\perp^{2}
y_\perp^{2}}\,
\frac{2}{\xi}
e^{-i \frac{1-\xi}{\xi} p_{J\perp} \cdot z_\perp -i p_{J\perp} \cdot y_\perp},
\eea 
respectively.

\bibliographystyle{JHEP}
\bibliography{references}

\end{document}